\documentclass[titlepage,11pt,amstex]{article}
\usepackage{amssymb}
\usepackage{amsmath}
\usepackage{eurosym}
\usepackage{comment}

\usepackage{multirow} 
\usepackage[utf8]{inputenc}
\usepackage{amssymb}
\usepackage{amsfonts}
\usepackage{amsmath}
\usepackage{mathptmx}
\usepackage{eurosym}
\usepackage{graphicx}
\usepackage{rotating}
\usepackage[justification=centering]{caption}
\usepackage{comment}

\usepackage{ifthen} 
\usepackage{xifthen} 

\usepackage{calc}  
\usepackage{time}  
\usepackage{clock}  

\usepackage{harvard}

\providecommand{\cite}{\citeasnoun}
\renewcommand{\cite}{\citeasnoun}

\usepackage{bibunits}

\usepackage{fancybox}
  
\usepackage{adjustbox} 

\usepackage[justification=centering]{caption} 

\usepackage{section}
\usepackage{theorem}
\usepackage{verbatim}
\providecommand{\bTentative}{}

\usepackage{eurosym} 
\usepackage{fix-cm}  
\usepackage{textcomp} 
\usepackage{pifont}  
\usepackage[T1]{fontenc}
    
\providecommand{\SpecialFonts}{
    \usepackage{calligra} 
	
    \usepackage[varumlaut]{yfonts}  
    }

\usepackage{pdfpages} 

\usepackage{gloss}
\usepackage{makeidx} 

\usepackage{enumerate}  
\usepackage{etaremune}  

\usepackage{amssymb}  
\usepackage{amsfonts}  
\usepackage{amsmath}  

\usepackage{array}
\usepackage{bbold} 
\usepackage{dsfont}  
\usepackage{fixmath} 
\usepackage{mathtools}  
\usepackage{mathrsfs} 

\usepackage{endnotes}

\usepackage{afterpage} 
\usepackage{layout} 
\usepackage{lscape} 

\usepackage{setspace} 

\usepackage{url}  

\usepackage{longtable} 

\usepackage{subfigure}

\providecommand{\paraNumbering}{\theoremstyle{change}}
\SpecialFonts
\excludecomment{Tentative}
\excludecomment{Plan}
\excludecomment{Abstract}
\excludecomment{Advanced}
\excludecomment{DP}
\excludecomment{graphicpack}
\excludecomment{ps2pdf}
\includecomment{pdflatex}
\includecomment{Alternative}
\includecomment{Old}
\includecomment{Elementary}
\includecomment{Intermediate}
\includecomment{French}
\includecomment{reversemargin}
\excludecomment{SlidesIn} \includecomment{SlidesOut} \includecomment{Slides}
\includecomment{SlidesLandscapeOut} \includecomment{SlidesLandscape}
\includecomment{AbstractF}
\includecomment{TOCLists}
\includecomment{Calculations}
\includecomment{Article}
\includecomment{Book}
\includecomment{DP-W}
\includecomment{DP-CIR}
\includecomment{DP-SSRN}
\includecomment{DP-arXiv}
\includecomment{Journal}
\includecomment{Lectures}
\includecomment{Organization}

\newcommand{\cbu}{,\xspace}
\newcommand{\dbu}{.\xspace}
\newcommand{\bbibuniteconometrica}{\begin{bibunit}[econometrica] 
    \renewcommand{\cite}{\nocite*}
    \renewcommand\refname{} \renewcommand{\cbu}{} 
    \renewcommand{\dbu}{} \renewcommand{\newline}{}
    \vspace{-3.5\baselineskip}}
\newcommand{\bbibunitagsm}{\begin{bibunit}[agsm] \renewcommand{\cite}{\nocite*}
    \renewcommand\refname{} \renewcommand{\cbu}{} \renewcommand{\dbu}{} 
    \renewcommand{\newline}{}
    \vspace{-3.5\baselineskip}}
\newcommand{\ebibunit}{\putbib[refer1] \end{bibunit}}

\newcommand{\bbibunitunsrt}{\begin{bibunit}[unsrt] \renewcommand{\cite}{\nocite*}
    \renewcommand\refname{} \renewcommand{\cbu}{} \renewcommand{\dbu}{} \renewcommand{\newline}{}
    \vspace{-3.5\baselineskip}}
\newcommand{\ebibunitunsrt}{\putbib[refer1] \end{bibunit}}

\makeindex  
\makeglossary  
\makegloss

\newcommand{\NumericNumberedLists}{
\def\labelenumi{\arabic{enumi}.}
\def\theenumi{\arabic{enumi}}
\def\labelenumii{\arabic{enumii}.}
\def\theenumii{\arabic{enumii}}
\def\p@enumii{\theenumi.}
\def\labelenumiii{\arabic{enumiii}.}
\def\theenumiii{\arabic{enumiii}}
\def\p@enumiii{\theenumi.\theenumii.}
\def\labelenumiv{\arabic{enumiv}.}
\def\theenumiv{\arabic{enumiv}}
\def\p@enumiv{\p@enumiii.\theenumiii}
}

\allowdisplaybreaks

\newlength{\totalhormargin}
\newlength{\totalvermargin}

\newcommand{\February}{February }
\newcommand{\March}{March }

\newcommand{\July}{July }

\def\todayMY{\ifcase\month\or
  January\or February\or March\or April\or May\or June\or
  July\or August\or September\or October\or November\or
  December\fi\ \number\year}

\providecommand{\vartitleadjust}{}
\providecommand{\vartitle}{}

\providecommand{\varAuthors}{Jean-Marie Dufour \thanks{\ \ \DufourAddress} \\
  McGill University}

\providecommand{\varnewline}{ \\ }
\providecommand{\firstversion}[1]{First version: #1 }
\providecommand{\revised}[1]{\varnewline Revised: #1 }
\providecommand{\thisversion}[1]{\varnewline This version: #1 }
\providecommand{\compiled}[1]{\varnewline Compiled: #1 }

\providecommand{\vardate}{\today, \texttime}

\providecommand{\DufourAddress}{William Dow Professor of Economics, McGill University,
  Centre interuniversitaire de recherche en analyse des
  organisations (CIRANO), and Centre interuniversitaire de recherche en
  \'{e}conomie quantitative (CIREQ). Mailing address:
  Department of Economics, McGill University, Leacock Building, Room 414,
  855 Sherbrooke Street West, Montr\'{e}al, Qu\'{e}bec H3A 2T7, Canada.
  TEL: (1) 514 398 6071; FAX: (1) 514 398 4800; e-mail: 
  \protect\url=jean-marie.dufour@mcgill.ca=\thinspace. Web page:
  \protect\url{http://www.jeanmariedufour.com} }

\newcommand{\sptha}{\hspace{-0.01em}}
\newcommand{\spthb}{\hspace{-0.01em}}

\newcommand{\sppr}{}
\newcommand{\theoremname}{Theorem}
\newcommand{\acknowledgementname}{Acknowledgement}
\newcommand{\algorithmname}{Algorithm}
\newcommand{\assumptionname}{Assumption}
\newcommand{\axiomname}{Axiom}
\newcommand{\casename}{Case}
\newcommand{\claimname}{Claim}
\newcommand{\conclusionname}{Conclusion}
\newcommand{\conditionname}{Condition}
\newcommand{\conjecturename}{Conjecture}
\newcommand{\corollaryname}{Corollary}
\newcommand{\criterionname}{Criterion}
\newcommand{\definitionname}{Definition}
\newcommand{\examplename}{Example}
\newcommand{\exercisename}{Exercise}
\newcommand{\lemmaname}{Lemma}
\newcommand{\notationname}{Notation}
\newcommand{\problemname}{Problem}
\newcommand{\proofname}{\sppr Proof}

\newcommand{\propertyname}{\sppr Property}
\newcommand{\propositionname}{Proposition}
\newcommand{\reflistname}{References}
\newcommand{\remarkname}{Remark}
\newcommand{\resultname}{Result}
\newcommand{\solutionname}{Solution}

\newcommand{\summaryname}{Summary}

\newenvironment{proof}[1][\sptha \proofname]{\par
  \normalfont
  \trivlist
  \item[\hskip\labelsep\scshape
    #1{.}]\ignorespaces}
    {\qed\endtrivlist\vspace{\baselineskip}}

\newenvironment{proofnoname}[1][\sptha \proofname]{\par
  \noindent
  \normalfont}%
  {\qed\vspace{\baselineskip}}

\newenvironment{proofflex}[1][\spthb \proofname]{\par
  \normalfont
  \trivlist
  \item[\hskip\labelsep\scshape
    #1{\ }]\ignorespaces}
    {\qed\endtrivlist\vspace{\baselineskip}}

\newenvironment{proofflexb}[1][\spthb \proofname]{\par
  \normalfont
  \trivlist
  \item[\hskip\labelsep\scshape
    #1{\ }]\ignorespaces}
    {\qed\endtrivlist\vspace{\baselineskip}}

\newenvironment{proofflexc}[1][\spthb \proofname]{\par
  \noindent
  \normalfont}
  {\qed\vspace{\baselineskip}}

\newenvironment{proofflexwc}[1][\spthb \proofname]{\par
  \normalfont
  \trivlist
  \item[\hskip\labelsep\scshape
    #1{\ }]\ignorespaces}

\newenvironment{proofflexd}[1][\spthb \proofname]{\par
  \noindent
  \normalfont}
  {\vspace{-0.5\baselineskip}}

\newenvironment{sec2subsec}{\renewcommand{\section}{\subsection}}{}
\newenvironment{reflist}{\quad \newline \noindent {\Large \bf \reflistname}
  \newline \quad \newline
  \begin{list}{}{\itemindent -.15in \leftmargin .15in \parsep 0in
                 \itemsep 0in} \item \vspace{-.35in} }{\end{list}}

\newenvironment{npar}{\noindent \bf{\thesection.\thenpar}}{}
\newenvironment{subnpar}{\noindent \bf{\thesubsection.\thesubnpar}}{}

\newcounter{paran}

\newcounter{npar}[section]
\newcounter{nparr}[section]

\newcounter{subnpar}[subsection]
\newcounter{subnparr}[subsection]

\DeclareRobustCommand{\qed}{%
  \ifmmode 
  \else \leavevmode\unskip\penalty9999 \hbox{}\nobreak\hfill
  \fi
  \quad\hbox{\qedsymbol}}
\newcommand{\openbox}{\leavevmode
  \hbox to.77778em{%
  \hfil\vrule
  \vbox to.675em{\hrule width.6em\vfil\hrule}%
  \vrule\hfil}}
\DeclareRobustCommand{\qeddirect}{%
  \ifmmode 
  \else \leavevmode\unskip\penalty9999 \hbox{}\nobreak\hfill
  \fi
  \quad\hbox{\qedsymbol}}
\providecommand{\qedsymbol}{\openbox} 

\newcommand{\bproofin}{\begin{proof}}
\newcommand{\eproofin}{\end{proof}}
\newcommand{\bproofend}{\begin{proofnoname}}
\newcommand{\eproofend}{\end{proofnoname}}
\newcommand{\bproofth}{\begin{proofth}}
\newcommand{\eproofth}{\end{proofth}}

\newcommand{\thsection}{\thesection}

\newcommand{\thsectioneq}{\thesection}

\newcommand{\thsec}{\thsection}
\newcommand{\thseceq}{\thsectioneq}

\renewcommand{\theequation}{{\rm \thseceq.\arabic{equation}}}

\makeatletter
   \long
\def\@makecaption#1#2{\vskip 0\p@
   \setbox\@tempboxa\hbox{#1 #2}}
\makeatother

\makeatletter

\providecommand{\l@theorem}{\@dottedtocline{1}{0em}{5em}}
\renewcommand{\l@theorem}{\@dottedtocline{1}{0em}{5em}}

\providecommand{\listttheoremnameb}{\sectitlesize List of Definitions, Assumptions, Propositions and Theorems}

\newcommand\listoftheorems{
 \section*{\listttheoremnameb
           \@mkboth{\MakeUppercase\listttheoremnameb}
           {\MakeUppercase\listttheoremnameb}}
           \@starttoc{lth}
           }
\newcommand{\listoftheoremscont}[1]{
 \providecommand{\listttheoremnameb}{#1}
 \renewcommand{\listttheoremnameb}{#1}
 \section*{\listttheoremnameb
           \@mkboth{\MakeUppercase\listttheoremnameb}
           {\MakeUppercase\listttheoremnameb}}
           \@starttoc{lth}
                      \addcontentsline{toc}{section}{\listttheoremnameb}
           }

\makeatother
\providecommand{\contentshift}{\hspace{-3.5em}}
\renewcommand{\contentshift}{\hspace{-3.5em}}
\providecommand{\contentshiftS}{\hspace{0em}}
\renewcommand{\contentshiftS}{\hspace{0em}}

\newcommand{\captionproofflex}[2]{}

\newcommand{\bAssumptionA}{\begin{assumption}}
\newcommand{\eAssumptionA}{\end{assumption}}

\providecommand{\paraNumbering}{}
\renewcommand{\paraNumbering}{}
\paraNumbering

\newtheorem{theorem}{\sptha \theoremname}[section]
\newtheorem{theoremseq}{\sptha \theoremname}
{\theorembodyfont{\normalfont}%
 \newtheorem{theoremSeqRom}{\sptha \theoremname}}
\newtheorem{acknowledgement}{\sptha \acknowledgementname}[section]
\newtheorem{algorithm}{\sptha \algorithmname}[section]
\newtheorem{assumption}{\sptha \assumptionname}[section]
\newtheorem{assumptionSec}{\sptha \assumptionname}[section]
\newtheorem{assumptionV}{\sptha \assumptionname}[section]
\newtheorem{assumptionLetter}{\sptha \assumptionname}

\newtheorem{axiom}{\sptha \axiomname}[section]
\newtheorem{case}{\sptha \casename}[section]
\newtheorem{claim}{\sptha \claimname}[section]
\newtheorem{conclusion}{\sptha \conclusionname}[section]
\newtheorem{condition}{\sptha \conditionname}[section]
\newtheorem{conjecture}{\sptha \conjecturename}[section]
\newtheorem{corollary}[theorem]{\sptha \corollaryname}
\newtheorem{criterion}{\sptha \criterionname}[section]
\newtheorem{definition}{\sptha \definitionname}[section]
\newtheorem{definitionSec}{\sptha \definitionname}[section]
{\theorembodyfont{\normalfont}%
    \newtheorem{example}{\sptha \examplename}[section]}
{\theorembodyfont{\normalfont}
    \newtheorem{exampleSec}{\sptha \examplename}[section]}
\newtheorem{exercise}{\sptha \exercisename}[section]
\newtheorem{lemma}[theorem]{\sptha \lemmaname}
\newtheorem{lemmaSec}{\sptha \lemmaname}[section]
\newtheorem{notation}{\sptha \notationname}[section]
\newtheorem{problem}{\sptha \problemname}[section]
{\theorembodyfont{\normalfont} 
    \newtheorem{proofth}{\sptha \proofname}[section]}
\newtheorem{property}[theorem]{\sptha \propertyname}
\newtheorem{proposition}[theorem]{\sptha \propositionname}
\newtheorem{propositionSec}{\sptha \propositionname}[section]
{\theorembodyfont{\normalfont} 
    \newtheorem{remark}{\sptha \remarkname}[section]}
    {\theorembodyfont{\normalfont}%
\newtheorem{result}{\sptha \resultname}[section]}
\newtheorem{solution}{\sptha \solutionname}[section]
\newenvironment{statement}{}{}
\newtheorem{summary}{\sptha \summaryname}[section]

\providecommand{\resetcountersSection}{\renewcommand{\thsec}{\thsection} 
  \renewcommand{\thseceq}{\thsectioneq} 
  \setcounter{theorem}{0} \setcounter{definition}{0} \setcounter{equation}{0}}

\newcommand{\captionassumption}[2]{\textsc{ #2}.
  \addcontentsline{lth}{theorem}{{\bf #1 \contentshiftS  \protect\numberline{\theassumption}}
  {\contentshift  : \contentshiftS #2}}}

\newcommand{\captionlemma}[2]{\textsc{ #2}.
 \addcontentsline{lth}{theorem}{{\bf #1 \contentshiftS \protect\numberline{\thelemma}}
 {\contentshift  : \contentshiftS #2}}}

\newcommand{\captionproofemptynocontent}[2]{}

\newcommand{\captionproofin}[2]{}

\providecommand{\vartitleadjust}{}
\renewcommand{\vartitleadjust}{}

\providecommand{\varnewline}{ \\ }
\renewcommand{\varnewline}{}

\providecommand{\vartitle}{\vartitleadjust Simple robust two-stage estimation and inference \\
 for generalized impulse responses and multi-horizon causality}
\renewcommand{\vartitle}{\vartitleadjust Simple robust two-stage estimation and inference \\
 for generalized impulse responses and multi-horizon causality}

\providecommand{\DufourAddress}{William Dow Professor of Economics, McGill University,
  Centre interuniversitaire de recherche en analyse des
  organisations (CIRANO), and Centre interuniversitaire de recherche en
  \'{e}conomie quantitative (CIREQ). Mailing address:
  Department of Economics, McGill University, Leacock Building, Room 414,
  855 Sherbrooke Street West, Montr\'{e}al, Qu\'{e}bec H3A 2T7, Canada.
  TEL: (1) 514 398 6071; FAX: (1) 514 398 4800; e-mail: 
  \protect\url=jean-marie.dufour@mcgill.ca=\thinspace. Web page:
  \protect\url{http://www.jeanmariedufour.com} }
  
\providecommand{\EndongAddress}{Department of Economics, McGill University, Montr\'eal, Qu\'ebec H3A2T7, Canada. TEL: (1) 514 772 7078; e-mail:
\protect\url=endong.wang@mail.mcgill.ca=\thinspace. Web page:
  \protect\url{http://www.endongwang.com}}
  
\renewcommand{\EndongAddress}{Department of Economics, McGill University, Montr\'eal, Qu\'ebec H3A2T7, Canada. TEL: (1) 514 772 7078; e-mail:
\protect\url=endong.wang@mail.mcgill.ca=\thinspace. Web page:
  \protect\url{http://www.endongwang.com}}

\providecommand{\varAuthors}{Jean-Marie Dufour \thanks{\ \ \DufourAddress} \\
 McGill University}
\renewcommand{\varAuthors}{Jean-Marie Dufour \thanks{\ \ \DufourAddress} \\
 McGill University \and Endong Wang\thanks{\ \ \EndongAddress} \\
  McGill University}

\providecommand{\vardate}{\today, \texttime}
\providecommand{\vardate}{\dateformat{\February 2023}{\revised{\March 2024}}{\thisversion{\July 2024}}}
\renewcommand{\vardate}{\firstversion{\February 2023} \\ \revised{\March 2024} 
     \\ \thisversion{\July 2024} \\ \compiled{\today, \texttime}}

\providecommand{\listttheoremnameb}{List of Definitions, Assumptions, Propositions 
     and Theorems}
\renewcommand{\listttheoremnameb}{List of Definitions, Assumptions, Propositions 
     and Theorems}
	 
\renewcommand{\thefootnote}{\alph{footnote}}

\def \beginenumerateT {\begin{enumerate}}
\def \endenumerateT {\end{enumerate}}

\def \beginenumerateTh[#1] {\begin{enumerate}[#1]}
\def \endenumerateTh {\end{enumerate}}

\begin{French}

\end{French}

\begin{graphicpack}

\end{graphicpack}

\begin{DP}

\end{DP}

\begin{DP-W}

\end{DP-W}

\begin{DP-CIR}

\end{DP-CIR}

\begin{DP-SSRN}

\end{DP-SSRN}

\begin{DP-arXiv}

\end{DP-arXiv}

\begin{Lectures} 

\end{Lectures}

\begin{Organization}

\end{Organization}

\begin{reversemargin}

\end{reversemargin}

\begin{Article} 

\end{Article}

\begin{Journal}

\end{Journal}

\input{tcilatex.sty}

\excludecomment{Tentative}
\input{tcilatex}
\begin{document}

\title{%
\vartitle%
\footnote{ \quad This work was supported by the William Dow Chair in Political
  Economy (McGill University), the Social Sciences and Humanities Research
  Council of Canada, and the Fonds de recherche sur la soci\'{e}t\'{e}
  et la culture (Qu\'{e}bec).} }
\author{%
\varAuthors%
}
\date{%
\vardate%
}

\maketitle

\date{\today}
\begin{abstract}
This paper introduces a novel two-stage estimation and inference procedure for \textit{generalized impulse responses} (GIRs). GIRs encompass all coefficients in a multi-horizon linear projection model of future outcomes of $y$ on lagged values (Dufour and Renault, 1998), which include the Sims’ impulse response. The conventional use of Least Squares (LS) with heteroskedasticity- and autocorrelation-consistent covariance estimation is less precise and often results in unreliable finite sample tests, further complicated by the selection of bandwidth and kernel functions. Our two-stage method surpasses the LS approach in terms of estimation efficiency and inference robustness. The robustness stems from our proposed covariance matrix estimates, which eliminate the need to correct for serial correlation in the multi-horizon projection residuals. Our method accommodates non-stationary data and allows the projection horizon to grow with sample size. Monte Carlo simulations demonstrate our two-stage method outperforms the LS method. We apply the two-stage method to investigate the GIRs, implement multi-horizon Granger causality test, and find that economic uncertainty exerts both short-run (1-3 months) and long-run (30 months) effects on economic activities.
\quad


\quad

\noindent \textbf{JEL \thinspace\ classification}: C3; C12; C15; C53.
\end{abstract}

\newpage

\tableofcontents%

\newpage

\listoftables%

\listoffigures%

\listoftheorems
\addcontentsline{toc}{section}{\listttheoremnameb}%

\newpage

\pagenumbering{arabic} \setcounter{section}{0} \setcounter{page}{1} 

\renewcommand{\thefootnote}{\arabic{footnote}}%

\pagestyle{headings}

\newpage
\section{Introduction \label{Sec: Introduction}}

\resetcountersSection
The concept of causality introduced by \cite{wiener1956theory} and \cite{granger1969investigating}, often referred to as Granger causality, constitutes a foundational approach for studying dynamic relationships between time series. While \cite{granger1969investigating} primarily focused on bivariate causality at a single horizon, the concept was generalized by \cite{dufour1998short} to account for multi-horizon causality.\footnote{For related work, see also \cite{sims1980macroeconomics}, \cite{hsiao1982autoregressive}, and \cite{lutkepohl1993testing}.} This extension serves two purposes. First, Granger causality at one horizon may not fully capture the causal dynamics in a multivariate system, as causality at one horizon does not necessarily imply causality at multiple horizons.\footnote{An example from \cite{dufour1998short}: In a trivariate system ($x,y,z$), $x_t$ may not Granger-cause $y_t$ at time $t$, but $x_t$ might still help predict $y_t$ several periods ahead through indirect effects mediated by $z_t$. For instance, $x_t$ could predict $z_t$ one period ahead, which in turn affects $y_t$ at a later time.} Second, the concept of impulse response proposed by \cite{sims1980macroeconomics} has limitations, as a zero impulse response is neither a necessary nor sufficient condition for non-causality in the sense of Wiener and Granger (see \cite{dufour1993relationship}). To address the issue of non-causality between two vectors of variables across multiple horizons in the presence of auxiliary variables, \cite{dufour1998short} introduced a necessary and sufficient condition on the coefficients in a multi-horizon linear projection model. These coefficients, referred to as ‘‘generalized impulse responses’’ (GIRs), include Sims’ impulse response as a special case. In this paper, we propose an innovative two-stage estimation method for the GIRs in a multi-horizon linear projection model\footnote{\cite{dufour2006short} refer to this model as ‘‘autoregression at horizon $h$’’ or ‘‘($p,h$)-autoregression’’ within a finite lag VAR($p$) framework. The autoregression at horizon $h$ model is also referred to as "Local Projection," which is widely used to estimate impulse response functions due to its straightforward inference [see \cite{jorda2005estimation}, \cite{montiel2021local}, and \cite{wolf2021same}].}, which complements two conventional methods: the recursive method and the Least Squares (LS) projection method.

The recursive method begins by estimating the VAR coefficients, followed by the recursive computation of GIRs, as these represent nonlinear transformations of the underlying VAR coefficients. Statistical tests for non-causality restrictions across multiple horizons (two or more) are expressed as zero constraints on multilinear forms involving the VAR coefficients. However, the Wald-type test criteria, can lead to asymptotically singular covariance matrices, making standard asymptotic theory inapplicable. This issue has been explored by \cite{lutkepohl1997modified}, \cite{benkwitz2000problems}, \cite{inoue2020uniform}, and \cite{dufour2023wald}. To address this challenge, \cite{pelletier2004problems} and \cite{dufour2006short} implemented a multi-horizon linear projection model with LS estimation.\footnote{The multi-horizon linear projection model has gained considerable attention in economics and finance, particularly for forecasting, impulse response estimation, causality testing, and causality measurement. For related work, see \cite{fama1988dividend}, \cite{campbell1988dividend}, \cite{dufour2010short}, \cite{dufour2012measuring}, \cite{zhang2016exchange},  \cite{salamaliki2019transmission}, \cite{shi2020causal}.}  Appropriate asymptotic inference for the coefficient estimators can be obtained through adjustments for serially correlated residuals, such as using heteroskedasticity and autocorrelation consistent (HAC) estimators. However, LS-based multi-horizon linear projection estimates may suffer from reduced estimation efficiency as the horizon lengthens, and the reliability of confidence intervals constructed with HAC standard error estimates can be fragile. For further discussion, see \cite{dufour2006short}, \cite{kilian2011reliable}, and \cite{montiel2021local}. Although \cite{dufour2006short} proposed a closed-form formula for estimating the covariance matrix, they also cautioned that the finite-sample estimate may not always be positive semi-definite. Therefore, in this study, we conduct a comprehensive reevaluation of the estimation method and inference procedures for GIR coefficients, aiming to improve the reliability of covariance matrix estimates and potentially achieve more efficient coefficient estimates.

Our two-stage estimation method for GIRs makes five key contributions to the existing literature on multi-horizon linear projection and causality studies.

First, we introduce an innovative identification method for GIRs that leverages second-moment conditions of observables and VAR innovations. This novel approach offers an alternative to the traditional LS method for estimating multi-horizon linear projection models, drawing on the well-established Two-Stage Least Squares (2SLS) framework but diverging in its application. The method consists of two distinct stages: in the first, VAR residuals are estimated using LS, and in the second, these residuals are employed as instrumental variables in a 2SLS procedure. To address potential issues related to unit roots, we also incorporate a lag-augmented version of the two-stage estimation. Our approach complements the LS method in multi-horizon projection models, enhancing both estimation precision and the robustness of inference.

Second, we compare the estimation efficiency of our two-stage approach with the LS method through both theoretical and empirical analyses. The theoretical discussion is based on an illustrative AR(2) process, where we compare the asymptotic variances of the two methods using explicit formulas and the true values of the underlying data-generating process. This analysis shows that LS estimates outperform the two-stage estimates only when the horizon is very short or the data is highly cyclical. Empirically, we conduct Monte Carlo simulations on a bivariate VAR(2) process, which demonstrate that the two-stage estimates are generally no less efficient than LS estimates. Our findings serve as a reminder to the econometrics community that instrumental variable estimates in time series may not always suffer from efficiency loss compared to LS. Furthermore, our theoretical results show that the two-stage estimates are asymptotically as efficient as the lag-augmented two-stage estimates, contradicting the common belief that lag augmentation in time series typically reduces efficiency.

Third, we develop a simple and robust method for estimating the covariance matrix of GIR estimates, which relies on the long-run variance of the regression score function. This approach eliminates the need to correct for serial correlation in the projection error. Our method removes the necessity for HAC estimates for all coefficients in a multi-horizon linear projection, including those for impulse response functions and lagged coefficients. By bypassing the complications associated with selecting bandwidths and kernel functions in HAC estimation, our approach simplifies the process. Monte Carlo simulations show that our robust covariance estimates lead to more reliable empirical test results compared to HAC estimates.

Fourth, we contribute to econometric theory by deriving the asymptotic normality of two-stage GIR estimates uniformly over the parameter space, allowing the projection horizon to grow with the sample size. Specifically, we consider the persistence of underlying data processes to range from stationary to integrated of order one or order two. The projection horizon is permitted to expand infinitely at varying rates relative to the sample size, depending on the parameter space. This uniformity across the parameter space and allowance for long-range horizons provide robust theoretical justification for empirical macroeconomic applications.

Fifth, our empirical analysis of the GIRs for economic activity in response to economic uncertainty reveals both short-run (1–3 months) and long-run (around 30 months) causal effects. The findings underscore the longer-than-expected impact of uncertainty on economic activity. Additionally, our empirical GIR studies highlight the limitations of impulse response analysis. In the long run, the Wald test on impulse responses is statistically insignificant, while the causality test remains significant, suggesting that impulse responses may fail to capture persistent causal effects over extended horizons.

\textit{Relevant Literature}--From an economic conceptual perspective, the multi-horizon linear projection model, as examined in this paper, forms a foundational framework in the study of causality and predictability \cite{lutkepohl1993testing}, \cite{dufour1998short}, \cite{dufour2006short}, \cite{dufour2010short}, \cite{diebold2014network}. The novel identification and estimation method we propose draws from innovation algorithms for (vector) ARMA processes \cite{hannan1982recursive}, \cite{brockwell1991time}, \cite{dufour2022practical}, and is closely related to recent literature on Local Projection-Instrumental Variable (LP-IV) methods, as explored by \cite{stock2018identification}. To the best of our knowledge, no existing studies have focused on using estimated VAR residuals as internal instruments to estimate GIRs within a multi-horizon linear projection model.

Regarding the statistical inference of two-stage estimates, the methodology for obviating HAC standard errors, even in the presence of serially correlated residuals, has been explored by \cite{montiel2021local}, \cite{breitung2023projection}, and \cite{xu2023local}. However, these studies primarily focus on Sims; impulse responses rather than all coefficients in the projection model. The concept of uniform inference, which controls the test level in the "worst" case scenario, is thoroughly discussed in \cite{dufour2003identification} and \cite{andrews2020generic}. Specific applications to time series can be found in \cite{mikusheva2007uniform}, \cite{mikusheva2012one}, \cite{inoue2020uniform}, \cite{montiel2021local}, and \cite{xu2023local}. Research on long (potentially infinite) projection horizons has also been conducted by \cite{montiel2021local} and \cite{xu2023local}. Our paper is the first to fully examine uniform inference for all coefficient estimates in a multi-horizon linear projection model, while also addressing restrictions on the projection horizon for stationary, I(1), and I(2) processes.

\textit{Outline}--Section \ref{section2} introduces the notation and outlines the data generation process. In Section \ref{section3}, we present the concept of two-stage identification and estimation. Section \ref{section4score} explores robust inference methods that eliminate the need to correct for serial correlation in the projection residuals. The main statistical results on uniform inference for two-stage estimates are detailed in Section \ref{section5result}. Section \ref{section6eff} compares the asymptotic efficiency of our two-stage estimates with standard LS estimates. Section \ref{section8MC} presents the results of Monte Carlo simulations. In Section \ref{section9Emp}, we apply the proposed statistical methodology for GIR estimation to analyze the dynamic causality of economic uncertainty on macroeconomic aggregates. Finally, we conclude the paper in Section \ref{section10conc}. 


\section{Framework\label{section2}}

\subsection{Model and parameter space}
In this paper, we consider a multivariate time series following a Vector Autoregressive (VAR) process,
\begin{align}
\label{equ1}
 y_t=\Phi_1 y_{t-1} + \Phi_2 y_{t-2} + \cdots + \Phi_p y_{t-p} + {u}_t 
\end{align}
where ${y}_t$ represents a $K$-dimensional vector of variables, and ${u}_t$ is a vector of innovations. The sequence of ${u}_t$ is defined on a probability space $(\Omega,\mathcal{F},\mathbb{P})$. We define the information set at time $t$, denoted as $\mathcal{F}_t$, which is generated by $ \{y_s\}_{s\leq  t}$. We consider ${u}_t$ as a white noise vector process with non-singular covariance matrix $\Sigma_u$. The order $p$ is a finite integer known a priori.\footnote{Two reasons support the assumption that the order $p$ is known a priori. First, the VAR coefficient estimates still have pointwise convergence to a standard Gaussian distribution as long as order $p$ can be consistently estimated by certain types of information criteria (see p62, \cite{kilian2017structural}). Second, in practice, order $p$ can be considered a known upper bound for the number of VAR lags. Thus, the assumption that $p$ is known a priori is not as restrictive as it might initially appear.} For the sake of notation simplicity, we omit the intercept and deterministic function in our notation, which generally do not impact the limiting results for the estimated coefficients.\footnote{In practice, researchers may incorporate dummy variables or polynomial functions of time $t$ to “detrend” the data, as discussed in \cite{toda1995statistical}.} The initial conditions are set as $y_t=0$ for $-p+1\leq t\leq 0$. 

We investigate a broad range of time series processes, including stationary processes, I(1), and I(2) processes. In addition, we account for the presence of a 'local-to-unit' root, aiming to ensure uniform inference across the specified parameter space while allowing the projection horizon to grow infinitely. Our framework also accommodates the potential presence of cointegration within the process, further enhancing its practical applicability. Notably, we highlight that processes conforming to a VARX model can also benefit from our findings. This is because VARX is a specific case of VAR, where the equation for the exogenous variable imposes constraints on the coefficients of the endogenous variables, setting them to zero (see p. 74 in \cite{kilian2017structural}).

We build up the following setup for the VAR parameter space definition: 
\begin{definition}
\label{defps}
VAR Parameter Space: Given $c=(c_0,c_1)'$,  $c_0,c_1>0$, $\epsilon \in (0,1)$, and integer $\delta\in \{0,1,2\}$, let $\mathcal{B}(\delta,c,\epsilon)$ denote the space of autoregressive coefficients such that the lag polynomial $\Phi_\delta(L)=I_K - \sum_{i=1}^p \Phi_i L^i$ admits the factorization
\begin{align}
    \Phi_\delta(L)=B_{\delta}(L)U_1(L) U_2(L)\Pi
\end{align}
where $\lambda_{min}(\Pi)>c_0$, $U_i(L)= I - \mathbb 1_{(\delta\geq i) }P_i L$, $P_i=\textup {diag}[\rho_{m,i}]$, $\rho_{m,i}\in [-1,1]$, for $1\leq m\leq K$ and $i=1,2$. In particular, it is restricted that $|\rho_{m,2}|=1, \text{ if }|\rho_{m,1}|\geq 1-\epsilon/2$. The polynomial $B_{\delta}(L)=\sum_{i=0}^{p-\delta} B_iL^i$ satisfies $B_0=\Pi^{-1}$ and $\|\mathbf B_{\delta}^h\| <c_1 (1-\epsilon)^h$, for all positive integer $h$, where $\mathbf B_{\delta}$ is the companion matrix defined as
\begin{align}
    \mathbf B_{\delta} = \left[\begin{array}{ccccc}
      B_0^{-1}B_1   & B_0^{-1}B_2 & \cdots& B_0^{-1}B_{p-\delta-1} & B_0^{-1}B_{p-\delta}   \\
      I   & 0 & \cdots&0  &0\\
      0 & I &&  \vdots&0\\
       \vdots& &\ddots &0&\vdots\\
       0& \cdots&0 &I&0
    \end{array} \right].
\end{align}
\end{definition}

The parameter space encompasses various stationary, I(1), and I(2) VAR processes, with the integer $\delta$ serving as a control parameter that represents the degree of data persistence. When all roots are believed to be distant from the unit circle, $\delta=0$. If at most one root is close to or on the unit circle, $\delta=1$, while in the case where two roots may be unity, $\delta=2$. The parameter $\rho_{m,i}$, which lies between -1 and 1, can be treated as local-to-unity (e.g., $\rho_{m,i}=1-c/T$). In contrast, the roots in $B_{\delta}(L)$ are constrained to remain strictly bounded away from the unit circle. Finally, the non-singular rotation matrix $\Pi$ allows the process to incorporate cointegration. 

Notably, although I(2) processes are rarely observed in practice, we include them in our framework to enhance the generalizability of our theory. Specifically, we illustrate the parameter space condition where $|\rho_{m,2}|=1$ if $|\rho_{m,1}|\geq 1-\epsilon/2$. This constraint implies that when $|\rho_{m,1}|$ is sufficiently close to the unit circle, $|\rho_{m,2}|$ must lie on the unit circle. However, this condition is not as restrictive as it may initially seem, as $\epsilon$ can be chosen arbitrarily small. This setup accommodates a wide range of persistent datasets, including cases where both roots equal unity, one root equals unity while the other is local-to-unity, or one root is local-to-unity while the other is strictly less than unity. The only exception is when both roots are local-to-unity, such as when $|\rho_{m,1}|=|\rho_{m,2}|=1-c/T$. This consideration arises because two local-to-unit roots offer limited practical meaning and would require additional notations in the derivation of the limiting distribution.

Our purpose is to establish a Gaussian limiting distribution for GIR estimates, covering cases where the data is persistent and integrated up to order two, while allowing the projection horizon to scale with the sample size. We show that as data persistence increases, the constraints on the projection horizon become more restrictive. This is an extension of traditional setups with a finite (fixed) horizon, the variation in the projection residual remains finite, regardless of the integration order.  

Under the data generating process of \eqref{equ1}, t he multi-horizon linear projection model is represented as:
\begin{align}
\label{hregression}
    {y}_{t+h}= \sum_{i=1}^p \Phi_i^{(h)} y_{t+1-i} + {u}_t^{(h)}, 
\end{align}
where $(\Phi_1^{(h)},\Phi_2^{(h)},\cdots,\Phi_p^{(h)})$ are GIRs defined by \cite{dufour1998short}. The recursive formula is detailed as follows,
\begin{align}
\label{equ10}
    &\Phi_{j}^{(h+1)}=\Phi_{j+h}+\sum_{l=1}^{h} \Phi_{h-l+1} \Phi_{j}^{(l)}=\Phi_{j+1}^{(h)}+\Phi_{1}^{(h)} \Phi_{j}, \quad  \Phi_{j}^{(1)}=\Phi_{j}, \quad \Phi_1^{(h)}=\Psi_h;
\end{align}
where the coefficient $\Psi_h$ represents the Sims' impulse response function at horizon $h$. The residual term ${u}_t^{(h)}$ follows an MA($h-1$) process:
\begin{align}
     {u}_t^{(h)}=y_{t+h} - \textup{P}_L(y_{t+h} \mid \mathcal{F}_t ) = \sum_{j=0}^{h-1}\Psi_j  {u}_{t+h-j}.
\end{align}
Generally, coefficients in \eqref{hregression} can be estimated through two methodologies. The review of estimation methods is presented in the next subsection.

\subsection{Existing estimation methods}
This subsection reviews two prominent approaches for estimating GIR coefficients: the Least Squares Projection (LS-Proj) method and the Recursive (RC) method. Given that equation \eqref{hregression} represents a simultaneous equation system with identical regressors across equations, we can, without loss of generality, focus on the first equation. The projection model is expressed as:
\begin{align}
\label{2.6}
    y_{1,t+h} =\beta_h' x_t + e_{t,h}
\end{align}
where $\beta_h$ denotes the $p$-lag coefficients in the first row of the system, where $\beta_h  = (\Phi_{1\bullet,1}^{(h)},\Phi_{1\bullet,2}^{(h)},\cdots,\Phi_{1\bullet,p}^{(h)})'$. The vector $x_t$ consists of lagged values of $y$, specifically $x_t := (y_t',\cdots,y_{t-p+1}')'$, and $e_{t,h}$ represents the first element of $u_t^{(h)}$.

\textbf{LS Projection (LS-Proj) method}: As demonstrated by \cite{dufour2006short}, the orthogonality between the residuals and the $p$-lag regressors allows the LS method to yield consistent coefficient estimates:
\begin{align}
\label{2.11}
    \hat{\beta}_h^{\text{LS}} = \underset{\beta_h}{\text{argmin}} \sum_{t=1}^{T-h} (y_{1,t+h} - \beta_h' x_t)^2
\end{align}
However, a key challenge arises in conducting statistical inference due to the presence of serially correlated residuals. It requires estimating the long-run variance (LRV) of the regression score function, $x_t e_{t,h}$, 
\begin{align}
    \text{LRV}(x_t e_{t,h})  := \sum_{k=-\infty}^{\infty} \mathbb{E}[ x_t  x_{t+k}' e_{t,h} e_{t+k,h} ]
\end{align}
The summation is truncated at order $h-1$ if $u_t$ satisfies a martingale difference sequence (m.d.s.) condition. When the forecast horizon exceeds one, the presence of serially correlated residuals often leads practitioners to rely on various HAC estimators for constructing confidence intervals or test statistics. \cite{dufour2006short} provide an explicit formula for $\text{LRV}(x_t e_{t,h})$. However, they also recognize that in small samples, the resulting estimates may not be positive semi-definite. The explicit covariance matrix estimate proposed by \cite{dufour2006short} corresponds to a special case of the HAC estimator with truncated lags and equal weights.

\textbf{Recursive VAR (RC-VAR) method}: The recursive method (RC) involves estimating coefficients using the recursive formula provided in equations (3.7) and (3.8) of \cite{dufour1998short}. In this approach, the first stage estimates the VAR slope coefficients, followed by the iterative computation of coefficients across different horizons using the recursive formula. Within the finite-order VAR framework, the RC method is typically implemented through a companion matrix, which transforms a VAR($p$) process into a VAR(1) model. This transformation simplifies the projection equations, as described in equation (2.2.9) of \cite{lutkepohl2005new}. Thus, the recursive method can be represented as follows:
\begin{align}
\label{recursive}
     \hat{\beta}_h^{\text{RC}} = \nu_{1}' \hat{\Phi}^h
\end{align}
where the vector $\nu_{1}$ is a unit-size vector with a conformable dimension, where the first element is one and all others are zero. The matrix $\hat{\Phi}$ represents the estimated companion matrix derived from the VAR coefficient estimates. It is important to note that the RC method is closely related to the approach used for computing impulse response functions, as demonstrated by \cite{lutkepohl1990asymptotic}. 

Compared to the LS projection method, applying the RC method requires caution due to its reliance on the delta method for statistical inference, given the non-linear transformation involved in the estimation process. As noted by \cite{dufour2006short}, standard Wald-type test statistics may result in asymptotically singular covariance matrices due to the reduced rank of the Jacobian matrix. This issue can arise, for instance, if the underlying process is white noise, as recognized by \cite{benkwitz2000problems}, \cite{dufour2015wald}, \cite{inoue2020uniform}, and \cite{dufour2023wald}, among others. Consequently, conventional asymptotic theory may not apply to such statistics without the appropriate rank condition. This limitation partly explains the preference for the LS projection method, despite its potential inefficiency.

\section{Identification and estimation}
\label{section3}
In this section, we present an innovative two-stage method for identifying and estimating the coefficients of multi-horizon linear projections.


\subsection{Moment-based identification}
We revisit the first row of the multi-horizon linear projection model, as defined in equation \eqref{2.6}:
\begin{align}
\label{4.1}
     {y}_{1,t+h}&=\beta_h' x_t + e_{t,h},
\end{align}
The conventional identification approach relies on the assumption of weak exogeneity, expressed as $\mathbb{E}[x_t({y}_{1,t+h}-\beta_h' x_t)] = 0$, which is typically satisfied when $u_t$ is orthogonal to past observables. As an alternative, we propose using VAR innovations as instruments for identifying the coefficient parameters:
\begin{align}
\label{3.12}
    z_t := ({u}_t',{u}_{t-1}',\cdots,{u}_{t-p+1}')'.
\end{align}
As the underlying process follows a VAR model, the system of equations can be written as:
\begin{align}
    &y_{1,t+h} = \beta_h' x_t + e_{t,h}, \\
    &x_t = \overline{\Psi}_p z_t + v_t.
\end{align}
This formulation provides an alternative identification strategy, utilizing the structure of the VAR model for efficient estimation.
where $\overline{\Psi}_p$ is a $pk\times pk$ upper triangular matrix, $\overline{\Psi}_p = [\overline{\Psi}_{ij,p}]_{1\leq i,j\leq p}$, $\overline{\Psi}_{ij,p}=\Psi_{j-i}$ if $j\geq i$ and zero otherwise,  $v_t=\text{P}(x_{t}|\mathcal{F}_{t-p})$ is a $pK\times 1$ vector of residuals, $v_t=(v_{1,t}',v_{2,t}',\cdots,v_{p,t}')'$, and  $v_{i,t} =y_{t-i+1} - \sum_{j=i}^{p}\Psi_{j-i} u_{t+1-j} $. 

The two-equation system described above can essentially be interpreted as a two-stage least squares (2SLS) procedure, where $z_t$ serves as the instrument. Thus, we can replace $x_t$ with its instrumental variable (IV) representation:

\begin{align}
    y_{1,t+h} = \beta_h' \overline{\Psi}_p z_t + (e_{t,h} + \beta_h' v_t).
\end{align}

Since the instrument $z_t$ comprises innovations from time $t$ to $t-p+1$, it satisfies the weak exogeneity condition: $\mathbb{E}[e_{t,h} z_t] = 0$. Moreover, as each component of $z_t$ corresponds to the innovation term of the related observable, $z_t$ is naturally correlated with the regressor $x_t$. This correlation ensures that the triangular matrix $\overline{\Psi}_p$ is non-singular. Specifically, the covariance matrix $\mathbb{E}[z_t x_t']$ is given by:
\begin{align}
\label{sigzx}
    \mathbb{E}[z_t x_t'] = (I_p \otimes \Sigma_u) \overline{\Psi}_p'.
\end{align}
The non-singularity of both the covariance matrix $\Sigma_u$ and the triangular matrix $\overline{\Psi}_p$ ensures the validity of the instrument. The coefficient $\beta_h$ can then be identified using the following moment conditions:
\begin{align}
    \beta_h = \mathbb{E}[z_t x_t']^{-1}  \mathbb{E}[z_t y_{1,t+h}].
\end{align}
For instance, in a simple VAR(1) model where $x_t = y_t$ and $z_t = u_t$, we have $\mathbb{E}[z_t x_t'] = \Sigma_u$ and $\mathbb{E}[z_t y_{1,t+h}] = \Sigma_u \Psi_{1\bullet,h}'$. Consequently, $\beta_h$ is identified as:
\begin{align}
    \beta_h = \mathbb{E}[z_t x_t']^{-1} \mathbb{E}[z_t y_{1,t+h}] = \Psi_{1\bullet,h}'.
\end{align}
This identification strategy provides an alternative estimation method using the instrument $z_t$, rather than the conventional use of $x_t$.

\subsection{Feasible and infeasible two-stage estimates}
In this subsection, we propose two types of two-stage estimators: infeasible and feasible. As previously discussed, the identification of $\beta_h$ relies on the VAR innovation $u_t$. Infeasible estimates are derived using the actual, though typically unobserved, innovations $u_t$. In contrast, feasible estimates are computed using the LS-estimated residuals, $\hat{u}_t$.

\subsubsection{Two-stage estimates}
Following the moment-based identification method discussed earlier, and assuming $z_t$ is available, the two-stage estimates can be computed as:
\begin{align}
\label{2sls}
    \tilde \beta_h^{2S} =\left( \sum_{t=p}^{T-h} z_t x_t' \right)^{-1} \left( \sum_{t=p}^{T-h}z_t y_{1,t+h} \right).
\end{align}
These estimates, $\tilde \beta_h^{2S}$, are termed infeasible because the innovation process $u_t$ is generally unobserved. In practice, it is common to replace $u_t$ with the estimated VAR residuals, $\hat{u}_t$, and derive an estimate for $z_t$ as follows:
\begin{align}
    \hat{z}_t = (\hat{u}_t', \hat{u}_{t-1}', \cdots, \hat{u}_{t-p+1}')',
\end{align}
where $\hat{u}_t = y_t - \sum_{i=1}^{p}\hat{\Phi}_i y_{t-i}$, and $\hat{\Phi}_i$ represents the coefficient matrices estimated through LS on a VAR($p$) model. This leads to the feasible estimator:
\begin{align}
\label{fea2sls}
     \hat{\beta}_h^{2S} =\left( \sum_{t=p}^{T-h} \hat{z}_t x_t' \right)^{-1} \left( \sum_{t=p}^{T-h}\hat{z}_t y_{1,t+h} \right).
\end{align}
A critical step in deriving the asymptotic distribution for the feasible estimator is establishing its asymptotic equivalence with the infeasible estimator. Specifically, it is necessary to demonstrate that incorporating the estimated $z_t$ does not introduce asymptotic bias at the order of the square root of the sample size. This formal result is presented in Proposition \ref{propiden}. It is important to note that using estimated variables as instruments in two-stage least squares (2SLS) differs fundamentally from employing estimated variables directly as regressors, as in the LS-based VARMA estimation method discussed by \cite{hannan1982recursive}. In the latter case, efficiency losses occur when estimated residuals are used as regressors, prompting the development of remedial procedures to improve efficiency. In contrast, our paper demonstrates that using estimated residuals as instruments does not impair efficiency. 

With both feasible and infeasible two-stage estimators established, an important question arises: why transition from the well-established, straightforward LS-based estimation to the two-stage method? This question is addressed by highlighting three key advantages.

First, two-stage estimators typically yield more efficient results across a broad spectrum of data-generating processes and projection horizons. A detailed analysis of asymptotic efficiency is provided in Section \ref{section6eff}, where an illustrative AR(2) model offers intuitive insights into the enhanced efficiency of two-stage estimators. Additionally, the section on Monte Carlo simulations examines these efficiency comparisons from an empirical perspective.

Second, the two-stage approach addresses serial correlation in projection residuals by providing robust covariance estimation, thereby eliminating the need for HAC corrections. This leads to more reliable inference, as HAC estimates are known to be unreliable in finite samples. In Section \ref{section4score}, we introduce a novel method for estimating the long-run variance of the regression score function, with detailed theoretical and methodological discussions.

Third, the two-stage method facilitates the analysis of infinite projection horizons, a significant advantage for macroeconometric research, where the projection horizon often represents a substantial portion of the sample. In contrast, the literature on standard LS estimation remains largely silent on infinite projection horizons. We address this gap by exploring long projection horizons and specifying the conditions under which the Gaussian limiting distribution holds for two-stage estimates.

\subsubsection{Lag-augmented two-stage estimates}
Our two-stage estimates encounter non-standard convergence issues when the data exhibit a unit root.\footnote{For example, in the case of an AR(1) random walk, the covariance matrix $T^{-1}\sum_{t=1}^T u_t y_t$ does not converge to the variance of $u_t$ but instead to a stochastic process. Although this process has finite mean and variance, it is almost surely bounded by a constant, as guaranteed by Chebyshev's inequality. This ensures consistency, as the sample mean of the score function, $u_t e_{t,h}$, converges in probability to zero. However, this invalidates the use of Wald-type tests with standard critical values.} To address this, we introduce a lag-augmented version of the two-stage estimates, as proposed by \cite{toda1995statistical} and \cite{dolado1996making}.


The regression model under consideration is augmented with one or two additional lags, represented as:
\begin{align}
\label{4.33}
     {y}_{1,t+h}&= (\beta_h', \gamma_\delta') x_{t,\delta} + e_{t,h}, 
\end{align}
for $\delta = 1, 2$, where the coefficient $\gamma_\delta$ is a $(\delta K)$-dimensional nuisance parameter associated with the extra lag(s), and $\gamma_\delta$ equals to zero in the population. Here, $x_{t,1} := (x_t', y_{t-p}')'$ and $x_{t,2} := (x_t', y_{t-p}', y_{t-p-1}')'$. Notably, the lag-augmented regression nests the standard regression when $\delta = 0$, in which case $x_{t,0} = x_t$ and $\gamma_0$ is an empty vector.

The inclusion of the extra lag(s) serves as a control variable, allowing the previous regressors to be transformed (e.g., first-differenced) into stationary variables. This transformation facilitates the derivation of the Gaussian limiting distribution. Instead of directly applying the Least Squares (LS) method, as done in the linear projection framework, we employ VAR residuals as instruments with lag augmentation: $z_{t,1}:= (z_t', y_{t-p}')'$ or $z_{t,2}:= (z_t', y_{t-p}', y_{t-p-1}')'$, using a two-stage least squares (2SLS) approach. Here, $z_t$ is defined as in equation \eqref{3.12}. This yields the infeasible lag-augmented estimates as follows:\footnote{When $\delta=0$, the estimate $\tilde \beta_h^{\text{LA(0)-2S}}$ coincides with the non-lag-augmented two-stage estimate $\tilde \beta_h^{\text{2S}}$. Thus, $\tilde \beta_h^{2S}$ ($\hat \beta_h^{2S}$) and $\tilde \beta_h^{LA(0)-2S}$ ($\hat \beta_h^{LA(0)-2S}$) are interchangeable in this paper.}:
\begin{align}
\label{lg-2sls}
    \tilde \beta_h^{\text{LA($\delta$)-2S}} =H_\delta  (\sum_{t=p+\delta}^{T-h} z_{t,\delta} x_{t,\delta}')^{-1} ( \sum_{t=p+\delta}^{T-h} z_{t,\delta} y_{1,t+h} ) ,
\end{align}
where $H_\delta$ represents a $pK \times (p + \delta)K$ selection matrix, defined as $H_\delta := (I_{pK}, 0_{pK \times \delta K})$. Following the rationale for lag augmentation as outlined in \cite{dolado1996making} and related literature, the first $p$ regressors in $x_{t,\delta}$ can be linearly transformed into a stationary process due to the lag augmentation. This leads to the convergence in probability of the sample covariance. Further details, along with the theoretical proof, are provided in Section \ref{section5result}.

Since the innovation process $u_t$ is not directly observed, we propose feasible lag-augmented estimates:
\begin{align}
\label{fealg-2sls}
    \hat \beta_h^{\text{LA($\delta$)-2S}} = H_\delta \left( \sum_{t=p+\delta}^{T-h} \hat z_{t,\delta} x_{t,\delta}' \right)^{-1} \left( \sum_{t=p+\delta}^{T-h} \hat z_{t,\delta} y_{1,t+h} \right),
\end{align}
where $\hat{z}_{t,\delta}$ is obtained by replacing $z_t$ in $z_{t,\delta}$ with $\hat{z}_t$, and $\hat{z}_t$ is calculated in the same manner as in the stationary case. The theoretical proof of the asymptotic equivalence between the feasible and infeasible lag-augmented two-stage estimates will be provided in Section \ref{section5result}.

\section{Regression score function and statistical inference}
\label{section4score}
In this section, we propose an alternative estimation method for the long run variance of the regression score function, which obviates the need to correct the serial correlation of the score function.

\subsection{HAC standard error}
\label{sec3.2}

In general, the necessity of HAC inference in conventional methods stems from the serial correlation in the regression score function, $x_t e_{t,h}$, particularly when the horizon $h$ exceeds one. This subsection introduces a simple method to obviate the need for HAC inference. 

Let the regression score function for the two-stage estimates be denoted as
\begin{align}
\label{3.13}
s_{t,h}:= z_t e_{t,h} =(e_{t,h} u_t', e_{t,h} u_{t-1}',\cdots,e_{t,h} u_{t-p+1}')',
\end{align}
Under certain regularity conditions on the innovation process, such as strict stationarity and ergodicity, or strong mixing, the scaled summation of the regression score function is expected to converge in law,
\begin{align}
\label{clt0}
\Bar{T}^{-1/2} \sum_{t=p}^{T-h} s_{t,h} \xrightarrow{d} N(0, \Omega_{s,h}),
\end{align}
where $\Bar{T}=T-h-p+1$, and $\Omega_{s,h}$ denotes the LRV of $s_{t,h}$,
\begin{align}
\label{omegash}
\Omega_{s,h}:=\sum_{k=-\infty}^{\infty} \mathbb{E}[s_{t,h} s_{t+k,h}']
\end{align}
Note the summation would be truncated at order $h-1$ if $u_t$ satisfies a m.d.s. condition. The sum of lead-lag covariance matrices stems from the serial correlation presenting in $e_{t,h}$, given its characterization as an MA($h-1$) process. Notably, the summation is truncated at order $h-1$ due to the property that $\mathbb{E}[s_{t,h} s_{t-k}']=\mathbf{0}$ for all $|k| \geq h$ under the m.d.s. assumption.

Consequently, to conduct statistical tests or establish confidence intervals, it becomes imperative to prove \eqref{clt0} and propose a consistent estimate of $\Omega_{s,h}$. Typically, due to the lead-lag summation described in \eqref{omegash} and the need of positive semi-definite LRV estimates in finite samples, researchers commonly resort to utilizing HAC estimators, such as the Newey-West estimator, along with specific bandwidth and weight selections. However, we have recognized that by imposing slightly more stringent assumptions on the innovation process $u_t$, it may be feasible to obviate the need for HAC estimation to achieve a positive semi-definite estimate of $\Omega_{s,h}$.

\subsection{Simple and robust heteroskedastic consistent standard error}
This subsection derive the limiting distribution for \eqref{clt0} along with a simple and robust approach to estimate the long-run variance $\Omega_{s,h}$.

\subsubsection{Reordered regression score function}
We perform an algebraic manipulation of the regression score function and introduce a new series, $s_{t,h}^{*}$, defined as:
\begin{align}
\label{4.19}
    s_{t,h}^{*}:=(e_{t,h} u_t', e_{t+1,h} u_{t}',\cdots,e_{t+p-1,h} u_{t}')'.
\end{align}
Specifically, the construction of $s_{t,h}^{*}$ involves replacing the $j$-th component in $s_{t,h}$ with its $(j-1)$-th leading value. For instance, the second component in $s_{t,h}$, $e_{t,h}u_{t-1}$, is replaced by $e_{t+1,h}u_{t}$. The introduction of $s_{t,h}^{*}$ serves two purposes: (1) it reorders the elements of $s_{t,h}$, ensuring that both series have the same long-run variance; and (2) under certain regularity conditions on the $u_t$ process, $s_{t,h}^{*}$ becomes serially uncorrelated, and its long-run variance is identical to its variance. To clarify this property, we explicitly express $s_{t,h}^{*}$ as:
\begin{align}
     s_{t,h}^{*}= (e_{t,h},e_{t+1,h},\cdots,e_{t+p-1,h})'\otimes u_t.
\end{align}
To avoid the need for HAC estimators and to propose positive semi-definite consistent estimates of $\Omega_{s,h}$, we introduce the following assumption:

\begin{assumption}
\captionassumption{\assumptionname}{Mean independence assumption on the innovation process}
\label{assimeanind}
    \item The process $u_t$ is mean-independent, i.e., $\mathbb{E}[u_t | \{u_\tau\}_{\tau\neq t}] = 0$ almost surely.
\end{assumption}

There are three main reasons for adopting the mean-independence assumption in our innovation process. First, it provides a sufficient condition for proposing a convenient method to estimate the long-run variance $\Omega_{s,h}$. Typically, when estimating $\Omega_{s,h}$, which involves summing lead-lag covariances, HAC estimators are required to ensure positive semi-definiteness in finite samples. However, under Assumption \ref{assimeanind}, $\Omega_{s,h}$ becomes algebraically equivalent to the variance of $s_{t,h}^*$. For illustration, consider the case where $t < \tau$, which leads to:
\begin{align}
\label{3.25}
\begin{split}
    \mathbb{E}[ s_{t,h}^{*} s_{\tau,h}^{*'}] 
    &= \mathbb{E}[ \mathbb{E}[s_{t,h}^{*} s_{\tau,h}^{*'} \mid u_{t+1}, u_{t+2}, \cdots ] ] \text{ (Law of Iterated Expectations)}\\
    &= \mathbb{E}[
    (e_{t,h},e_{t+1,h},\cdots,e_{t+p-1,h})'\otimes
    \underbrace{\mathbb{E}[u_t \mid u_{t+1},u_{t+2},\cdots ]}_{=0}
    s_{\tau,h}^{*'} ] = 0.
\end{split}
\end{align}
where the second equality holds because the explicit form of $s_{t,h}^*$ reveals that $s_{\tau,h}^*$ and $(e_{t,h}, e_{t+1,h}, \cdots, e_{t+p-1,h})$ are measurable with respect to the information set $\sigma(u_{t+1}, u_{t+2}, \cdots)$, since $e_{t+i,h}$ is a linear combination of $u_{t+1+i}, \cdots, u_{t+h+i}$. This framework, previously applied by \cite{montiel2021local} in impulse response estimation, is extended in our study to encompass all coefficients within a multi-horizon linear projection model.\footnote{\cite{montiel2021local}, footnote 7, notes that inference for lagged coefficients in Local Projection models typically relies on HAC estimators. We address this by demonstrating that using VAR-estimated residuals as instruments yields a standard limiting distribution for these coefficients without HAC estimators.}

It is important to note, as \cite{xu2023local} highlights, that the mean-independence assumption is sufficient but not necessary to achieve a zero-correlation result for the regression score function. Moreover, this assumption may not hold in all contexts, especially in the presence of skewed distributions or conditional heteroskedasticity, which could violate it. While we acknowledge these limitations, the second and third motivations for using the mean-independence assumption may outweigh these concerns, making it a valuable assumption in this context.

Second, the mean-independence assumption facilitates the application of the Central Limit Theorem (CLT) for martingale difference sequences to establish convergence in distribution. In conventional frameworks with finite horizons, various versions of the CLT can be employed. The key point here is that a finite horizon guarantees a finite variance for the regression score function, ensuring that $\lambda_{\text{max}}(\Omega_{s,h}) < c < \infty$. However, when the horizon grows relative to the sample size, as often occurs in macroeconomic applications where the projection horizon constitutes a significant portion of the sample size, the variance may become unbounded, particularly in processes with unit roots. This issue is observed in Monte Carlo simulations, where increasing the horizon degrades the empirical size of statistical tests based on asymptotic variance and critical values from a standard Gaussian distribution, especially in highly persistent processes. Therefore, we rely on the CLT for martingale difference sequences, which accommodates unbounded variance.

Third, while convergence can still be established under weaker conditions, such as assuming a martingale difference sequence and mixing, these conditions impose restrictions on the horizon $h$. \cite{xu2023local} proposed an alternative method to derive asymptotic results by restructuring the regression score function in a more complex manner. This approach introduces two asymptotically negligible terms, with an order dependent on the horizon $h$, implying that the theorem would only hold if $h/T \to 0$ in the limit. In our view, this condition applies even for finite-order stationary VAR processes, which are commonly encountered in practice. While this restructuring allows for relaxing the mean-independence assumption to a martingale difference sequence assumption, it limits the horizon $h$ from growing toward infinity at the rate permitted under stationarity. Our results, by contrast, allow the horizon $h$ to represent a non-trivial proportion of the sample size $T$. Therefore, we adopt the mean-independence assumption in this paper.

In light of these considerations, the algebraic manipulation and the result presented in Equation \eqref{3.25} demonstrate that the variance of $s_{t,h}^*$ is equal to the long-run variance (LRV) of $s_{t,h}$. This insight allows us to bypass the use of HAC standard errors, as the LRV of $s_{t,h}$—which involves summing lead-lag covariances—typically requires HAC estimation to ensure positive semi-definiteness in finite samples. The natural estimate of the variance of $s_{t,h}^*$, however, satisfies this criterion directly.

We formalize these findings in the following lemma:

\begin{lemma}
\captionlemma{\lemmaname}{Lemma of the LRV matrix}
\label{lemmauncorr}
Suppose Assumption \ref{assimeanind} holds, then 
\begin{align}
    \text{Var}(s_{t,h}^*)=\Omega_{s,h},
\end{align}
where $s_{t,h}^{*}$ and $\Omega_{s,h}$ are defined in \eqref{4.19} and \eqref{omegash}, respectively.
\end{lemma}

The proof of Lemma \eqref{lemmauncorr} is provided in Appendix \eqref{prooflemma4.1}. Lemma \eqref{lemmauncorr} states that $\Omega_{s,h}$ is identical to the variance matrix of $s_{t,h}^*$, offering a straightforward method for estimating the LRV $\Omega_{s,h}$. Empirical researchers need only estimate the sample variance of $s_{t,h}^*$, rather than the LRV of $s_{t,h}$. As the sample variance is naturally positive semi-definite, this approach eliminates the need for HAC correction of the serial correlation.

\subsubsection{Asymptotic normality}
The asymptotic convergence to normality will be established in two steps. First, we decompose the summation of $s_{t,h}$ into three parts: the primary component, which is the summation of the reordered regression score function $s_{t,h}^*$, and two additional asymptotically negligible terms, $\overline{s}_{1,h}$ and $\overline{s}_{2,h}$. Second, we demonstrate the convergence in distribution of the summation of $s_{t,h}^*$ using the martingale Central Limit Theorem, thereby confirming the convergence of the regression score function. The primary challenge arises from the potentially unbounded variance of $e_{t,h}$ as the horizon $h$ increases, particularly in cases of persistent data.

The summation of $s_{t,h}$ is expanded as follows:
\begin{align}
\label{4.32}
\sum_{t=p}^{T-h} s_{t,h} = \overline{s}_{1,h} + \overline{s}_{2,h} + \sum_{t=p}^{T-h-p+1} s_{t,h}^*,
\end{align}
where $\overline{s}_{1,h}$ and $\overline{s}_{2,h}$ are two terms of order ($p-1$), defined as $\overline{s}_{1,h} = \sum_{i=1}^{p-1} s_{i,h}^{**}$ and $\overline{s}_{2,h} = \sum_{i=1}^{p-1} s_{i,h}^{***}$, with $s_{i,h}^{**} = J_{(-i)} s_{i,h}^{*}$ and $s_{i,h}^{***} = J_{(p-i)} s_{\bar{T}+i,h}^{*}$. Here, $\bar{T} = T - h - p + 1$, and $J_{(i)}$ and $J_{(-i)}$ denote the first $iK$ rows and the last $(p-i)K$ rows of an identity matrix of dimension $pK$, respectively, such that $[J_{(i)}', J_{(-i)}'] = I_{pK}$. The terms $\overline{s}_{1,h}$ and $\overline{s}_{2,h}$ arise from the reordering process. When the VAR order $p$ is finite, these terms are bounded by a constant scaled by the standard error of $s_{t,h}^*$, almost surely, as guaranteed by Chebyshev's inequality.

To establish the limiting result, we show that the scaled summations of $\sum_{t=p}^{T-h} s_{t,h}$ and $\sum_{t=p}^{T-h-p+1} s_{t,h}^*$ are asymptotically equivalent.

\begin{lemma}
\captionlemma{\lemmaname}{Asymptotic equivalence}
\label{lemmaneg}
    Let the autoregressive coefficients be in the parameter space $\mathcal{B}(\delta,c,\epsilon)$. Let $c=(c_0,c_1)$, $c_0,c_1>0$, $\epsilon \in (0,1)$. If one of the following conditions holds, 
    \begin{enumerate}[(i)]
        \item $\delta=0$ and $0< h/T\leq \alpha$ for $\alpha\in (0,1)$,
        \item $\delta=1$ and $h/T\xrightarrow{p}0$, 
        \item $\delta=2$ and $h^3/T\xrightarrow{p}0$, 
    \end{enumerate}
    then 
    \begin{align}
    \label{4.39}
        \bar T^{-1/2} w'(\sum_{t=p}^{T-h} s_{t,h} -  \sum_{t=p}^{\bar T} s_{t,h}^*) \xrightarrow{p} 0,
    \end{align}
    where $\bar T=T-h-p+1$,  $w\in\mathbb R^{pK}$, and $\|w\|=1$.
\end{lemma}

The proof of Lemma \ref{lemmaneg} is provided in Appendix \eqref{prooflemma4.2}. In essence, Lemma \ref{lemmaneg} asserts that the Euclidean norm of $\overline{s}_{1,h} + \overline{s}_{2,h}$ is asymptotically negligible at the scale of $\bar T^{-1/2}$. Although the summation involves a finite number of terms (with finite order $p$), the variance is not guaranteed to be constant when the data exhibits persistence and the horizon $h$ approaches infinity. The proof hinges on showing that the variance of each term becomes negligible relative to $\bar T$, a condition that depends on the boundedness of the horizon $h$ and the persistence of the data.

Building on Lemma \ref{lemmaneg}, proving the convergence of $\bar T^{-1/2} w' \sum_{t=p}^{T-h} s_{t,h}$ is equivalent to establishing the convergence of $\bar T^{-1/2} w' \sum_{t=p}^{\bar T} s_{t,h}^*$. Next, we impose the following regularity conditions on the innovation process, followed by the result on distributional convergence.

\begin{assumption}
\label{assmoment}
    Assume:
    \begin{enumerate}
        \item $\mathbb{E}[\|u_t\|^8] < \infty$, and $\lambda_{\min}(\mathbb{E}[u_t u_t' | \mathcal{F}_{t-1}]) > c > 0$ for some constant $c$, almost surely.
        \item The process $u_t \otimes u_t$ has absolutely summable cumulants up to order four.
    \end{enumerate}
\end{assumption}

\begin{proposition}
\label{propclt}
    Suppose Assumptions \ref{assimeanind} and \ref{assmoment} hold. Let the conditions of Lemma \ref{lemmaneg} be satisfied. Then:
    \begin{align}
    \label{clt}
        \bar T^{-1/2} \sum_{t=p}^{T-h} w' s_{t,h} / (w' \Omega_{s,h} w)^{1/2} \xrightarrow{d} N(0,1),
    \end{align}
    where $\bar T = T-h-p+1$, $w \in \mathbb{R}^{pK}$, and $\|w\|=1$.
\end{proposition}

The proof of Proposition \ref{propclt} is provided in Appendix \ref{proofprop4.3}. This proposition represents a key step in establishing asymptotic results for our two-stage estimates. It states that, for data processes of orders I(0), I(1), or I(2), and under certain conditions regarding the projection horizon $h$, the scaled summation of the regression score function converges to a Gaussian distribution when normalized by $\bar T^{1/2}$. By combining this result with the earlier finding that the LRV matrix $\Omega_{s,h}$ is equivalent to the covariance matrix of $s_{t,h}^*$, we conclude that $\Omega_{s,h}$ can be consistently estimated using a sample covariance matrix.

Proposition \ref{propclt} is one of econometrics contributions to the statistical inference of multi-horizon projection estimates. It demonstrates that, in a general projection (or prediction) model, the estimates can bypass the need for HAC inference, even in the presence of serial correlation in the residuals. Instead, the two-stage estimates can rely on White’s heteroskedasticity-robust inference, reconstructed from the regression score function.

\section{Asymptotic results}
\label{section5result}
In this section, we present asymptotic uniform inference for our two-stage estimates. Recognizing the potential presence of a unit root in the process and the need to establish the corresponding asymptotic distributional theory, we introduce a square matrix of dimension $pK$, denoted as $G_{\delta, T}$. This matrix is a function of both the transformation matrix and the probability scaling matrix, and is defined as follows:
\begin{align}
\label{g}
    \begin{split}
        & G_{0,T}=I_{pK}, \quad  G_{\delta,T}= \bar \Upsilon_\delta \bar P_\delta (I_{p} \otimes \Pi) \text{ for } \delta=1,2
    \end{split}
\end{align}
where $\bar P_\delta, \bar \Upsilon_\delta$ are square matrices of dimension $pK$,
\begin{align*}
\begin{split}
    &\bar P_1=\left[ \begin{array}{cccc}
     I    & -P_1    &&     0   \\
           &\ddots &\ddots\\
          && I & - P_1\\
   0       &&&  I
    \end{array} \right],
    \quad 
    \bar \Upsilon_1=    \left[\begin{array}{cc}
       I_{(p-1)K}  & 0 \\
       0  & \Upsilon_1
    \end{array} \right],\\
    &\bar P_2=\left[ \begin{array}{cccccc}
     I    & -(P_1+P_2)   &   P_1P_2    &       &       & 0  \\
          &  I    & -(P_1+P_2)    & P_1P_2     &       & \\
          &       & \ddots &\ddots &\ddots &\\
          &       &        &I      &  -(P_1+P_2)   &P_1P_2 \\
          &&& & I &  -P_2 \\
    0       &&&  & &  I
    \end{array} \right],\quad \bar \Upsilon_2= 
    \left[\begin{array}{ccc}
      I_{(p-2)K}  &  &0\\
         & \Upsilon_1&\\
       0  &  &\Upsilon_2 \\
    \end{array} \right].
\end{split}
\end{align*}
The matrices $\Upsilon_1$ and $\Upsilon_2$ represent $K$-dimensional diagonal probability scaling matrices, defined as $\Upsilon_{\delta} = \text{diag}[g_{m,\delta}^{-1/2}]_{1 \leq m \leq K}$, where $g_{m,1} = \min(T, \frac{1}{1 - |\rho_{m,1}|})$ and $g_{m,2} = g_{m,1}^2 \min(T, \frac{1}{1 - |\rho_{m,2}|})$. In the case of $\delta = 1$, when $\rho_{m,1}$ is bounded away from the unit disk, $g_{m,1}$ is constant. However, when $\rho_{m,1}$ is local to unity (e.g., $1 - c/T$) or lies on the unit circle, $g_{m,1}$ becomes proportional to $T$. For $\delta = 2$, the situation is more complex. If only one root is local to unity or a unit root, then $\bar{g}_{m,2}$ is proportional to $T$. If two roots are on the unit circle, then $\bar{g}_{m,2}$ scales as $T^3$. This result aligns with the convergence rate of the sample variance required by integrated time series variables, which is $T^2$ and $T^4$ for I(1) and I(2) processes, respectively (see \cite{park1988statistical}, \cite{phillips1988testing}, \cite{toda1995statistical}, among others).

The matrix $\bar{P}_\delta$ is a differenced matrix designed to prevent perfect multicollinearity in the limit for cointegration, following standard procedures for autoregressive time series models with unit roots (\cite{sims1990inference}). If all diagonal elements of $P_1$ are equal to one, $\bar{P}_1$ simply applies the first difference to $\Pi y_t$. The transformation matrix $G_{\delta,T}$, combined with the probability scaling matrix, provides the necessary high-level condition for our asymptotic results.

To derive the asymptotic distribution, we propose the following limiting conditions:
\begin{assumption}
\label{ass4}
    For any $\delta\in \{0,1,2\}$, $c=(c_0,c_1)'$, $c_0,c_1>0$, and $\epsilon\in (0,1)$,
    \begin{align}
    \underset{N\rightarrow \infty}{\lim}
    \underset{T\rightarrow \infty}{\lim}
    \underset{\Phi \in \mathcal{B}(\delta,c,\epsilon)}{\inf}
    \textup{P}_\Phi
    \left( 
    \lambda_{min}
    \left(
    \frac{1}{T}\sum_{t=1}^T G_{\delta,T}  x_{t} x_{t}' G_{\delta,T}'\right)\geq 1/N \right)=1
    \end{align}
    where $G_{\delta,T}$ is defined in \eqref{g}.
\end{assumption}

Assumption \ref{ass4} serves as a commonly required intermediate step in establishing the limiting distribution of OLS estimates in autoregressive models. It states that the scaled covariance matrix is non-singular with probability one. It is a high level assumption seen in Local Projection literature, e.g., see \cite{montiel2021local} and \cite{xu2023local}.


\subsection{Limiting results for feasible and infeasible estimates}
In this subsection, we first present the asymptotic normality results for the infeasible estimates, which rely on the unobserved innovation process. We then establish the asymptotic equivalence between feasible and infeasible estimates, demonstrating that the two estimators are identical at the scale of $\bar T^{1/2}$.

To illustrate the asymptotic variance of both feasible and infeasible estimates, we define $\Omega_{\beta,h}$ as follows:
\begin{align}
\label{omegabeta}
    \Omega_{\beta,h} := \Sigma_{zx}^{-1} \Omega_{s,h} \Sigma_{zx}^{'-1},
\end{align}
where $\Sigma_{zx} = \mathbb{E}[z_t x_t']$, and $\mathbb{E}[z_t x_t']$ and $\Omega_{s,h}$ are defined in \eqref{sigzx} and \eqref{omegash}, respectively.

The limiting distribution of the infeasible estimates is provided below.
\begin{proposition}
\label{propinfea}
    Suppose Assumption \ref{assimeanind}, \ref{assmoment}, and \ref{ass4} hold. Let the autoregressive coefficients be in the parameter space $\mathcal{B}(\delta,c,\epsilon)$ and $w\in\mathbb R^{pK}$, $\|w\|=1$. 
    \begin{enumerate}[(i)]
        \item Let $\alpha\in (0,1)$. Then 
        \begin{align}
        \underset{T \rightarrow \infty}{\lim}\sup_{\Phi\in \mathcal{B}(0,c,\epsilon)} 
        \sup_{0<h\leq \alpha T }
        \bar T^{1/2} 
        \frac{w' 
        (\tilde{\beta}_h^{2S}-\beta_h)}{(w'\Omega_{\beta,h} w)^{1/2}} \xrightarrow{d} N(0,1), 
    \end{align}
       where $\tilde{\beta}_h^{2S}$ is defined in \eqref{2sls}.
    \item Let a sequence $\{\bar h_T\}$ of positive integers satisfied $\frac{\bar h}{T}\max (1, \frac{\bar h}{w'\Omega_{\beta,h} w }) \xrightarrow{p}0$. Then 
        \begin{align}
        \underset{T \rightarrow \infty}{\lim}\sup_{\Phi\in \mathcal{B}(1,c,\epsilon)} 
        \sup_{0<h\leq \bar h_T }
        \bar T^{1/2}  \frac{w' 
        (\tilde {\beta}_h^{\text{LA(1)-2S}}-\beta_h) }{(w'\Omega_{\beta,h} w)^{1/2}} \xrightarrow{p} 0 .
    \end{align}
      where $\tilde {\beta}_h^{\text{LA(1)-2S}}$ is defined in \eqref{lg-2sls}.
    \item Let a sequence $\{\bar h_T\}$ of positive integers satisfied $\frac{\bar h^3}{T}\max (1, \frac{\bar h}{w'\Omega_{\beta,h} w })\xrightarrow{p}0$. Then
        \begin{align}
        \underset{T \rightarrow \infty}{\lim}\sup_{\Phi\in \mathcal{B}(2,c,\epsilon)} 
        \sup_{0<h\leq \bar h_T }
        \bar T^{1/2} \frac{w'  
        (\tilde {\beta}_h^{\text{LA(2)-2S}}-\beta_h) }{(w'\Omega_{\beta,h} w)^{1/2}}  \xrightarrow{p} 0 .
    \end{align}
      where $\tilde {\beta}_h^{\text{LA(2)-2S}}$ is defined in \eqref{lg-2sls}.
    \end{enumerate}
\end{proposition}

Proposition \ref{propinfea} provides asymptotic statistical inference for both lag-augmented and none lag-augmented two-stage infeasible estimates. It is important to note that increasingly restrictive conditions on the projection horizon are necessary to account for the persistence in the underlying data process. This restriction primarily stems from the requirements of the martingale Central Limit Theorem and the asymptotic negligibility of the bias introduced by lag-augmentation.

A key insight from Proposition \ref{propinfea} is that lag-augmented and none lag-augmented two-stage infeasible estimates share the same asymptotic variance, $\Omega_{\beta,h}$. This reveals a surprising constancy in asymptotic efficiency, regardless of whether lag-augmentation is employed. This finding is a contradiction with the conventional view in standard LS estimation, where asymptotic efficiency is expected to vary with the lag-augmentation.

Moreover, we establish a crucial result confirming the asymptotic equivalence of infeasible and feasible estimates. 

\begin{proposition}
\label{propiden}
    Suppose Assumption \ref{assimeanind}, \ref{assmoment}, and \ref{ass4} hold. Let the vector autoregressive coefficients be in the parameter space $\mathcal{B}(\delta,c,\epsilon)$ and $w\in\mathbb R^{pK}$, $\|w\|=1$. 
    \begin{enumerate}[(i)]
        \item Let $\alpha\in (0,1)$. Then 
        \begin{align}
        \underset{T \rightarrow \infty}{\lim}\sup_{\Phi\in \mathcal{B}(0,c,\epsilon)} 
        \sup_{0<h\leq \alpha T }
        \bar T^{1/2} 
        \frac{w' ( 
        \hat {\beta}_h^{2S} 
        - 
        \tilde{\beta}_h^{2S})}{(w'\Omega_{\beta,h} w)^{1/2}} \xrightarrow{p} 0. 
    \end{align}
      where $\hat{\beta}_h^{2S}$ and $\tilde{\beta}_h^{2S}$ are respectively defined in \eqref{fea2sls} and \eqref{2sls}.
    \item Let a sequence $\{\bar h_T\}$ of positive integers satisfied $\frac{\bar h}{T}\max (1, \frac{\bar h}{w'\Omega_{\beta,h} w }) \xrightarrow{p}0$. Then
        \begin{align}
        \underset{T \rightarrow \infty}{\lim}\sup_{\Phi\in \mathcal{B}(1,c,\epsilon)} 
        \sup_{0<h\leq \bar h_T }
        \bar T^{1/2}  \frac{w' ( 
        \hat {\beta}_h^{\text{LA(1)-2S}} 
        - 
        \tilde {\beta}_h^{\text{LA(1)-2S}} )}{(w'\Omega_{\beta,h} w)^{1/2}} \xrightarrow{p} 0 .
    \end{align}
      where $\hat {\beta}_h^{\text{LA(1)-2S}} $ and $\tilde {\beta}_h^{\text{LA(1)-2S}} $ are respectively defined in \eqref{fealg-2sls} and \eqref{lg-2sls}.
    \item Let a sequence $\{\bar h_T\}$ of positive integers satisfied $\frac{\bar h^3}{T}\max (1, \frac{\bar h}{w'\Omega_{\beta,h} w })\xrightarrow{p}0$. Then
        \begin{align}
        \underset{T \rightarrow \infty}{\lim}\sup_{\Phi\in \mathcal{B}(2,c,\epsilon)} 
        \sup_{0<h\leq \bar h_T }
        \bar T^{1/2} \frac{w' ( 
        \hat {\beta}_h^{\text{LA(2)-2S}} 
        - 
        \tilde {\beta}_h^{\text{LA(2)-2S}} )}{(w'\Omega_{\beta,h} w)^{1/2}}  \xrightarrow{p} 0 .
    \end{align}
       where $\hat {\beta}_h^{\text{LA(2)-2S}} $ and $\tilde {\beta}_h^{\text{LA(2)-2S}} $ are respectively defined in \eqref{fealg-2sls} and \eqref{lg-2sls}.
    \end{enumerate}
\end{proposition}
The proof of Proposition \eqref{propiden} is in Appendix \eqref{proofpropinfea}. This demonstrates that incorporating estimated residuals as instruments does not introduce asymptotic bias.

\subsection{Uniform inference on feasible two-stage estimates}
\label{secasym}

This subsection presents the uniform inference for both lag-augmented and non-lag-augmented two-stage estimates, scaled by the estimated covariance matrix. First, we outline the method for estimating the covariance matrix. The estimator for $\Omega_{\beta,h}$ is given by:

\begin{align}
\label{hatomegabeta}
    \hat{\Omega}_{\beta,h} = \hat{\Sigma}_{zx}^{-1} \hat{\Omega}_{s,h} \hat{\Sigma}_{zx}^{'-1},
\end{align}
where:
\begin{align}
    \begin{split}
        & \hat{\Sigma}_{zx} = (I_p \otimes \hat{\Sigma}_u) \hat{\overline \Psi}_p', \\
        & \hat{\Omega}_{s,h} = (\bar T - p + 1)^{-1} \sum_{t=p}^{\bar T} \hat{s}_{t,h}^{*} \hat{s}_{t,h}^{*'}, \\
        & \hat{s}_{t,h}^{*} = (\hat{e}_{t,h}, \hat{e}_{t+1,h}, \cdots, \hat{e}_{t+p-1,h})' \otimes \hat{u}_t,
    \end{split}
\end{align}
where $\hat{\Sigma}_u = T^{-1} \sum_{t=1}^T \hat{u}_t \hat{u}_t'$, and $\hat{\overline \Psi}_p'$ is constructed from the recursive VAR-based impulse response functions. Here, $\hat{e}_{t,h}$ represents the LS residual of the multi-horizon linear projection with order $p$, and $\hat{u}_t$ is the standard LS residual of the VAR($p$) model.

We deliberately choose the explicit formula for $\Sigma_{zx}$ in \eqref{sigzx} for estimation rather than relying on the sample average, such as $\bar{T}^{-1} \sum_{t=p}^{T-h} \hat{z}_t x_t'$, commonly used in the stationary case. This choice is based on three key considerations: (1) The sample average method loses $h$ observations due to projection. While asymptotically consistent as long as $T-h \rightarrow \infty$, this approach may suffer from efficiency loss and finite sample bias. Recursive VAR-based estimates, in contrast, are well-known for their efficiency. Practitioners can also apply finite sample bias correction for LS VAR slope coefficient estimates (\cite{pope1990biases}). (2) The closed-form expression of $\Sigma_{zx}$ reveals that $\overline{\Psi}_p$ is a block lower triangular matrix, allowing for greater precision by setting zeros in the upper triangular blocks and ones on the main diagonal. By contrast, the sample average method estimates the entire matrix directly, with blocks in the upper triangular area expected to be zero in the population but not necessarily so in finite samples. Although practitioners could manually set these blocks to zero to enhance precision, this adds an extra step to the algorithm. (3) The closed-form formula is flexible, applying to both lag-augmented and non-lag-augmented estimates, as they share the same structure. In contrast, the sample average method requires the use of $\hat{z}_{t,\delta}$ and $x_{t,\delta}$ depending on $\delta$. Our approach simplifies implementation in computer algorithms and facilitates theoretical proof of consistency.

We summarize the uniform inference for feasible estimates with the robust covariance matrix estimates in the following proposition.
\begin{proposition}
\label{unitheo}
    Suppose Assumption \ref{assimeanind}, \ref{assmoment}, and \ref{ass4} hold. Let the autoregressive coefficients be in the parameter space $\mathcal{B}(\delta,c,\epsilon)$, $Z\sim N(0,1)$, $x\in \mathbb R$, $w\in\mathbb R^{pK}$, and $\|w\|=1$.
    \begin{enumerate}[(i)]
        \item Let $\alpha\in (0,1)$. Then 
        \begin{align}
        \underset{T \rightarrow \infty}{\lim}\sup_{\Phi\in \mathcal{B}(0,c,\epsilon)} 
        \sup_{0<h\leq \alpha T }
        \left |
        \textup P_{\Phi} \left(
        \bar T^{1/2} \frac{ w' ( 
        \hat {\beta}_h^{\textup{2S}}         - 
       \beta_h ) }{ (w'\hat \Omega_{\beta,h}w)^{1/2}}
        <x \right) - \textup P(Z<x) \right |\xrightarrow{p} 0. 
    \end{align}
    where $\hat {\beta}_h^{\textup{2S}}$ and $\hat \Omega_{\beta,h}$ are respectively defined in \eqref{lg-2sls} and \eqref{hatomegabeta}.
    \item Let a sequence $\{\bar h_T\}$ of positive integers satisfied $\frac{\bar h}{T}\max (1, \frac{\bar h}{w'\Omega_{\beta,h} w }) \xrightarrow{p}0$. Then
        \begin{align}
        \underset{T \rightarrow \infty}{\lim}\sup_{\Phi\in \mathcal{B}(1,c,\epsilon)} 
        \sup_{0<h\leq \bar h_T }
        \left |
        \textup P_{\Phi} \left(
        \bar T^{1/2} \frac{ w' ( 
        \hat {\beta}_h^{\textup{LA(1)-2S} }         - 
       \beta_h ) }{ (w'\hat \Omega_{\beta,h}w)^{1/2}}
        <x \right) - \textup P(Z<x) \right | \xrightarrow{p} 0.
    \end{align}
    where $\hat {\beta}_h^{\textup{LA(1)-2S} }$ and $\hat \Omega_{\beta,h}$ are respectively defined in \eqref{fealg-2sls} and \eqref{hatomegabeta}.
    \item Let a sequence $\{\bar h_T\}$ of positive integers satisfied $\frac{\bar h^3}{T}\max (1, \frac{\bar h}{w'\Omega_{\beta,h} w }) \xrightarrow{p}0$. Then,
        \begin{align}
        \underset{T \rightarrow \infty}{\lim}\sup_{\Phi\in \mathcal{B}(2,c,\epsilon)} 
        \sup_{0<h\leq \bar h_T }
        \left |
        \textup P_{\Phi} \left(
        \bar T^{1/2} \frac{ w' ( 
        \hat {\beta}_h^{\textup{LA($2$)-2S} }         - 
       \beta_h ) }{ (w'\hat \Omega_{\beta,h}w)^{1/2}}
        <x \right) - \textup P(Z<x) \right | \xrightarrow{p} 0.
    \end{align}
    where $\hat {\beta}_h^{\textup{LA(2)-2S} }$ and $\hat \Omega_{\beta,h}$ are respectively defined in \eqref{fealg-2sls} and \eqref{hatomegabeta}.
    \end{enumerate}
\end{proposition}
See the proof of Proposition \ref{unitheo} in Appendix \ref{proofunitheo}. It is noteworthy that the condition on the horizon $h$ becomes more restrictive as the underlying process becomes more persistent. Readers may wonder about the intuition behind the term $\max(1, \frac{\bar h}{w'\Omega_{\beta,h} w})$. This term suggests that the horizon $h$ depends on the scale of $w'\Omega_{\beta,h} w$: if $w'\Omega_{\beta,h} w$ is proportional to $h$, the condition on $h$ can be relaxed to $h/T, h^3/T \xrightarrow{p} 0$ for the cases of $\delta = 1, 2$, respectively. Conversely, in the worst case where $w'\Omega_{\beta,h} w$ is constant, the condition on $h$ tightens to $h^2/T, h^4/T \xrightarrow{p} 0$ for $\delta = 1, 2$, respectively.

This difference arises from the long-run variance (LRV) of the regression score function, $\Omega_{s,h}$, which is embedded within the asymptotic variance matrix $\Omega_{\beta,h} = \Sigma_{zx}^{-1} \Omega_{s,h} \Sigma_{zx}^{'-1}$. The maximum eigenvalue of $\Omega_{s,h}$ can be proportional to the horizon $h$ when unit roots are present. However, the minimum eigenvalue is only guaranteed to be bounded below by a constant.\footnote{Consider the following example: in a scalar autoregressive process with $u_t \overset{i.i.d.}{\sim} (0, 1)$, the upper-left $2 \times 2$ block of $\Omega_{s,h}$ has elements $\sum_{i=0}^{h-1} \psi_i^2$ on the main diagonal and $\sum_{i=0}^{h-2} \psi_i \psi_{i+1}$ off the diagonal. If the AR process is a random walk, the impulse response functions are $\psi_h = 1$ for all $h$. This results in $\sum_{i=0}^{h-1} \psi_i^2 = h$ and $\sum_{i=0}^{h-2} \psi_i \psi_{i+1} = h-1$. Thus, the maximum eigenvalue of this block equals $2h - 1$, while the minimum eigenvalue remains 1.} Consequently, the scale of $w'\Omega_{\beta,h} w$ is crucial in determining the condition on $h$ when the data may be non-stationary.

Remarks:
\begin{enumerate}[1.]
    \item Our theorem addresses three cases ($\delta = 0, 1, 2$) for data processes exhibiting varying degrees of persistence. In the case of stationarity, where researchers often work with processed (first-differenced) data and reject the unit root null hypothesis using tests like the Dickey–Fuller test, our asymptotic result for $\delta = 0$ demonstrates that the projection horizon can represent a non-trivial portion of the sample size. Conversely, for persistent data, potentially containing unit roots, the asymptotic normality results hold as long as the projection horizon remains small relative to the sample size. This constraint on the projection horizon serves as a caution for empirical research, suggesting that assuming a fixed horizon $h$ in small samples may be inappropriate for level data, which typically exhibits substantial persistence.
    \item Our result generalizes the work of \cite{montiel2021local}, who focused on Sims' impulse response functions under conditions of stationarity and processes with at most one unit root or a local-to-unit root. Notably, they do not include the term $\max (1, \frac{\bar h}{w'\Omega_{\beta,h} w })$ in their conditions. This omission arises because, when only the Sims' impulse responses are of interest, the variance of Local Projection impulse response estimates is proportional to $h$ in the presence of a unit root. Thus, for potentially non-stationary data processes, the condition $\bar h/T \xrightarrow{p} 0$ is sufficient. However, if researchers are interested in lagged Generalized Impulse Responses (GIRs) and encounter ${w'\Omega_{\beta,h} w} \ll h$, a more restrictive condition ($h^2/T \xrightarrow{p} 0$ or $h^4/T \xrightarrow{p} 0$) is required. Nevertheless, if the underlying process is stationary, the projection horizon can represent a non-trivial portion of the sample size for both Sims' impulse responses and lagged GIRs.
\end{enumerate}

\section{Efficiency comparisons}
\label{section6eff}
In this section, we aim to compare the asymptotic efficiency of our two-stage GIR estimates with LS GIR estimates using an illustrative AR(2) process:
\begin{align}
y_t = \phi_1 y_{t-1} + \phi_2 y_{t-2} + u_t ,\quad u_t \overset{i.i.d.}{\sim} N(0,1),
\end{align}
where $\phi_1 = \rho_1 + \rho_2$ and $\phi_2 = -\rho_1 \rho_2$, with $\rho_1$ and $\rho_2$ representing the autoregressive roots. The parameter $\rho_1$ is varied through a grid search over the interval $[0.01, 0.99]$ in increments of 0.01, while $\rho_2$ takes values from the set $\{0.8, 0.5, 0.2, -0.5\}$. We focus on the AR(2) process instead of AR(1) because it captures a wider range of empirical data characteristics, including stationarity, persistence, and cyclicality.

The asymptotic efficiency is evaluated by comparing the asymptotic variance, which is computed using a closed-form formula. For LS projection method estimates in \eqref{2.11}, the asymptotic variance, denoted as $\Omega_{\beta,h}^{LS} := \text{AsymVar}(\hat{\beta}_h^{LS})$, is given by:
\begin{align}
\Omega_{\beta,h}^{LS} = \mathbb{E}[x_t x_t']^{-1} \left(\sum_{k=-h+1}^{h-1} \mathbb{E}[x_t x_{t+k}' e_{t,h} e_{t+k,h}]\right) \mathbb{E}[x_t x_t']^{-1}.
\end{align}
The matrix $\mathbb{E}[x_t x_t']$ contains the autocovariance of $y_t$. The truncation of the long-run variance at order $h-1$ reflects the i.i.d. assumption on $u_t$. This assumption also simplifies computation, as $\mathbb{E}[x_t x_{t+k}' e_{t,h} e_{t+k,h}] = \mathbb{E}[x_t x_{t+k}'] \mathbb{E}[e_{t,h} e_{t+k,h}]$. For the two-stage projection estimates, we use the formula for $\Omega_{\beta,h}$ provided in \eqref{omegabeta}.
\begin{figure}[hp]
    \centering
     \includegraphics[width=0.7 \textwidth]{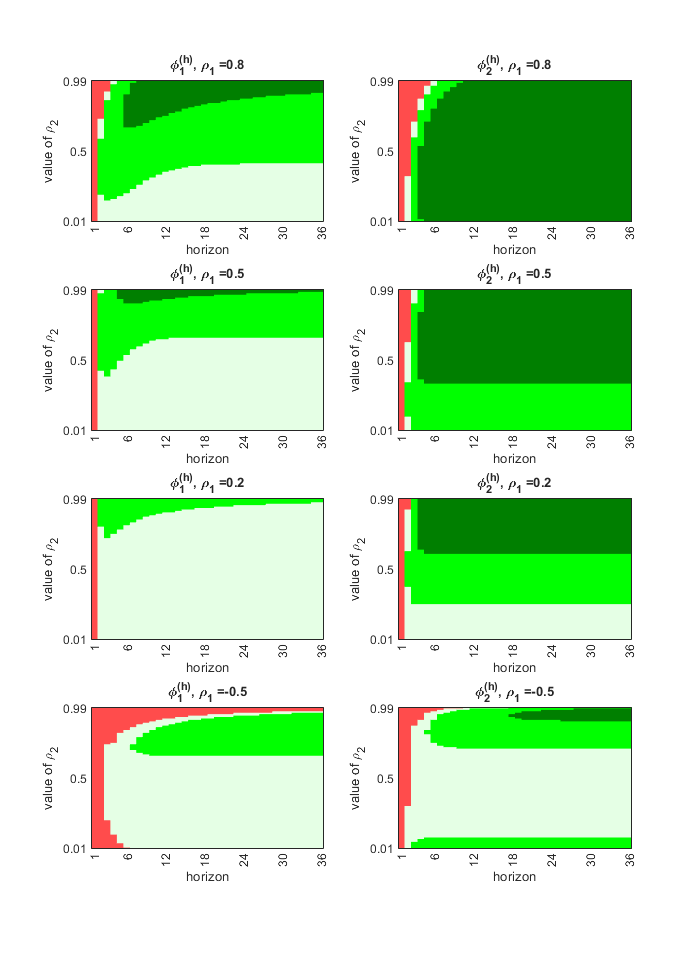}
     \caption{Efficiency comparison}
    \label{figure1}
    \footnotesize{Figure 1: Efficiency comparison for estimand $\phi_1^{(h)}$ and $\phi_2^{(h)}$, where $h=1,2,\cdots,36$.} 
\end{figure}

In Figure \ref{figure1}, we compare the efficiency of two estimators, $\phi_1^{(h)}$ and $\phi_2^{(h)}$, across horizons $h = 1, 2, \dots, 36$. Green/red coloring is used to indicate whether the two-stage estimate is more/less efficient than the LS projection estimate. Varying shades of green represent the degree to which the two-stage estimate outperforms LS projection estimates in terms of standard error. Standard error, rather than variance, is chosen as it directly reflects the narrowing of the confidence interval. The lightest green indicates that the confidence interval of the two-stage estimate is up to 10\% narrower than that of the LS projection estimates. The medium green shows a 10\% to 30\% narrowing, while the darkest green reflects a confidence interval more than 30\% narrower.

Figure \ref{figure1} demonstrates that two-stage estimates only exhibit less efficiency than LS estimates for short horizons, mostly at $h = 1$. In this case, the LS projection estimate coincides with the pseudo maximum likelihood estimate, yielding minimal variance. However, as $h$ increases, it becomes increasingly rare for LS projection estimates to outperform two-stage projection estimates. This occurs in only two cases, both at short horizons: (1) when the data is highly persistent, with $\rho_1 = 0.8$ and $\rho_2$ approaching 1, and (2) when the data is cyclical, with $\rho_1 = -0.5$.

In the first scenario, characterized by highly persistent data where both roots approach one, the outperformance of LS projection estimates at short horizons can be attributed to the well-known phenomenon of super-consistency. However, this advantage diminishes quickly as the horizon $h$ increases. This is because the variance matrix of LS projection estimates requires summing $h-1$ lead-lag autocovariances of $e_{t,h}$, resulting in large values due to persistence. In contrast, the variance matrix of two-stage estimates only depends on the variance of $e_{t,h}$, contributing to its greater efficiency.

In the second scenario, where the data exhibits cyclical patterns, the LS projection estimate's efficiency advantage arises from the behavior of the LRV matrix. The summation of $h-1$ lead-lag autocovariances in the LS projection variance matrix is sensitive to the cyclical nature of the data. When the autocovariances of $e_{t,h}$ differ in sign, they can offset one another, leading to a smaller LRV for $e_{t,h}$ compared to cases where both roots have the same sign and autocovariances decay gradually to zero.

In summary, two-stage estimates typically outperform LS projection estimates across a broad range of parameter spaces and projection horizons, particularly when the horizon is moderately large or the data exhibits moderate stationarity. The only cases where LS projection estimates have an advantage are (1) for a horizon of $h = 1$, (2) for extremely persistent data with a short horizon ($h \leq 3$), or (3) for persistent and cyclical data with a short horizon. Overall, two-stage estimates offer significant efficiency gains in a wide range of practical applications.

\section{Monte Carlo simulation}
\label{section8MC}

In this section, we conduct Monte Carlo simulations to assess the performance of our estimation and inference methods for GIR coefficient estimates. The data generating process (DGP) is based on a bivariate VAR(2) model, with initial observations set to $y_1 = y_2 = \mathbf{0}$. The DGP is specified in terms of autoregressive roots, allowing us to control the stationarity or non-stationarity of the process by adjusting the eigenvalues of the root matrices. We consider a range of processes, including white noise, stationary, I(1), and I(2) models, to provide a comprehensive evaluation of the finite sample performance of our two-stage estimation models with and without lag-augmentation.

\begin{table}[htbp]
 \caption[Monte Carlo simulations: stationary process]{}
 \label{table1}
\begin{center}
Table \thetable

Monte Carlo simulations: stationary process
\begin{align}
     {y}_t = \left[\begin{array}{cc}
     1.1    &  -0.2 \\
     0.2    &  1.1 
    \end{array}\right]  {y}_{t-1} + \left[\begin{array}{cc}
    -0.24    &  0.08 \\
    -0.14    &  -0.28 
    \end{array}\right]  {y}_{t-1} +  {u}_t.
\end{align}
\end{center}

  \begin{center}
  \resizebox{\textwidth}{!}{
    \begin{tabular}{lccccccccccccc}
    \hline
        & \multicolumn{6}{c}{$\phi_{12,1}^{(h)}$}         &     & \multicolumn{6}{c}{$\phi_{12,2}^{(h)}$} \\
    \hline
    $h$   & 1 & 3  & 6   & 12  & 24  & 36   &     & 1 & 3  & 6   & 12  & 24  & 36  \\
    value & -0.200 & -0.438 & -0.370 & -0.098 & -0.003 & -6.13E-05 &     & 0.080 & 0.175 & 0.148 & 0.039 & 0.001 & 2.45E-05 \\
    \hline
        & \multicolumn{13}{c}{bias of estimates} \\
    \hline
    RC-VAR & 0.002 & -0.001 & 0.012 & 0.016 & -0.002 & -4.80E-04 &     & -0.010 & -0.009 & -0.003 & -0.005 & 0.001 & 1.76E-04 \\
    LS-Proj & 0.002 & 0.014 & 0.024 & 0.023 & 0.023 & 0.020 &     & -0.010 & -0.028 & -0.027 & -0.025 & -0.015 & -0.017 \\
    2S-Proj & -0.003 & 0.006 & 0.013 & 0.014 & 0.016 & 0.012 &     & 0.004 & -0.009 & -0.005 & -0.007 & -0.006 & 0.001 \\
    2S(1)-Proj & -0.003 & 0.006 & 0.013 & 0.014 & 0.017 & 0.012 &     & 0.004 & -0.009 & -0.005 & -0.008 & -0.006 & 0.001 \\
    2S(2)-Proj & -0.003 & 0.006 & 0.012 & 0.014 & 0.016 & 0.012 &     & 0.004 & -0.009 & -0.004 & -0.008 & -0.006 & 0.001 \\
     \hline
        & \multicolumn{13}{c}{rmse of estimates} \\
    \hline  
    RC-VAR & 0.071 & 0.107 & 0.119 & 0.061 & 0.010 & 0.002 &     & 0.074 & 0.082 & 0.052 & 0.024 & 0.004 & 0.001 \\
    LS-Proj & 0.071 & 0.138 & 0.163 & 0.161 & 0.162 & 0.174 &     & 0.074 & 0.139 & 0.165 & 0.167 & 0.164 & 0.179 \\
    2S-Proj & 0.077 & 0.126 & 0.147 & 0.151 & 0.153 & 0.158 &     & 0.121 & 0.114 & 0.109 & 0.108 & 0.108 & 0.110 \\
    2S(1)-Proj & 0.077 & 0.126 & 0.147 & 0.151 & 0.153 & 0.159 &     & 0.121 & 0.113 & 0.109 & 0.108 & 0.109 & 0.110 \\
    2S(2)-Proj & 0.077 & 0.126 & 0.148 & 0.151 & 0.152 & 0.160 &     & 0.122 & 0.113 & 0.110 & 0.108 & 0.109 & 0.112 \\
     \hline
        & \multicolumn{13}{c}{empirical size of tests (5\% nominal size)} \\
    \hline    
    RC-VAR & 0.052 & 0.056 & 0.075 & 0.148 & 0.139 & 0.130 &     & 0.055 & 0.047 & 0.089 & 0.152 & 0.128 & 0.113 \\
    LS-Proj & 0.055 & 0.087 & 0.100 & 0.074 & 0.095 & 0.107 &     & 0.050 & 0.080 & 0.095 & 0.086 & 0.077 & 0.127 \\
    2S(0)-Proj & 0.053 & 0.065 & 0.085 & 0.064 & 0.050 & 0.058 &     & 0.059 & 0.061 & 0.055 & 0.054 & 0.058 & 0.048 \\
    $\textup{2S(0)-Proj}_b$ & 0.047 & 0.049 & 0.068 & 0.048 & 0.031 & 0.039 &     & 0.050 & 0.059 & 0.051 & 0.047 & 0.046 & 0.040 \\
    2S(1)-Proj & 0.052 & 0.068 & 0.086 & 0.062 & 0.052 & 0.064 &     & 0.060 & 0.063 & 0.056 & 0.057 & 0.059 & 0.048 \\
    $\textup{2S(1)-Proj}_b$ & 0.046 & 0.053 & 0.069 & 0.048 & 0.034 & 0.041 &     & 0.054 & 0.056 & 0.055 & 0.051 & 0.049 & 0.044 \\
    2S(2)-Proj & 0.055 & 0.063 & 0.087 & 0.064 & 0.053 & 0.064 &     & 0.062 & 0.066 & 0.065 & 0.052 & 0.056 & 0.056 \\
    $\textup{2S(2)-Proj}_b$ & 0.047 & 0.051 & 0.069 & 0.051 & 0.031 & 0.041 &     & 0.054 & 0.056 & 0.054 & 0.044 & 0.046 & 0.042 \\
    \hline
        & \multicolumn{13}{c}{coverage ratio of nominal 95\% confidence interval} \\
    \hline   
    RC-VAR & 0.948 & 0.944 & 0.925 & 0.852 & 0.861 & 0.870 &     & 0.945 & 0.953 & 0.911 & 0.848 & 0.872 & 0.887 \\
    LS-Proj & 0.945 & 0.913 & 0.900 & 0.926 & 0.905 & 0.893 &     & 0.950 & 0.920 & 0.905 & 0.914 & 0.923 & 0.873 \\
    2S(0)-Proj & 0.947 & 0.935 & 0.915 & 0.936 & 0.950 & 0.942 &     & 0.941 & 0.939 & 0.945 & 0.946 & 0.942 & 0.952 \\
    $\textup{2S(0)-Proj}_b$ & 0.953 & 0.951 & 0.932 & 0.952 & 0.969 & 0.961 &     & 0.950 & 0.941 & 0.949 & 0.953 & 0.954 & 0.960 \\
    2S(1)-Proj & 0.948 & 0.932 & 0.914 & 0.938 & 0.948 & 0.936 &     & 0.940 & 0.937 & 0.944 & 0.943 & 0.941 & 0.952 \\
    $\textup{2S(1)-Proj}_b$ & 0.954 & 0.947 & 0.931 & 0.952 & 0.966 & 0.959 &     & 0.946 & 0.944 & 0.945 & 0.949 & 0.951 & 0.956 \\
    2S(2)-Proj & 0.945 & 0.937 & 0.913 & 0.936 & 0.947 & 0.936 &     & 0.938 & 0.934 & 0.935 & 0.948 & 0.944 & 0.944 \\
    $\textup{2S(2)-Proj}_b$ & 0.953 & 0.949 & 0.931 & 0.949 & 0.969 & 0.959 &     & 0.946 & 0.944 & 0.946 & 0.956 & 0.954 & 0.958 \\
     \hline
        & \multicolumn{13}{c}{average width of nominal 95\% confidence interval} \\
    \hline 
    RC-VAR & 0.274 & 0.408 & 0.446 & 0.236 & 0.035 & 0.005 &     & 0.279 & 0.317 & 0.196 & 0.091 & 0.013 & 0.002 \\
    LS-Proj & 0.275 & 0.495 & 0.582 & 0.595 & 0.593 & 0.592 &     & 0.280 & 0.500 & 0.588 & 0.607 & 0.604 & 0.594 \\
    2S(0)-Proj & 0.293 & 0.473 & 0.540 & 0.565 & 0.581 & 0.596 &     & 0.458 & 0.443 & 0.412 & 0.408 & 0.418 & 0.430 \\
    $\textup{2S(0)-Proj}_b$ & 0.305 & 0.508 & 0.590 & 0.619 & 0.637 & 0.660 &     & 0.470 & 0.460 & 0.428 & 0.420 & 0.432 & 0.446 \\
    2S(1)-Proj & 0.293 & 0.473 & 0.540 & 0.565 & 0.581 & 0.596 &     & 0.458 & 0.443 & 0.412 & 0.408 & 0.418 & 0.430 \\
    $\textup{2S(1)-Proj}_b$ & 0.305 & 0.508 & 0.590 & 0.619 & 0.637 & 0.659 &     & 0.471 & 0.460 & 0.428 & 0.420 & 0.432 & 0.446 \\
    2S(2)-Proj & 0.293 & 0.473 & 0.540 & 0.565 & 0.581 & 0.596 &     & 0.458 & 0.443 & 0.412 & 0.408 & 0.418 & 0.430 \\
    $\textup{2S(2)-Proj}_b$ & 0.306 & 0.511 & 0.593 & 0.623 & 0.640 & 0.664 &     & 0.471 & 0.462 & 0.432 & 0.425 & 0.437 & 0.452 \\
    \hline 
    \end{tabular}}
    \end{center}
    
      \medskip
    \noindent
   \footnotesize Note - The DGP follows stationary VAR(2) process, $u_t\overset{i.i.d.}{\sim} N(0, [1,0.5 ; 0.5,1] )$. Two $2\times 2$ root matrices are $[0.7,-0.2;0,0.7]$ and $[0.4,0;0.2,0.4]$, whose eigenvalues are $0.7,0.7$ and $0.4,0.4$, respectively. The parameters of interest, $\phi_{12,1}^{(h)}$ and $\phi_{12,2}^{(h)}$, are the first component in second column of coefficient matrix $\Phi_1^{(h)}$ and $\Phi_2^{(h)}$, respectively. The number of replication is 1000. For each replication, the number of bootstrap simulation is 2000. The "value" row shows the true value of the parameters. 
\end{table}%

\begin{table}[htbp]
  \caption[Monte Carlo simulations: I(1) process]{}
 \label{table2}
\begin{center}
Table \thetable

Monte Carlo simulations: I(1) process
\begin{align}
     {y}_t = \left[\begin{array}{cc}
     1.1    &  -0.2 \\
     0.2    &  1.4 
    \end{array}\right]  {y}_{t-1} + \left[\begin{array}{cc}
     -0.24    &  0.08 \\
     -0.2    &  -0.4
    \end{array}\right]  {y}_{t-1} +  {u}_t.
\end{align}
\end{center}

  \begin{center}
  \resizebox{\textwidth}{!}{
    \begin{tabular}{lccccccccccccc}
    \hline
      & \multicolumn{6}{c}{$\phi_{12,1}^{(h)}$}         &     & \multicolumn{6}{c}{$\phi_{12,2}^{(h)}$} \\
    \hline
    $h$   & 1 & 3  & 6   & 12  & 24  & 36   &     & 1 & 3  & 6   & 12  & 24  & 36  \\
    value & -0.200 & -0.606 & -0.930 & -1.090 & -1.111 & -1.111 &     & 0.080 & 0.242 & 0.372 & 0.436 & 0.444 & 0.444 \\
    \hline
        & \multicolumn{13}{c}{bias of estimates} \\
    \hline
 RC-VAR & 0.069 & 0.093 & 0.091 & 0.198 & 0.424 & 0.576 &     & 0.078 & 0.098 & 0.084 & 0.128 & 0.193 & 0.245 \\
    LS-Proj & 0.069 & 0.134 & 0.175 & 0.309 & 0.564 & 0.766 &     & 0.078 & 0.159 & 0.207 & 0.327 & 0.514 & 0.622 \\
    2S(1)-Proj & 0.076 & 0.125 & 0.154 & 0.284 & 0.520 & 0.726 &     & 0.133 & 0.142 & 0.148 & 0.205 & 0.275 & 0.353 \\
    2S(2)-Proj & 0.077 & 0.128 & 0.164 & 0.283 & 0.534 & 0.706 &     & 0.136 & 0.144 & 0.156 & 0.213 & 0.300 & 0.367 \\
    \hline
        & \multicolumn{13}{c}{rmse of estimates} \\
    \hline       
    RC-VAR & 0.074 & 0.183 & 0.319 & 0.442 & 0.546 & 0.646 &     & 0.074 & 0.152 & 0.172 & 0.182 & 0.223 & 0.263 \\
    LS-Proj & 0.074 & 0.208 & 0.356 & 0.541 & 0.792 & 0.968 &     & 0.071 & 0.204 & 0.341 & 0.497 & 0.736 & 0.880 \\
    2S(1)-Proj & 0.077 & 0.196 & 0.346 & 0.569 & 0.871 & 1.072 &     & 0.108 & 0.166 & 0.201 & 0.232 & 0.290 & 0.344 \\
    2S(2)-Proj & 0.075 & 0.198 & 0.336 & 0.555 & 0.865 & 1.066 &     & 0.105 & 0.156 & 0.199 & 0.240 & 0.310 & 0.362 \\  
      \hline
        & \multicolumn{13}{c}{empirical size of tests (5\% nominal size)} \\
    \hline       
    RC-VAR & 0.058 & 0.060 & 0.060 & 0.242 & 0.452 & 0.540 &     & 0.048 & 0.060 & 0.066 & 0.118 & 0.337 & 0.480 \\
    LS-Proj & 0.055 & 0.074 & 0.092 & 0.189 & 0.334 & 0.483 &     & 0.053 & 0.088 & 0.092 & 0.114 & 0.169 & 0.227 \\
    2S(1)-Proj & 0.060 & 0.061 & 0.060 & 0.148 & 0.273 & 0.398 &     & 0.052 & 0.067 & 0.056 & 0.095 & 0.144 & 0.233 \\
    $\textup{2S(1)-Proj}_b$ & 0.051 & 0.054 & 0.037 & 0.051 & 0.110 & 0.194 &     & 0.047 & 0.064 & 0.036 & 0.060 & 0.087 & 0.146 \\
    2S(2)-Proj & 0.058 & 0.079 & 0.077 & 0.130 & 0.281 & 0.389 &     & 0.051 & 0.067 & 0.071 & 0.101 & 0.159 & 0.225 \\
    $\textup{2S(2)-Proj}_b$ & 0.054 & 0.068 & 0.046 & 0.057 & 0.103 & 0.159 &     & 0.040 & 0.056 & 0.041 & 0.068 & 0.101 & 0.142 \\
    \hline
        & \multicolumn{13}{c}{coverage ratio of nominal 95\% confidence interval} \\
    \hline
    RC-VAR & 0.942 & 0.940 & 0.940 & 0.758 & 0.548 & 0.460 &     & 0.952 & 0.940 & 0.934 & 0.882 & 0.663 & 0.520 \\
    LS-Proj & 0.945 & 0.926 & 0.908 & 0.811 & 0.666 & 0.517 &     & 0.947 & 0.912 & 0.908 & 0.886 & 0.831 & 0.773 \\
    2S(1)-Proj & 0.940 & 0.939 & 0.940 & 0.852 & 0.727 & 0.602 &     & 0.948 & 0.933 & 0.944 & 0.905 & 0.856 & 0.767 \\
    $\textup{2S(1)-Proj}_b$ & 0.949 & 0.946 & 0.963 & 0.949 & 0.890 & 0.806 &     & 0.953 & 0.936 & 0.964 & 0.940 & 0.913 & 0.854 \\
    2S(2)-Proj & 0.942 & 0.921 & 0.923 & 0.870 & 0.719 & 0.611 &     & 0.949 & 0.933 & 0.929 & 0.899 & 0.841 & 0.775 \\
    $\textup{2S(2)-Proj}_b$ & 0.946 & 0.932 & 0.954 & 0.943 & 0.897 & 0.841 &     & 0.960 & 0.944 & 0.959 & 0.932 & 0.899 & 0.858 \\
    \hline
        & \multicolumn{13}{c}{average width of nominal 95\% confidence interval} \\
    \hline
     RC-VAR & 0.261 & 0.348 & 0.354 & 0.499 & 0.712 & 0.872 &     & 0.301 & 0.378 & 0.332 & 0.440 & 0.433 & 0.441 \\
    LS-Proj & 0.263 & 0.482 & 0.621 & 0.938 & 1.268 & 1.372 &     & 0.303 & 0.554 & 0.719 & 1.119 & 1.561 & 1.690 \\
    2S(1)-Proj & 0.294 & 0.474 & 0.589 & 0.878 & 1.272 & 1.475 &     & 0.528 & 0.536 & 0.589 & 0.730 & 0.874 & 0.965 \\
    $\textup{2S(1)-Proj}_b$ & 0.306 & 0.509 & 0.645 & 1.019 & 1.651 & 2.008 &     & 0.541 & 0.563 & 0.626 & 0.784 & 0.968 & 1.083 \\
    2S(2)-Proj & 0.294 & 0.478 & 0.592 & 0.889 & 1.296 & 1.510 &     & 0.527 & 0.540 & 0.591 & 0.736 & 0.892 & 0.983 \\
    $\textup{2S(2)-Proj}_b$ & 0.308 & 0.517 & 0.654 & 1.037 & 1.688 & 2.069 &     & 0.544 & 0.568 & 0.632 & 0.801 & 1.010 & 1.134 \\
    \hline
    \end{tabular}  }
    \end{center}
        
      \medskip
    \noindent
   \footnotesize Note - The DGP follows a VAR(2) process integrated at order one, $u_t\overset{i.i.d.}{\sim} N(0, [1,0.5 ; 0.5,1] )$. Two $2\times 2$ root matrices are $[0.7,-0.2;0,1]$ and $[0.4,0;0.2,0.4]$, whose eigenvalues are $1,0.7$ and $0.4,0.4$, respectively. The parameters of interest, $\phi_{12,1}^{(h)}$ and $\phi_{12,2}^{(h)}$, are the first component in second column of coefficient matrix $\Phi_1^{(h)}$ and $\Phi_2^{(h)}$, respectively. The number of replication is 1000. For each replication, the number of bootstrap simulation is 2000. The "value" row shows the true value of the parameters. Since the DGP contains one unit root, we present results of two-stage estimation with one/two lag-augmentation.
\end{table}%

We examine three versions of our two-stage estimation model: the standard two-stage model without lag-augmentation, the two-stage model with one additional lag, and the two-stage model with two additional lags. As benchmarks, we include the recursive VAR-based estimation method with delta-method inference and the standard least squares (LS) linear projection model with HAC inference. Additionally, we provide $t$-interval bootstrap results for the three versions of our two-stage estimates.\footnote{The algorithm for simulation-based inference is presented in Appendix \ref{section7simul}.} The horizons considered are $h = 1, 3, 6, 12, 24, 36$. The goal of these simulations is to evaluate the finite sample performance of the estimates in terms of bias, root mean squared error (RMSE), empirical test size, confidence interval coverage ratio, and average confidence interval width. For the benchmark LS projection model with HAC inference, we use MATLAB’s "hac" function with a Bartlett kernel and bandwidth equal to $h$ to estimate HAC standard errors. The recursive VAR-based method uses the closed-form formula for covariance matrix estimation as described in \cite{lutkepohl1990asymptotic}.

Simulation results for stationary, I(1), and I(2) processes are presented in Tables \ref{table1} and \ref{table2}. The abbreviations for each model are as follows: (1) RC-VAR: recursive VAR-based method with delta-method inference, utilizing the explicit formula for the Jacobian matrix reported in Proposition 3.6 (p. 110) of \cite{lutkepohl2005new}; (2) LS-Proj: LS projection method with HAC inference, using critical values from the $z$-table (MATLAB: ‘hac’ command, kernel = Bartlett, bandwidth = h); (3) 2S-Proj($\delta$): two-step method with $\delta$ lag-augmentation, using critical values from the $z$-table; and (4) $\textup{2S-Proj}(\delta)_b$: two-step method with equal-tailed percentile t-interval bootstrap (Wild bootstrap, as per \cite{gonccalves2004bootstrapping}). The results for white noise and I(2) processes are shown in Tables \ref{table0} and \ref{table3} in Appendix \ref{mcsim}. Specific parameter values and eigenvalues of the two polynomial root matrices are detailed in the table footnotes. These tables highlight the performance of two coefficients, $\phi_{12,1}^{(h)}$ and $\phi_{12,2}^{(h)}$, which represent the causality of the second variable on the first variable at horizon $h$.

Tables \ref{table1} and \ref{table2} offer a comprehensive evaluation of the performance of GIR coefficient estimates across different estimation methods. A comparison between the two-stage projection estimates and the LS projection estimates reveals that the two-stage estimates generally exhibit satisfactory finite sample performance. These estimates are characterized by more accurate coverage ratios and, in certain cases, greater efficiency, as reflected in the narrower average width of the confidence intervals. In contrast, the  LS projection estimates with Newey-West standard errors show distorted coverage ratios, with the distortion becoming more pronounced for persistent data and at longer horizons.


In terms of efficiency, measured by the average width of confidence intervals, the two-stage estimates generally outperform multi-horizon linear projection estimates, irrespective of whether lag-augmentation is applied. As expected from econometric theory, iterated VAR estimates are more efficient than either LS or two-stage projection methods. However, iterated VAR estimates exhibit the poorest coverage ratios due to reliance on delta-method inference.

In summary, the simulations demonstrate that GIR estimates obtained from the two-stage multi-horizon linear projection, utilizing our robust covariance matrix estimates, show relatively reliable finite sample performance compared to recursive estimates based on the explicit formula of delta-method inference or LS projection estimates with HAC standard error adjustments. While the two-stage estimates experience some efficiency loss due to the multi-horizon projection residual, they generally improve efficiency relative to LS projection estimates. 

\section{Empirical application: causality of economic uncertainty}
\label{section9Emp}
In this section, we examine the bidirectional Granger causality between economic uncertainty and various macroeconomic indices across multiple horizons. The concept of economic uncertainty has garnered significant attention in recent literature, as highlighted by studies such as \cite{baker2016measuring}, \cite{jurado2015measuring}, and \cite{rossi2015macroeconomic}. Empirical investigations into the causality of economic uncertainty can be found in works like \cite{salamaliki2019transmission}, \cite{gao2022oil}, and \cite{fernandez2023cross}, among others. We implement the Granger causality test over multiple horizons, report the Wald test statistics and corresponding $p$-values, and present both tables and figures for the GIR coefficient estimates, clarifying the causal relationship between economic uncertainty and macroeconomic indices.

\begin{table}[htbp]
  \caption[Table of multi-horizon causality]{}
\begin{center}
Table \thetable

Robust multi-horizon causality results
\end{center}

  \begin{center}
  \resizebox{\textwidth}{!}{
    \begin{tabular}{c|ccccc}
    \multicolumn{6}{c}{full sample 1974-2023} \\
    \hline
        & \multicolumn{1}{c}{CFNAI} & \multicolumn{1}{c}{JNL} & \multicolumn{1}{c}{Unemp} & \multicolumn{1}{c}{Inflation} & \multicolumn{1}{c}{FFR} \\
    \hline
    CFNAI & \multicolumn{1}{c}{1-6,31-32} & \multicolumn{1}{c}{25,28-36} & \multicolumn{1}{c}{1-21,28-36} & \multicolumn{1}{c}{10,21,28-35} & \multicolumn{1}{c}{1-13,15-22,25,28-36} \\
    JNL & \multicolumn{1}{c}{1,32} & \multicolumn{1}{c}{1-15,29-36} & \multicolumn{1}{c}{3-6,8-13,29-30,32-36} & \multicolumn{1}{c}{29-32,35} & \multicolumn{1}{c}{2-3,6,31} \\
    Unemp & \multicolumn{1}{c}{1-6,25,31-36} & \multicolumn{1}{c}{1-4,27-36} & \multicolumn{1}{c}{1-13,28-36} & \multicolumn{1}{c}{9-26,28-36} & \multicolumn{1}{c}{1-6,8-26,28-36} \\
    Inflation & \multicolumn{1}{c}{-} & \multicolumn{1}{c}{3-13,27-36} & \multicolumn{1}{c}{-} & \multicolumn{1}{c}{1-21,23,26,29-30,34} & \multicolumn{1}{c}{12-13,16-17} \\
    FFR & \multicolumn{1}{c}{10-21,} & \multicolumn{1}{c}{1-5,27-36} & \multicolumn{1}{c}{11,13,15-16,18,20,23,28-31,33-36} & \multicolumn{1}{c}{28,30-35} & \multicolumn{1}{c}{1-12,16-18,36} \\
    \hline
    \hline
    \multicolumn{1}{r}{} &     &     &     &     &  \\
    \multicolumn{6}{c}{post-volker 1987-2023} \\
    \hline
        & \multicolumn{1}{c}{CFNAI} & \multicolumn{1}{c}{JNL} & \multicolumn{1}{c}{Unemp} & \multicolumn{1}{c}{Inflation} & \multicolumn{1}{c}{FFR} \\
    \hline
    CFNAI & \multicolumn{1}{c}{1-6,29,31,33-36} & \multicolumn{1}{c}{24-36} & \multicolumn{1}{c}{1-22,24-25,28-30,32-36} & \multicolumn{1}{c}{3,5-6,9-10,18,28-36} & \multicolumn{1}{c}{1,4,9-12,31-36} \\
    JNL & \multicolumn{1}{c}{1,2,31-32,34} & \multicolumn{1}{c}{1-10,29-36} & \multicolumn{1}{c}{1-14,16,18,30-36} & \multicolumn{1}{c}{10-14,22,24,27,29-36} & \multicolumn{1}{c}{1-2,29-36} \\
    Unemp & \multicolumn{1}{c}{1-6,26,28-31.33-36,} & \multicolumn{1}{c}{27,29-36} & \multicolumn{1}{c}{1-3,6,11-12,16-22,28-36} & \multicolumn{1}{c}{28-36} & \multicolumn{1}{c}{1-2,4,6-15,25-36} \\
    Inflation & \multicolumn{1}{c}{-} & \multicolumn{1}{c}{9-10,12,27,29-36} & \multicolumn{1}{c}{32-36} & \multicolumn{1}{c}{1-16,19-20,22-24,28-33,35-36} & \multicolumn{1}{c}{-} \\
    FFR & \multicolumn{1}{c}{28} & \multicolumn{1}{c}{2-9,30-36} & \multicolumn{1}{c}{4-5,7,31-36} & \multicolumn{1}{c}{29-36} & \multicolumn{1}{c}{1-14,30} \\
    \hline
    \hline
    \multicolumn{1}{r}{} &     &     &     &     &  \\
    \multicolumn{6}{c}{pre-pandemic 1987-2019} \\
    \hline
        & \multicolumn{1}{c}{CFNAI} & \multicolumn{1}{c}{JNL} & \multicolumn{1}{c}{Unemp} & \multicolumn{1}{c}{Inflation} & \multicolumn{1}{c}{FFR} \\
    \hline
    CFNAI & 1-7,27-29,32,33-36 & -   & 1-12& -   & 1-14 \\
    JNL & 1-8,14,25-26,31-32 & 1-13& 1-36& 30,32,34-36 & 1,36 \\
    Unemp & 1-14,16-21,25 & -   & 1-11& -   & - \\
    Inflation & -   & 10 & -   & 1-14,16-17,20-22 & - \\
    FFR & - & -   & -   & -   & 1-14 \\
    \hline
    \hline
    \end{tabular}%
  \label{GCtable}
  }
  \end{center}
  \medskip
    \noindent
   \footnotesize  Note - This table present the months that the column variables are significantly causal on row variables. For each pair of variables, we test multi-horizon non-causality at level of $95\%$, $\mathcal{H}_0:\phi_{ij,k}^{(h)}=0$ for $k=1,2,\cdots,12$. The horizon considered is $h=1,2,\cdots,36$. We implement two-stage estimation with order 12, 15, and 18 for robustness check. The results indicate that the months demonstrate significant causality across all three orders.
\end{table}%

The dataset used in this analysis consists of five monthly variables spanning from January 1974 to June 2023, totaling 593 observations. These variables include the Chicago Fed National Activity Index (CFNAI), the economic uncertainty index (sourced from \cite{jurado2015measuring}, JLN 3-month Ahead Macroeconomic Uncertainty, hereafter referred to as JLN), the unemployment rate (Unemp), inflation, and the Federal Funds Rate (FFR)\footnote{See data source in Appendix}. The focus of this investigation is the multi-horizon Granger causality from the uncertainty index to the macroeconomic indicators.

The VAR model employed in this analysis includes 12 lags and an intercept term. This lag order selection is based on the information criteria computed using the ‘VARselect’ command in the R package ‘vars’ (version 1.6-1), with \textit{lag.max = 20} and \textit{type = ‘const’}. The results of the information criteria selection are as follows: AIC = 13, HQ = 3, SC = 2, and FPE = 13. Given that our dataset consists of typical monthly macroeconomic aggregates, we select 12 lags. This choice aligns with the argument made by \cite{hamilton2004comment} for impulse response estimation in monthly VAR models. However, we acknowledge the potential sensitivity of empirical results to the VAR lag order selection. To address this, we perform a robustness check by extending our estimations to lag orders of 15 and 18, as recommended by \cite{kilian2011reliable} in their empirical exercises.

In this study, we apply the Wald test on a set of GIRs to assess causality from one variable to another across specific horizons, as expressed by the null hypothesis:
\[
\mathcal{H}_0: \Phi_{ij,k}^{(h)} = 0, \text{ for } k = 1, 2, \dots, 12.
\]
The coefficient estimates and covariance matrix are derived using the two-stage estimation method outlined in this paper. Table \ref{GCtable} presents the Granger causality from the row variables to the column variables across multiple horizons. Notably, the cells along the main diagonal demonstrate the predictability of each variable concerning its own future output. It is evident that all five macroeconomic variables exhibit significant persistence.

\begin{figure}[htbp]
    \centering
    \includegraphics[width=0.5 \textwidth]{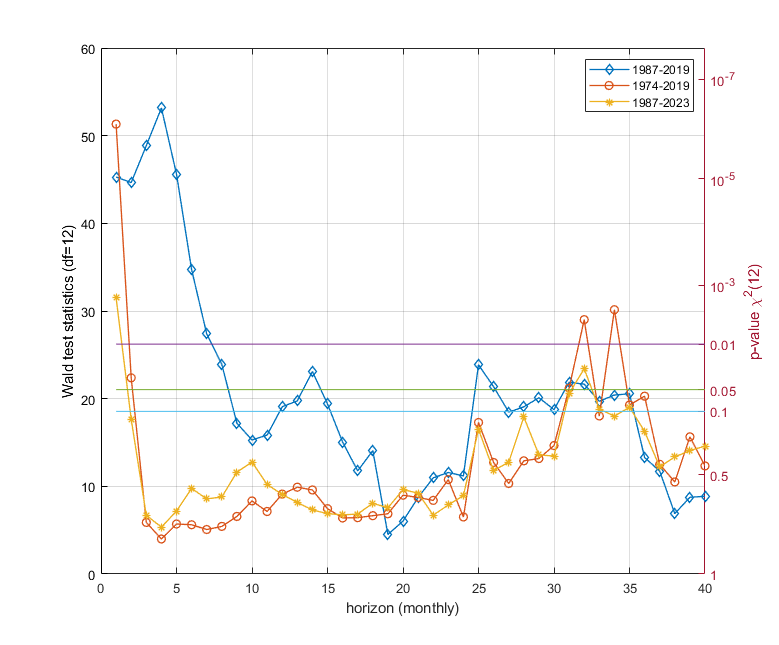}
    \caption[Multi-horizon causality test (JNL$\xrightarrow{h}$CFNAI)]{}
    \label{GCfigure}
    Figure \thefigure. Multi-horizon Granger causality test from  JNL to CFNAI. \footnotesize  The X-axis displays the monthly horizon, the Y-axis displays the $p$-value. We conduct the Wald test on the first twelve GIR coefficients from JNL to CFNAI. We consider three samples: 1987-2019, 1987-2023, and 1974-2023. For robustness checks, we consider three VAR models with lag 12, 15, and 18. For each sub-sample and horizon, the presented $p$-value is the minimum one from the three VAR models.  The three horizontal lines denote the significant levels of 90\%, 95\%, and 99\%, respectively.     
\end{figure}

\begin{figure}[htbp]
    \centering
    \caption{Two-stage estimates of generalized impulse responses to JNL}
    \includegraphics[width=0.7 \textwidth]{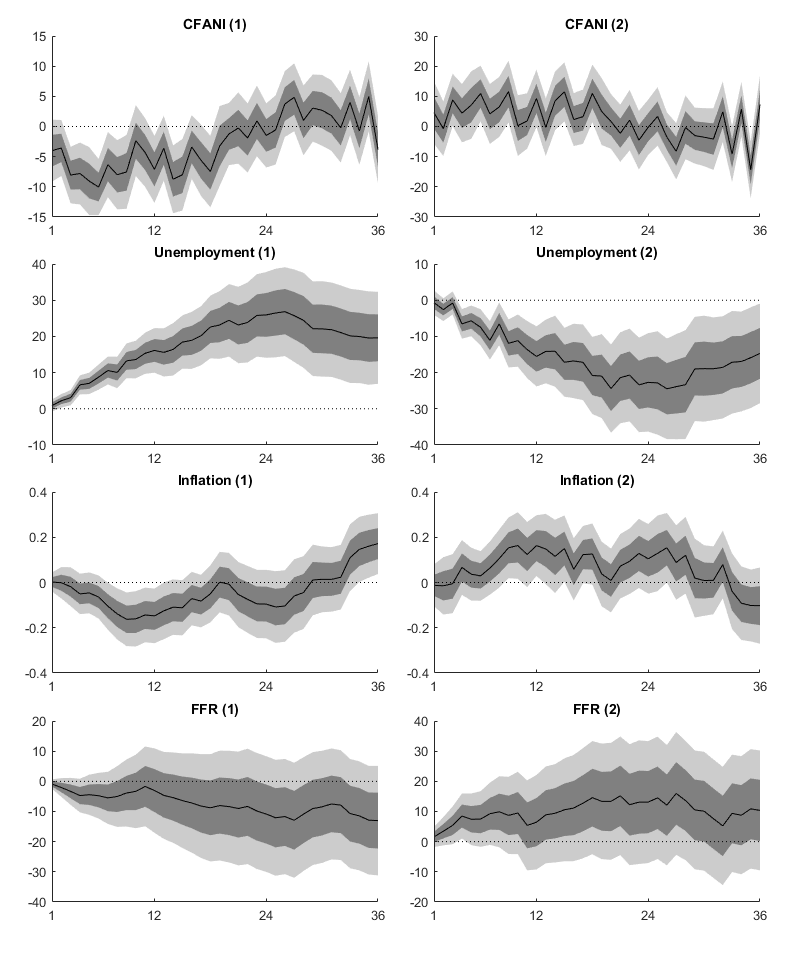}
    
     Figure \thefigure. Two-stage estimates of GIRs to Economic uncertainty index (JNL). \footnotesize{Note: The figure shows first two GIRs of macro-variables to economic uncertainty index (JNL). For instance, CFANI($1$) subplot displays the first GIR (the reduced-form Sims impulse response) of CFANI to JNL over 36 months; and CFANI($2$) subplot displays the second order GIR of CFANI to JNL over 36 months. The confidence interval for each estimator is built upon the asymptotic variance computed with the explicit formula of \eqref{omegabeta} and 68\% and 95\% $z$-score. The comparable GIR figures with LS projection method and recursive methods are presented in Appendix \ref{empres}.} 
    \label{figuregir}
    
\end{figure}

\begin{table}[p]
  \caption[Table of generalized impulse response estimates]{}
\begin{center}
Table \thetable

Two-stage estimates of generalized impulse responses
\end{center}

  \begin{center}
  \resizebox{\textwidth}{!}{
       \begin{tabular}{c|c|cccccccccccc}
    \hline
    \hline
    horizon & Causality test & \multicolumn{12}{c}{two-stage estimates of twelve-lag generalized impulse responses} \\
    \hline
    \multirow{2}[2]{*}{1} & 45.281* & -3.999 & 4.584 & -5.968 & 5.738 & -4.035 & 0.639 & 4.870 & -6.462 & 1.632 & 6.022 & -7.614 & 2.496 \\
        & (9.22E-06) & (0.160) & (0.423) & (0.284) & (0.269) & (0.453) & (0.910) & (0.385) & (0.198) & (0.764) & (0.300) & (0.152) & (0.619) \\
    \hline
    \multirow{2}[2]{*}{2} & 44.647* & -3.565 & -0.729 & 4.074 & -2.894 & -0.025 & 4.686 & -5.894 & 1.182 & 7.435 & -9.532 & 3.741 & -0.025 \\
        & (1.18E-05) & (0.167) & (0.885) & (0.425) & (0.570) & (0.996) & (0.380) & (0.209) & (0.818) & (0.171) & (0.061) & (0.444) & (0.996) \\
    \hline
    \multirow{2}[2]{*}{3} & 48.862* & -8.056* & 8.818 & -6.188 & 3.975 & 0.443 & -2.980 & 0.990 & 5.211 & -7.770 & 4.774 & -2.803 & -0.731 \\
        & (2.21E-06) & (0.002) & (0.067) & (0.244) & (0.487) & (0.938) & (0.583) & (0.864) & (0.387) & (0.172) & (0.405) & (0.660) & (0.913) \\
    \hline
    \multirow{2}[2]{*}{4} & 55.728* & -7.790* & 4.423 & 0.050 & 3.422 & -6.011 & 3.872 & 3.239 & -6.996 & 4.210 & -1.891 & -2.332 & 2.022 \\
        & (1.34E-07) & (0.006) & (0.410) & (0.993) & (0.571) & (0.277) & (0.490) & (0.585) & (0.222) & (0.465) & (0.761) & (0.721) & (0.761) \\
    \hline
    \multirow{2}[2]{*}{5} & 45.654* & -9.052* & 7.099 & 0.370 & -1.166 & -1.963 & 7.439 & -8.739 & 2.957 & 0.822 & -3.120 & 0.418 & 4.700 \\
        & (7.96E-06) & (0.004) & (0.254) & (0.954) & (0.840) & (0.744) & (0.240) & (0.142) & (0.618) & (0.899) & (0.647) & (0.953) & (0.501) \\
    \hline
    \multirow{2}[2]{*}{6} & 34.729* & -10.043* & 10.923* & -4.996 & 2.004 & 2.445 & -3.832 & -0.441 & 1.371 & -1.741 & -0.337 & 3.848 & -2.664 \\
        & (0.001) & (0.000) & (0.033) & (0.365) & (0.737) & (0.687) & (0.501) & (0.940) & (0.825) & (0.797) & (0.962) & (0.578) & (0.673) \\
    \hline
    \multirow{2}[2]{*}{7} & 27.394* & -6.338* & 4.179 & -1.441 & 6.309 & -8.394 & 3.218 & 0.129 & -3.369 & 1.789 & 2.672 & -4.081 & -2.652 \\
        & (6.78E-03) & (0.035) & (0.449) & (0.808) & (0.302) & (0.170) & (0.593) & (0.984) & (0.623) & (0.799) & (0.693) & (0.515) & (0.652) \\
    \hline
    \multirow{2}[2]{*}{8} & 23.908* & -7.998* & 6.476 & 3.519 & -4.893 & -2.146 & 6.265 & -8.511 & 3.942 & 3.728 & -5.984 & -1.943 & 6.785 \\
        & (0.021) & (0.012) & (0.243) & (0.541) & (0.392) & (0.706) & (0.300) & (0.186) & (0.549) & (0.562) & (0.323) & (0.735) & (0.259) \\
    \hline
    \multirow{2}[2]{*}{9} & 17.201 & -7.529* & 11.565* & -8.009 & 1.495 & 1.563 & -4.204 & 1.646 & 3.110 & -3.638 & -3.805 & 6.487 & -2.021 \\
        & (0.142) & (0.027) & (0.042) & (0.162) & (0.799) & (0.799) & (0.506) & (0.793) & (0.623) & (0.549) & (0.519) & (0.307) & (0.757) \\
    \hline
    \multirow{2}[2]{*}{12} & 19.073 & -7.109* & 9.303 & -8.101 & 3.384 & 3.520 & -4.736 & -4.037 & 8.813 & -4.169 & -0.005 & -0.152 & -0.071 \\
        & (0.087) & (0.021) & (0.080) & (0.137) & (0.556) & (0.564) & (0.432) & (0.480) & (0.129) & (0.493) & (0.999) & (0.979) & (0.991) \\
    \hline
    \multirow{2}[2]{*}{15} & 19.398 & -8.002* & 11.491* & -7.263 & 0.161 & 3.262 & 0.742 & -4.052 & 0.343 & 1.776 & -6.274 & 6.524 & 0.552 \\
        & (0.079) & (0.015) & (0.035) & (0.208) & (0.978) & (0.588) & (0.907) & (0.521) & (0.957) & (0.787) & (0.327) & (0.309) & (0.936) \\
    \hline
    \multirow{2}[2]{*}{18} & 14.073 & -7.460* & 10.968* & -4.234 & 0.754 & -3.853 & 4.695 & -7.847 & 6.297 & 0.614 & -1.952 & -1.080 & 0.177 \\
        & (0.296) & (0.024) & (0.040) & (0.485) & (0.902) & (0.527) & (0.475) & (0.215) & (0.289) & (0.919) & (0.735) & (0.858) & (0.980) \\
    \hline
    \multirow{2}[2]{*}{21} & 10.314 & -0.254 & -2.242 & 2.473 & -3.694 & 1.567 & 2.814 & -3.088 & 0.208 & -1.724 & 5.549 & -7.879 & 3.465 \\
        & (0.588) & (0.939) & (0.672) & (0.675) & (0.537) & (0.789) & (0.640) & (0.589) & (0.973) & (0.808) & (0.416) & (0.228) & (0.597) \\
    \hline
    \multirow{2}[2]{*}{24} & 11.192 & -1.477 & -0.225 & 4.034 & -2.677 & -2.833 & 1.510 & 2.949 & -5.683 & 2.678 & 4.300 & -10.677 & 13.937 \\
        & (0.513) & (0.619) & (0.965) & (0.493) & (0.637) & (0.638) & (0.828) & (0.662) & (0.373) & (0.670) & (0.507) & (0.098) & (0.029) \\
    \hline
    \multirow{2}[2]{*}{27} & 21.768* & 4.833 & -8.216 & 4.579 & -1.107 & -2.055 & -0.544 & 6.275 & -11.425* & 13.966* & -19.946* & 18.488* & -14.127* \\
        & (0.040) & (0.125) & (0.115) & (0.451) & (0.854) & (0.719) & (0.927) & (0.305) & (0.044) & (0.013) & (0.001) & (0.004) & (0.027) \\
    \hline
    \multirow{2}[2]{*}{30} & 18.752 & 2.676 & -3.537 & -1.199 & 6.857 & -11.838 & 13.102* & -17.167* & 15.749* & -11.753 & 3.664 & -1.679 & 1.953 \\
        & (0.095) & (0.415) & (0.511) & (0.844) & (0.299) & (0.058) & (0.024) & (0.003) & (0.015) & (0.073) & (0.590) & (0.810) & (0.784) \\
    \hline
    \multirow{2}[2]{*}{33} & 21.628* & 3.991 & -9.134 & 10.953 & -16.104* & 13.919* & -9.328 & 2.190 & -0.576 & 0.805 & -5.175 & 5.434 & -4.660 \\
        & (0.042) & (0.189) & (0.077) & (0.051) & (0.005) & (0.031) & (0.160) & (0.735) & (0.930) & (0.906) & (0.443) & (0.394) & (0.500) \\
    \hline
    \multirow{2}[2]{*}{36} & 16.861 & -3.830 & 7.315 & -8.338 & 1.420 & 1.529 & -2.438 & -0.442 & 0.549 & -2.429 & 1.348 & 3.158 & -1.953 \\
        & (0.155) & (0.221) & (0.176) & (0.161) & (0.811) & (0.799) & (0.703) & (0.945) & (0.931) & (0.730) & (0.854) & (0.639) & (0.758) \\
    \hline
    \hline
    \end{tabular}%
  \label{gir12}
  }
  \end{center}
  \medskip
    \noindent
   \footnotesize  Note - This table present two-stage estimation of twelve-lag generalized impulse response at horizon 1-9,12,15,18,21,24,27,30,33, and 36. The column of causality test displays the $p$-value of non-causality test at given horizon. The pivotal function is $\chi^2(12)$. The numbers with sign $*$ indicate these test statistics or the coefficient estimates are statistically significant at level 95\%. The comparable GIR estimates of LS projection method and recursive methods are presented in Appendix \ref{empres}.
   \label{girtable}
\end{table}%

Of particular interest is the finding that economic uncertainty (JNL) exhibits a causal effect on the CFNAI both in the short run (1–7 months) and the long run (approximately 2.5 years). While the short-term impact of uncertainty on economic activity is expected, its prolonged influence on the economy surpasses conventional expectations. Conversely, economic uncertainty has a sustained effect on the labor market, with JNL exerting causal influence on unemployment over a span of forty months. Surprisingly, JNL does not significantly affect the Federal Funds Rate (FFR), suggesting that the central bank may not adjust interest rates in direct response to fluctuations in economic uncertainty. Instead, the CFNAI shows causality on the FFR, which aligns with the widely held view that central banks use interest rate policy to stimulate or slow down the economy based on economic activity.

Table \ref{girtable} also provides empirical evidence that statistical insignificance of impulse responses from zero does not necessarily imply zero causality over multiple horizons, and, conversely, statistically significant impulse responses from zero do not necessarily imply causality. For instance, Sims' impulse response (the first column next to the causality test) is significant for horizons 9, 12, 15, and 18, yet none of these horizons exhibit significant multi-horizon causality tests. This discrepancy may be attributed to multicollinearity. On the other hand, at horizons 27 and 33, the causality test is statistically significant, but Sims' impulse response is not significantly different from zero. Thus, relying solely on Sims' impulse response for assessing multi-horizon causality could yield misleading results.

In our analysis, we specifically investigate the causality from the JNL index to CFNAI by visualizing the Wald test statistics for the null hypothesis of non-causality. Recognizing the potential sensitivity of empirical results to order selection, as highlighted by \cite{hamilton2004comment} in the context of \cite{bernanke1997systematic}, we conduct tests using three estimation models with lag orders of 12, 15, and 18. We report the minimum Wald test statistic across these models to ensure robustness. This approach ensures that if any Wald test statistic exceeds the critical value threshold in the plot, the results remain robust irrespective of the chosen lag order. The plot clearly illustrates that the uncertainty index does indeed exhibit causality toward economic activity, particularly around two and a half years, across all three sample periods and lag order selections.

\section{Conclusion}
\label{section10conc}
This paper introduces a novel estimation and inference method for GIRs within a multi-horizon linear projection model. Our proposed two-stage estimation method with heteroskedasticity-robust inference complements the standard LS approach with HAC inference. The proposed covariance matrix estimation offers a robust alternative, eliminating the need for correcting serial correlation in the projection residuals. This method provides researchers with a valuable tool for applying multi-horizon linear projections in causality testing, impulse response estimation, or predictability analysis, particularly for those concerned about the efficiency of the LS projection method and the finite sample performance of HAC covariance estimates.

We derive uniform inference for two-stage estimates across the parameter space and extensively highlight the advantages of two-stage estimates over LS estimates from two key perspectives: (1) two-stage estimates generally exhibit greater efficiency than LS estimates across a wide range of parameter spaces and projection horizons, and (2) two-stage estimates can be implemented without relying on HAC standard error estimates. Instead, our explicit formula for the covariance matrix provides a positive semi-definite, heteroskedasticity-robust estimate. By allowing the projection horizon to grow with the sample size, we emphasize the crucial dependence of asymptotic normality of projection coefficient estimates on both the projection horizon and data persistence.

Our research on GIRs within a linear projection model underscores the importance of multi-horizon Granger causality in economics and finance for capturing the full dynamics of causal relationships. It also highlights the limitations of relying solely on Granger causality at horizon one or impulse response analysis when examining multi-horizon causality.

For future research, extending the finite-order VAR to an infinite-order VAR remains a promising area of interest from both theoretical and empirical perspectives. Inference for infinite-lag VAR models may require imposing restrictions on the norm of the lagged VAR coefficient matrices and carefully selecting a lag length that grows with the sample size, as discussed by \cite{lewis1985prediction}.

Another potential extension involves exploring high-dimensional VAR models instead of finite-dimensional ones, as investigated by \cite{basu2015regularized}, \cite{adamek2023lasso}, \cite{hecq2023granger}, and others. With datasets expanding exponentially, the exploration of high-dimensionality has gained considerable attention, particularly regarding the need for robust inference on regularized coefficients. Developing a high-dimensional version of the two-stage estimation method would thus be of significant interest for future work.

\bibliographystyle{agsm}
\bibliography{references}

\newpage

\begin{appendix}

\newpage
\section{Appendix: More simulation results}
\label{mcsim}
\begin{table}[htbp]
\caption[Monte Carlo simulations: white noise process]{}
 \label{table0}
\begin{center}
Table \thetable

Monte Carlo simulations: white noise process
\end{center}

  \begin{center}
  \resizebox{\textwidth}{!}{
    \begin{tabular}{lccccccccccccc}
    \hline
        & \multicolumn{6}{c}{$\phi_{12,1}^{(h)}$}         &     & \multicolumn{6}{c}{$\phi_{12,2}^{(h)}$} \\
    \hline
    $h$   & 1 & 3  & 6   & 12  & 24  & 36   &     & 1 & 3  & 6   & 12  & 24  & 36  \\
     value & 0.000 & 0.000 & 0.000 & 0.000 & 0.000 & 0.000 &     & 0.000 & 0.000 & 0.000 & 0.000 & 0.000 & 0.000 \\
    \hline
        & \multicolumn{13}{c}{bias of estimates} \\
    \hline
RC-VAR & 0.002 & 1.30E-04 & -1.20E-05 & 4.39E-07 & 7.22E-11 & 1.19E-14 &     & 0.001 & -9.78E-06 & -2.62E-06 & 4.91E-08 & 2.24E-11 & 7.43E-15 \\
    LS-Proj & 0.002 & 0.001 & -0.003 & -0.001 & 0.002 & 0.002 &     & 0.001 & -0.004 & 0.003 & -0.006 & -0.001 & -0.003 \\
    2S(0)-Proj & 0.002 & 0.001 & -0.002 & -0.001 & 0.002 & 0.002 &     & 0.001 & -0.004 & 0.003 & -0.006 & -0.001 & -0.003 \\    
    2S(1)-Proj & -0.001 & 0.000 & -0.002 & -0.002 & 0.001 & -0.001 &     & 0.000 & 0.002 & 0.000 & 0.000 & -0.003 & -0.002 \\
    2S(2)-Proj & -0.003 & 0.000 & -0.002 & 0.002 & -0.004 & -0.001 &     & 0.000 & 0.001 & 0.009 & -0.003 & 1.30E-04 & -2.80E-04 \\
    \hline
        & \multicolumn{13}{c}{rmse of estimates} \\
    \hline
    RC-VAR & 0.075 & 0.010 & 0.001 & 9.03E-06 & 1.82E-09 & 3.78E-13 &     & 0.075 & 0.007 & 1.94E-04 & 1.92E-06 & 5.92E-10 & 2.33E-13 \\    
    LS-Proj & 0.075 & 0.073 & 0.078 & 0.077 & 0.078 & 0.082 &     & 0.075 & 0.076 & 0.081 & 0.077 & 0.081 & 0.085 \\
    $\textup{2S(0)-Proj}$ & 0.075 & 0.073 & 0.079 & 0.077 & 0.078 & 0.082 &     & 0.075 & 0.077 & 0.081 & 0.078 & 0.081 & 0.086 \\
    2S(1)-Proj & 0.077 & 0.077 & 0.079 & 0.076 & 0.081 & 0.083 &     & 0.077 & 0.077 & 0.078 & 0.077 & 0.080 & 0.081 \\
    2S(2)-Proj & 0.079 & 0.078 & 0.076 & 0.079 & 0.082 & 0.085 &     & 0.079 & 0.078 & 0.080 & 0.080 & 0.080 & 0.083 \\
     \hline
        & \multicolumn{13}{c}{empirical size of tests (5\% nominal size)} \\
    \hline
    RC-VAR & 0.054 & 0.001 & 0.000 & 0.000 & 0.000 & 0.000 &     & 0.047 & 0.001 & 0.000 & 0.000 & 0.000 & 0.000 \\
    LS-Proj & 0.057 & 0.036 & 0.059 & 0.050 & 0.049 & 0.054 &     & 0.048 & 0.056 & 0.073 & 0.041 & 0.050 & 0.065 \\
    $\textup{2S(0)-Proj}$ & 0.049 & 0.039 & 0.062 & 0.043 & 0.044 & 0.044 &     & 0.045 & 0.055 & 0.074 & 0.042 & 0.048 & 0.071 \\
    $\textup{2S(0)-Proj}_b$ & 0.056 & 0.041 & 0.063 & 0.049 & 0.046 & 0.043 &     & 0.044 & 0.056 & 0.076 & 0.039 & 0.051 & 0.066 \\
    2S(1)-Proj & 0.064 & 0.056 & 0.068 & 0.051 & 0.069 & 0.058 &     & 0.054 & 0.057 & 0.067 & 0.051 & 0.064 & 0.061 \\
    $\textup{2S(1)-Proj}_b$ & 0.058 & 0.045 & 0.054 & 0.047 & 0.050 & 0.040 &     & 0.047 & 0.047 & 0.051 & 0.042 & 0.051 & 0.042 \\
    2S(2)-Proj & 0.065 & 0.065 & 0.062 & 0.069 & 0.066 & 0.066 &     & 0.064 & 0.062 & 0.077 & 0.062 & 0.059 & 0.062 \\
    $\textup{2S(2)-Proj}_b$ & 0.050 & 0.049 & 0.049 & 0.052 & 0.058 & 0.056 &     & 0.052 & 0.051 & 0.060 & 0.054 & 0.051 & 0.057 \\
    \hline
        & \multicolumn{13}{c}{coverage ratio of nominal 95\% confidence interval} \\
    \hline
    RC-VAR & 0.946 & 0.999 & 1.000 & 1.000 & 1.000 & 1.000 &     & 0.953 & 0.999 & 1.000 & 1.000 & 1.000 & 1.000 \\   
    LS-Proj & 0.943 & 0.964 & 0.941 & 0.950 & 0.951 & 0.946 &     & 0.952 & 0.944 & 0.927 & 0.959 & 0.950 & 0.935 \\
    2S(0)-Proj & 0.951 & 0.961 & 0.938 & 0.957 & 0.956 & 0.956 &     & 0.955 & 0.945 & 0.926 & 0.958 & 0.952 & 0.929 \\
    $\textup{2S(0)-Proj}_b$ & 0.944 & 0.959 & 0.937 & 0.951 & 0.954 & 0.957 &     & 0.956 & 0.944 & 0.924 & 0.961 & 0.949 & 0.934 \\
    2S(1)-Proj & 0.936 & 0.944 & 0.932 & 0.949 & 0.931 & 0.942 &     & 0.946 & 0.943 & 0.933 & 0.949 & 0.936 & 0.939 \\
    $\textup{2S(1)-Proj}_b$ & 0.942 & 0.955 & 0.946 & 0.953 & 0.950 & 0.960 &     & 0.953 & 0.953 & 0.949 & 0.958 & 0.949 & 0.958 \\
    2S(2)-Proj & 0.935 & 0.935 & 0.938 & 0.931 & 0.934 & 0.934 &     & 0.936 & 0.938 & 0.923 & 0.938 & 0.941 & 0.938 \\
    $\textup{2S(2)-Proj}_b$ & 0.950 & 0.951 & 0.951 & 0.948 & 0.942 & 0.944 &     & 0.948 & 0.949 & 0.940 & 0.946 & 0.949 & 0.943 \\
    \hline
        & \multicolumn{13}{c}{average width of nominal 95\% confidence interval} \\
    \hline    
    RC-VAR & 0.294 & 0.051 & 0.005 & 2.07E-05 & 1.77E-09 & 3.87E-13 &     & 0.294 & 0.034 & 0.001 & 6.47E-06 & 7.00E-10 & 1.95E-13 \\
    LS-Proj & 0.295 & 0.297 & 0.298 & 0.302 & 0.311 & 0.319 &     & 0.294 & 0.298 & 0.298 & 0.303 & 0.311 & 0.322 \\
    $\textup{2S(0)-Proj}$ & 0.295 & 0.296 & 0.298 & 0.302 & 0.310 & 0.319 &     & 0.296 & 0.298 & 0.299 & 0.303 & 0.312 & 0.321 \\
    $\textup{2S(0)-Proj}_b$ & 0.295 & 0.303 & 0.298 & 0.307 & 0.310 & 0.324 &     & 0.297 & 0.303 & 0.301 & 0.309 & 0.313 & 0.328 \\
    2S(1)-Proj & 0.291 & 0.292 & 0.293 & 0.298 & 0.304 & 0.313 &     & 0.292 & 0.293 & 0.295 & 0.300 & 0.307 & 0.316 \\
    $\textup{2S(1)-Proj}_b$ & 0.304 & 0.305 & 0.307 & 0.314 & 0.322 & 0.334 &     & 0.305 & 0.306 & 0.309 & 0.315 & 0.325 & 0.337 \\
    2S(2)-Proj & 0.292 & 0.293 & 0.295 & 0.299 & 0.307 & 0.315 &     & 0.293 & 0.294 & 0.297 & 0.299 & 0.308 & 0.318 \\
    $\textup{2S(2)-Proj}_b$ & 0.305 & 0.307 & 0.311 & 0.315 & 0.327 & 0.337 &     & 0.306 & 0.307 & 0.311 & 0.315 & 0.326 & 0.339 \\
    \hline
    \end{tabular}
    }
    \end{center}
    \medskip
    \noindent
   \footnotesize  Note - The DGP follows a white noise process, $u_t\overset{i.i.d.}{\sim} N(0, [1,0.5 ; 0.5,1] )$. 
   
  
   
\end{table}%

\begin{table}[htbp]
  \caption[Monte Carlo simulations: I(2) process]{}
 \label{table3}
\begin{center}
Table \thetable

Monte Carlo simulations: I(2) process
\begin{align}
     {y}_t = \left[\begin{array}{cc}
     1.7    &  -0.2 \\
     0.2    &  1.4 
    \end{array}\right]  {y}_{t-1} + \left[\begin{array}{cc}
     -0.66    &  0.08 \\
     -0.2    &  -0.4
    \end{array}\right]  {y}_{t-1} +  {u}_t.
\end{align}
\end{center}

  \begin{center}
  \resizebox{\textwidth}{!}{
    \begin{tabular}{lccccccccccccc}
    \hline
        & \multicolumn{6}{c}{$\phi_{12,1}^{(h)}$}         &     & \multicolumn{6}{c}{$\phi_{12,2}^{(h)}$} \\
    \hline
    $h$   & 1 & 3  & 6   & 12  & 24  & 36   &     & 1 & 3  & 6   & 12  & 24  & 36  \\    
    value & -0.200 & -0.978 & -2.627 & -6.466 & -14.445 & -22.444 &     & 0.080 & 0.391 & 1.051 & 2.586 & 5.778 & 8.978 \\
    \hline
        & \multicolumn{13}{c}{bias of estimates} \\
    \hline
  RC-VAR & -0.003 & -0.018 & 0.015 & 0.492 & 3.033 & 7.087 &     & -0.013 & -0.034 & -0.065 & -0.264 & -1.295 & -2.942 \\
    LS-Proj & -0.003 & 0.006 & 0.124 & 0.804 & 4.004 & 9.242 &     & -0.013 & -0.068 & -0.224 & -0.708 & -2.375 & -4.874 \\
    2S(2)-Proj & -0.006 & -0.007 & 0.090 & 0.739 & 3.934 & 9.182 &     & -0.005 & -0.012 & -0.079 & -0.387 & -1.762 & -3.966 \\
    \hline
        & \multicolumn{13}{c}{rmse of estimates} \\
    \hline
    RC-VAR & 0.072 & 0.200 & 0.310 & 0.849 & 3.753 & 8.254 &     & 0.081 & 0.216 & 0.309 & 0.624 & 1.885 & 3.699 \\
    LS-Proj & 0.072 & 0.233 & 0.477 & 1.320 & 5.031 & 10.746 &     & 0.081 & 0.263 & 0.547 & 1.288 & 3.900 & 7.362 \\
    2S(2)-Proj & 0.084 & 0.211 & 0.393 & 1.132 & 4.713 & 10.365 &     & 0.149 & 0.254 & 0.352 & 0.750 & 2.380 & 4.776 \\
     \hline
        & \multicolumn{13}{c}{empirical size of tests (5\% nominal size)} \\
    \hline  
    RC-VAR & 0.066 & 0.064 & 0.075 & 0.194 & 0.471 & 0.593 &     & 0.070 & 0.069 & 0.081 & 0.119 & 0.298 & 0.462 \\
    LS-Proj & 0.065 & 0.084 & 0.106 & 0.277 & 0.520 & 0.639 &     & 0.071 & 0.095 & 0.137 & 0.238 & 0.323 & 0.420 \\
    2S(2)-Proj & 0.078 & 0.061 & 0.080 & 0.224 & 0.489 & 0.636 &     & 0.086 & 0.064 & 0.090 & 0.172 & 0.327 & 0.417 \\
    $\textup{2S(2)-Proj}_b$ & 0.062 & 0.048 & 0.050 & 0.068 & 0.084 & 0.114 &     & 0.076 & 0.056 & 0.069 & 0.068 & 0.106 & 0.122 \\
    \hline
        & \multicolumn{13}{c}{coverage ratio of nominal 95\% confidence interval} \\
    \hline       
    RC-VAR & 0.934 & 0.936 & 0.925 & 0.806 & 0.529 & 0.407 &     & 0.930 & 0.931 & 0.919 & 0.881 & 0.702 & 0.538 \\
    LS-Proj & 0.935 & 0.916 & 0.894 & 0.723 & 0.480 & 0.361 &     & 0.929 & 0.905 & 0.863 & 0.762 & 0.677 & 0.580 \\
    2S(2)-Proj & 0.922 & 0.939 & 0.920 & 0.776 & 0.511 & 0.364 &     & 0.914 & 0.936 & 0.910 & 0.828 & 0.673 & 0.583 \\
     $\textup{2S(2)-Proj}_b$ & 0.938 & 0.952 & 0.950 & 0.932 & 0.916 & 0.886 &     & 0.924 & 0.944 & 0.931 & 0.932 & 0.894 & 0.878 \\
    \hline
        & \multicolumn{13}{c}{average width of nominal 95\% confidence interval} \\
    \hline       
    RC-VAR & 0.265 & 0.730 & 1.143 & 2.369 & 6.405 & 11.614 &     & 0.297 & 0.789 & 1.146 & 2.167 & 4.821 & 7.303 \\
    LS-Proj & 0.266 & 0.831 & 1.614 & 3.362 & 8.823 & 14.233 &     & 0.298 & 0.917 & 1.749 & 3.632 & 9.703 & 15.937 \\
    2S(2)-Proj & 0.294 & 0.795 & 1.424 & 2.987 & 8.192 & 14.296 &     & 0.521 & 0.909 & 1.254 & 2.372 & 5.669 & 10.164 \\
     $\textup{2S(2)-Proj}_b$ & 0.307 & 0.859 & 1.593 & 3.694 & 12.819 & 28.260 &     & 0.537 & 0.954 & 1.382 & 2.772 & 7.611 & 14.729 \\
    \hline
    \end{tabular}  }
  \end{center}
      
      \medskip
    \noindent
   \footnotesize Note - The DGP follows a VAR(2) process integrated at order two, $u_t\overset{i.i.d.}{\sim} N(0, [1,0.5 ; 0.5,1] )$. Two $2\times 2$ root matrices are $[0.7,-0.2;0,1]$ and $[1,0;0.2,0.4]$, whose eigenvalues are $1,0.7$ and $1,0.4$, respectively. Since the DGP contains two unit roots, only two stage results estimated with two extra lags are presented. 
  \normalsize
\end{table}

\newpage
\section{Appendix: More empirical results}
\label{empres}
\begin{figure}[htbp]
    \centering
    \caption{Least Squares estimates of generalized impulse responses to JNL}
    \includegraphics[width=0.7 \textwidth]{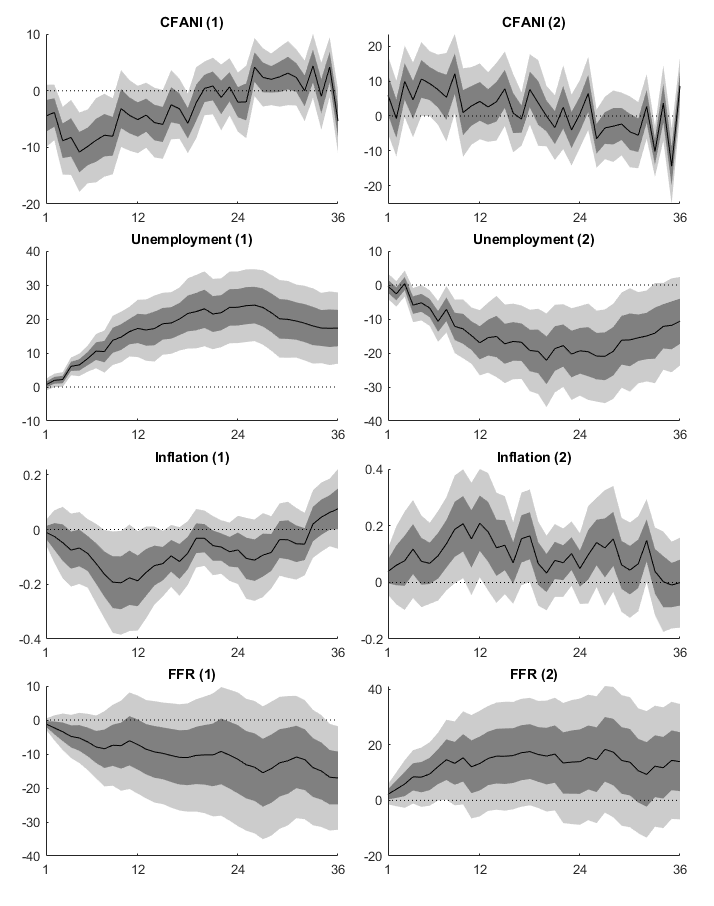}
    
     Figure \thefigure. Projection-based Least Squares estimates of generalized impulse responses to Economic uncertainty index (JNL)
\end{figure}

\begin{figure}[htbp]
    \centering
    \caption{Recursive estimates of generalized impulse responses to JNL}
    \includegraphics[width=0.7 \textwidth]{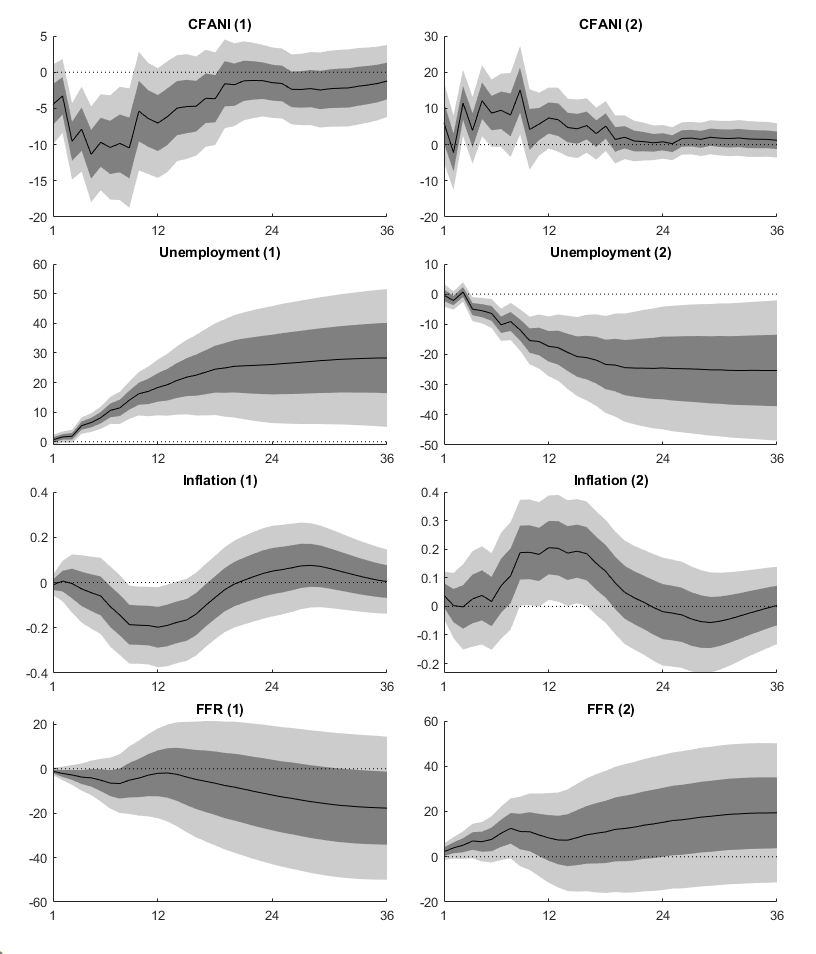}
    
     Figure \thefigure. Recursive estimates of generalized impulse responses to Economic uncertainty index (JNL)
\end{figure}

\begin{table}[htbp]
 \caption[Table of generalized impulse response estimates]{}
\begin{center}
Table \thetable

Projection Least Squares estimates of generalized impulse responses
\end{center}

  \begin{center}
  \resizebox{\textwidth}{!}{
    \begin{tabular}{c|c|cccccccccccc}
    \hline
    \hline
    horizon & Causality test & \multicolumn{12}{c}{Least Squares estimates of twelve-lag generalized impulse responses} \\
    \hline
    \multirow{2}[1]{*}{1} & 28.458* & -4.440 & 6.015 & -8.599 & 8.689 & -6.165 & 3.866 & -0.232 & -2.469 & -0.420 & 6.176 & -6.335 & 1.914 \\
        & (0.005)  & (0.117) & (0.284) & (0.141) & (0.135) & (0.239) & (0.413) & (0.961) & (0.576) & (0.932) & (0.275) & (0.210) & (0.413) \\
    \hline\multirow{2}[0]{*}{2} & 22.572* & -3.832 & -0.753 & 4.700 & -3.740 & 2.033 & 1.681 & -4.248 & 1.354 & 5.557 & -8.140 & 4.985 & -1.614 \\
        & (0.032) & (0.127) & (0.892) & (0.423) & (0.456) & (0.660) & (0.728) & (0.315) & (0.784) & (0.330) & (0.151) & (0.320) & (0.491) \\
    \hline\multirow{2}[0]{*}{3} & 25.244* & -8.820* & 9.894 & -7.114 & 6.373 & -2.769 & -0.911 & -0.603 & 6.119 & -9.216 & 8.045 & -5.838 & 2.398 \\
        & (0.014) & (0.004) & (0.058) & (0.190) & (0.217) & (0.581) & (0.828) & (0.901) & (0.308) & (0.132) & (0.168) & (0.249) & (0.359) \\
    \hline\multirow{2}[0]{*}{4} & 30.403* & -8.240* & 4.607 & 1.691 & 0.218 & -2.473 & 1.009 & 4.362 & -8.468 & 7.489 & -5.289 & 2.080 & -0.120 \\
        & (0.002) & (0.015) & (0.400) & (0.739) & (0.962) & (0.586) & (0.849) & (0.491) & (0.173) & (0.209) & (0.343) & (0.678) & (0.963) \\
    \hline\multirow{2}[0]{*}{5} & 25.477* & -10.833* & 10.622* & -3.119 & 1.891 & -3.733 & 7.751 & -10.411 & 7.358 & -4.475 & 0.929 & 1.040 & -0.286 \\
        & (0.013) & (0.002) & (0.049) & (0.546) & (0.678) & (0.489) & (0.251) & (0.107) & (0.238) & (0.459) & (0.870) & (0.847) & (0.920) \\
    \hline\multirow{2}[0]{*}{6} & 24.932* & -9.825* & 9.291* & -2.141 & 0.160 & 3.240 & -6.370 & 4.403 & -3.459 & 1.333 & 0.368 & -0.699 & 0.487 \\
        & (0.015) & (0.003) & (0.050) & (0.647) & (0.978) & (0.633) & (0.302) & (0.487) & (0.564) & (0.818) & (0.955) & (0.912) & (0.886) \\
        \hline
        \multirow{2}[0]{*}{7} & 25.086* & -8.747* & 7.527 & -2.682 & 6.346 & -8.989 & 6.521 & -5.152 & 1.090 & 1.355 & -0.530 & -1.500 & 1.226 \\
        & (0.014) & (0.021) & (0.144) & (0.606) & (0.302) & (0.139) & (0.313) & (0.412) & (0.859) & (0.835) & (0.931) & (0.763) & (0.676) \\
        \hline
    \multirow{2}[0]{*}{8} & 18.441 & -7.849* & 5.353 & 5.442 & -6.392 & 1.819 & -1.272 & -0.802 & 0.771 & 2.006 & -3.851 & 1.269 & 0.496 \\
        & (0.103) & (0.036) & (0.409) & (0.411) & (0.297) & (0.766) & (0.830) & (0.896) & (0.906) & (0.734) & (0.473) & (0.790) & (0.862) \\
    \hline\multirow{2}[0]{*}{9} & 14.931 & -8.043* & 12.102* & -8.692 & 4.421 & -2.844 & 0.116 & -0.365 & 2.409 & -3.430 & -0.712 & 2.146 & -0.181 \\
        & (0.245) & (0.019) & (0.038) & (0.133) & (0.502) & (0.662) & (0.985) & (0.953) & (0.679) & (0.511) & (0.880) & (0.661) & (0.952) \\
    \hline\multirow{2}[0]{*}{12} & 16.636 & -5.090 & 4.170 & -1.115 & -2.494 & 6.760 & -6.805 & 0.394 & 1.142 & 4.324 & -8.061 & 5.103 & -1.873 \\
        & (0.164) & (0.094) & (0.453) & (0.850) & (0.685) & (0.289) & (0.257) & (0.943) & (0.820) & (0.430) & (0.167) & (0.288) & (0.509) \\
    \hline\multirow{2}[0]{*}{15} & 17.578 & -5.949 & 7.779 & -4.613 & -0.790 & 3.534 & 1.338 & -4.672 & 0.811 & 1.366 & -6.866 & 10.936 & -6.618 \\
        & (0.129) & (0.055) & (0.153) & (0.404) & (0.866) & (0.432) & (0.823) & (0.423) & (0.878) & (0.773) & (0.196) & (0.081) & (0.081) \\
    \hline\multirow{2}[0]{*}{18} & 19.057 & -5.751* & 7.590 & -0.420 & -2.846 & 0.845 & 0.259 & -4.743 & 3.627 & 2.909 & -6.947 & 6.948 & -5.289 \\
        & (0.087) & (0.033) & (0.141) & (0.950) & (0.661) & (0.872) & (0.961) & (0.359) & (0.486) & (0.554) & (0.179) & (0.189) & (0.077) \\
    \hline\multirow{2}[0]{*}{21} & 26.188* & 0.841 & -3.358 & 4.070 & -6.062 & 4.064 & 2.161 & -5.100 & 0.555 & -0.416 & 3.354 & -0.911 & -3.688 \\
        & (0.010) & (0.742) & (0.480) & (0.448) & (0.294) & (0.494) & (0.691) & (0.294) & (0.924) & (0.953) & (0.599) & (0.882) & (0.390) \\
    \hline\multirow{2}[0]{*}{24} & 36.041* & -2.013 & 0.565 & 5.288 & -7.049 & 1.757 & -0.671 & 4.526 & -9.130 & 5.883 & 0.843 & -0.219 & -4.953 \\
        & (3.19E-04) & (0.450) & (0.909) & (0.335) & (0.150) & (0.740) & (0.921) & (0.481) & (0.117) & (0.276) & (0.866) & (0.965) & (0.252) \\
    \hline\multirow{2}[0]{*}{27} & 22.563* & 2.362 & -3.452 & 0.670 & 1.055 & -3.678 & 1.510 & 4.549 & -9.399* & 10.970* & -15.232* & 11.331* & -6.461 \\
        & (0.032) & (0.369) & (0.450) & (0.898) & (0.850) & (0.512) & (0.780) & (0.366) & (0.038) & (0.025) & (0.026) & (0.037) & (0.060) \\
    \hline\multirow{2}[0]{*}{30} & 26.887* & 3.114 & -4.569 & -1.186 & 7.513 & -12.262* & 12.328* & -14.308* & 10.747 & -7.208 & 0.193 & -1.453 & 1.782 \\
        & (0.008) & (0.280) & (0.379) & (0.832) & (0.204) & (0.005) & (0.010) & (0.024) & (0.095) & (0.226) & (0.973) & (0.788) & (0.531) \\
    \hline\multirow{2}[0]{*}{33} & 35.363* & 4.380 & -10.237* & 10.306* & -13.489* & 10.865 & -8.055 & 1.414 & -0.165 & 1.270 & -4.913 & 2.114 & 1.479 \\
        & (4.10E-04) & (0.128) & (0.005) & (0.013) & (0.017) & (0.083) & (0.170) & (0.806) & (0.979) & (0.839) & (0.457) & (0.700) & (0.638) \\
   \hline \multirow{2}[1]{*}{36} & 25.538* & -5.377 & 8.616* & -9.267* & 3.290 & -1.275 & 0.038 & -2.596 & 2.825 & -2.501 & 1.987 & -0.607 & 0.453 \\
        & (0.012) & (0.057) & (0.038) & (0.048) & (0.526) & (0.816) & (0.994) & (0.706) & (0.696) & (0.676) & (0.748) & (0.894) & (0.787) \\
    \hline
    \hline
    \end{tabular}%
    }
  \end{center}
  \medskip
    \noindent
   \footnotesize  Note - This table present Least Squares estimation of twelve-lag generalized impulse response at horizon 1-9,12,15,18,21,24,27,30,33, and 36. The coefficient and covariance matrix estimation is computed through ‘hac’ command in MATLAB R2020a with setting \textit{‘Bandwidth’=h+1}. The column of ‘causality test’ displays the Wald test statistics with its $p$-value of non-causality test at given horizon. The pivotal function is $\chi^2(12)$. 
  \label{lstable}%
\end{table}%

\begin{table}[htbp]
 \caption[Table of generalized impulse response estimates]{}
\begin{center}
Table \thetable

Recursive VAR estimates of generalized impulse responses
\end{center}

  \begin{center}
  \resizebox{\textwidth}{!}{
    \begin{tabular}{c|c|cccccccccccc}
    \hline
    \hline
    horizon & Causality test & \multicolumn{12}{c}{Recursive VAR estimates of twelve-lag generalized impulse responses} \\
    \hline
    \multirow{2}[2]{*}{1} & 33.113* & -4.440 & 6.015 & -8.599 & 8.689 & -6.165 & 3.866 & -0.232 & -2.469 & -0.420 & 6.176 & -6.335 & 1.914 \\
        & (0.001) & (0.114) & (0.288) & (0.147) & (0.101) & (0.181) & (0.370) & (0.957) & (0.548) & (0.929) & (0.232) & (0.169) & (0.369) \\
    \hline
    \multirow{2}[2]{*}{2} & 29.642* & -3.281 & -2.221 & 6.299 & -4.485 & 2.635 & 0.733 & -3.447 & 0.323 & 6.502 & -6.876 & 1.468 & 0.333 \\
        & (0.003) & (0.207) & (0.672) & (0.220) & (0.318) & (0.530) & (0.859) & (0.382) & (0.944) & (0.197) & (0.126) & (0.505) & (0.554) \\
    \hline
    \multirow{2}[2]{*}{3} & 31.534* & -9.550* & 11.486* & -8.191 & 7.037 & -3.289 & -0.712 & -0.523 & 5.515 & -6.453 & 3.385 & -2.333 & 1.099 \\
        & (0.002) & (0.000) & (0.014) & (0.107) & (0.161) & (0.488) & (0.872) & (0.917) & (0.319) & (0.209) & (0.320) & (0.194) & (0.189) \\
    \hline
    \multirow{2}[2]{*}{4} & 29.867* & -7.865* & 3.929 & 2.562 & 0.127 & -3.488 & 2.358 & 3.033 & -5.843 & 3.255 & -1.263 & -0.444 & 0.584 \\
        & (0.003) & (0.009) & (0.404) & (0.595) & (0.978) & (0.436) & (0.648) & (0.593) & (0.262) & (0.334) & (0.513) & (0.770) & (0.487) \\
    \hline
    \multirow{2}[2]{*}{5} & 28.181* & -11.340* & 12.100* & -4.361 & 2.067 & -3.287 & 6.970 & -7.806 & 2.809 & -0.421 & 0.904 & -2.467 & 1.672 \\
        & (0.005) & (0.001) & (0.016) & (0.360) & (0.654) & (0.535) & (0.231) & (0.148) & (0.448) & (0.855) & (0.650) & (0.092) & (0.034) \\
    \hline
    \multirow{2}[2]{*}{6} & 25.327* & -9.685* & 8.702 & -2.082 & 1.024 & 2.405 & -3.991 & 0.061 & 0.023 & 1.806 & -2.550 & 0.106 & 1.043 \\
        & (0.013) & (0.004) & (0.067) & (0.644) & (0.849) & (0.685) & (0.462) & (0.987) & (0.993) & (0.402) & (0.145) & (0.932) & (0.175) \\
    \hline
    \multirow{2}[2]{*}{7} & 22.133* & -10.394* & 9.474 & -2.834 & 6.399 & -8.155 & 3.668 & -2.277 & 1.754 & -1.608 & 0.149 & -0.228 & 0.684 \\
        & (0.036) & (0.005) & (0.067) & (0.590) & (0.270) & (0.129) & (0.311) & (0.344) & (0.420) & (0.382) & (0.925) & (0.860) & (0.401) \\
    \hline
    \multirow{2}[2]{*}{8} & 19.355 & -9.835* & 8.194 & 4.026 & -4.954 & -0.577 & 1.161 & -0.530 & -1.351 & 1.879 & -2.046 & 0.441 & 0.655 \\
        & (0.080) & (0.014) & (0.167) & (0.490) & (0.368) & (0.881) & (0.628) & (0.813) & (0.505) & (0.303) & (0.217) & (0.749) & (0.430) \\
    \hline
    \multirow{2}[2]{*}{9} & 14.901 & -10.469* & 15.060* & -8.241 & 3.382 & -3.387 & 3.051 & -3.621 & 2.001 & -0.744 & -0.204 & -0.528 & 0.944 \\
        & (0.247) & (0.013) & (0.016) & (0.129) & (0.365) & (0.160) & (0.142) & (0.044) & (0.247) & (0.657) & (0.890) & (0.622) & (0.199) \\
    \hline
    \multirow{2}[2]{*}{12} & 7.996 & -7.018 & 7.344 & -1.122 & -0.829 & -0.934 & 0.766 & -1.062 & 1.763 & -1.272 & -0.463 & -0.022 & 0.464 \\
        & (0.785) & (0.070) & (0.087) & (0.667) & (0.721) & (0.642) & (0.639) & (0.500) & (0.229) & (0.365) & (0.739) & (0.983) & (0.519) \\
    \hline
    \multirow{2}[2]{*}{15} & 5.399 & -4.744 & 4.358 & -0.108 & -0.549 & 0.285 & -0.262 & -0.301 & -0.244 & 0.460 & -1.340 & 0.445 & 0.305 \\
        & (0.943) & (0.199) & (0.291) & (0.959) & (0.758) & (0.861) & (0.861) & (0.831) & (0.825) & (0.637) & (0.146) & (0.528) & (0.590) \\
    \hline
    \multirow{2}[2]{*}{18} & 7.140 & -3.663 & 5.121 & -2.310 & 1.096 & -1.502 & 0.857 & -1.171 & 0.497 & -0.438 & 0.115 & -0.193 & 0.321 \\
        & (0.848) & (0.262) & (0.144) & (0.194) & (0.486) & (0.256) & (0.411) & (0.180) & (0.507) & (0.504) & (0.871) & (0.720) & (0.484) \\
    \hline
    \multirow{2}[2]{*}{21} & 4.377 & -1.190 & 0.984 & -0.358 & -0.170 & -0.487 & -0.121 & -0.077 & 0.149 & -0.022 & -0.051 & -0.015 & 0.179 \\
        & (0.976) & (0.668) & (0.731) & (0.774) & (0.858) & (0.597) & (0.871) & (0.911) & (0.811) & (0.970) & (0.933) & (0.973) & (0.639) \\
    \hline
    \multirow{2}[2]{*}{24} & 3.914 & -1.448 & 0.762 & -0.844 & 0.632 & -0.461 & 0.073 & -0.205 & 0.090 & -0.025 & 0.341 & -0.291 & 0.281 \\
        & (0.985) & (0.559) & (0.748) & (0.356) & (0.426) & (0.569) & (0.910) & (0.743) & (0.862) & (0.953) & (0.484) & (0.507) & (0.446) \\
    \hline
    \multirow{2}[2]{*}{27} & 2.969 & -2.384 & 1.791 & -0.653 & 0.397 & -0.567 & 0.171 & -0.335 & 0.224 & 0.079 & 0.271 & -0.254 & 0.355 \\
        & (0.996) & (0.343) & (0.435) & (0.463) & (0.595) & (0.434) & (0.749) & (0.539) & (0.639) & (0.846) & (0.581) & (0.590) & (0.376) \\
    \hline
    \multirow{2}[2]{*}{30} & 3.437 & -2.283 & 1.778 & -0.540 & 0.241 & -0.544 & 0.260 & -0.311 & 0.396 & -0.006 & 0.163 & -0.132 & 0.357 \\
        & (0.992) & (0.394) & (0.466) & (0.475) & (0.686) & (0.405) & (0.582) & (0.542) & (0.362) & (0.986) & (0.689) & (0.738) & (0.368) \\
    \hline
    \multirow{2}[2]{*}{33} & 2.742 & -1.939 & 1.552 & -0.331 & 0.089 & -0.240 & 0.150 & -0.154 & 0.368 & -0.072 & 0.078 & -0.010 & 0.282 \\
        & (0.997) & (0.472) & (0.539) & (0.642) & (0.867) & (0.704) & (0.735) & (0.752) & (0.362) & (0.811) & (0.839) & (0.976) & (0.440) \\
    \hline
    \multirow{2}[2]{*}{36} & 2.543 & -1.221 & 1.183 & -0.143 & 0.000 & -0.143 & 0.111 & -0.056 & 0.273 & -0.025 & -0.044 & 0.096 & 0.192 \\
        & (0.998) & (0.631) & (0.624) & (0.823) & (1.000) & (0.798) & (0.779) & (0.899) & (0.433) & (0.910) & (0.892) & (0.723) & (0.533) \\
    \hline
    \hline
    \end{tabular}%
    }
  \end{center}
  \medskip
    \noindent
   \footnotesize  Note - This table present Recursive VAR estimation of twelve-lag generalized impulse response at horizon 1-9,12,15,18,21,24,27,30,33, and 36. The coefficient and covariance matrix estimation is computed through Delta-method inference. The column of ‘causality test’ displays the Wald test statistics with its $p$-value of non-causality test at given horizon. The pivotal function is $\chi^2(12)$. 
  \label{VARtable}%
\end{table}%

\newpage
\section{Appendix: Lemmas}

Prior to delving into the proofs, we present a few notations and lemmas. Consider two extended transformation matrices, $\bar G_{1,T}$ and $\bar G_{2,T}$, the purpose of which is to deal with lag-augmented estimation. $\bar G_{1,T}$ represents a square matrix with dimensions $(p+1)K$, while $\bar G_{2,T}$ denotes a square matrix with dimensions $(p+2)K$, 
\begin{align}
\label{barG}
\begin{split}
    &\bar G_{1,T}:=\left[\begin{array}{cc}
      I_K   & 0 \\
       0  & \bar \Upsilon_1
    \end{array} \right] \left[\begin{array}{cc}
      I_K & [-P_1, 0_{K\times (p-1)K}]  \\
      0_{pK\times K}   & \bar P_1
    \end{array} \right]  (I_{p+1}\otimes \Pi),\\
    & \bar G_{2,T}:=\left[\begin{array}{cc}
      I_{2K}   & 0 \\
       0  & \bar \Upsilon_2
    \end{array} \right] \left[\begin{array}{ccc}
      I_K & -(P_1+P_2) & [ P_1P_2, 0_{K\times (p-2)K}]  \\
      I_K & -(P_1+P_2) & [ P_1P_2, 0_{K\times (p-2)K}]   \\
      0_{pK\times K} &  0_{pK\times K}  & \bar P_2
    \end{array} \right]  (I_{p+2}\otimes \Pi).
    \end{split}
\end{align}
The matrices $\bar G_{1,T}$ and $\bar G_{2,T}$ are composed of three distinct components, each serving a crucial role. First, the matrix $\Pi$ is to eliminate potential cointegration within the vector of observables $y_t$. Second, the matrix containing $P_1,P_2$ is to derive Dickey-Fuller type regressors. Finally, the component involving $\bar\Upsilon_i$ represents a probability scaling matrix, accommodating varying convergence speeds for potential non-stationary processes. 

Then, we introduce notations for rotated and differenced $y_t$ and $x_t$:
\begin{align}
& \tilde y_t = \Pi y_t \\
& \Delta_{10}\tilde y_t = (I_K- P_1L)\tilde y_t \\
& \Delta_{21}\tilde y_t = (I_K- P_2L)\tilde y_t \\
& \Delta_{20}\tilde y_t = (I_K- P_1L) (I_K- P_2L)\tilde y_t 
\end{align}
and 
\begin{align}
& \tilde x_t =(I_p\otimes \Pi) x_t \\
& \Delta_{10}\tilde x_t = (I_{pK}- I_p\otimes P_1 L)\tilde x_t \\
& \Delta_{21}\tilde x_t = (I_{pK}- I_p\otimes P_2L)\tilde x_t \\
& \Delta_{20}\tilde x_t = (I_{pK}- I_p\otimes P_1L) (I_{pK} - I_p\otimes P_2L) \tilde x_t .
\end{align}

\newpage

\begin{lemma}[Boundedness of impulse responses]
\label{lemmabound}
        Let the autoregressive coefficients be in the parameter space $\mathcal{B}(\delta,c,\epsilon)$. Let $c=(c_0,c_1)$, $c_0,c_1>0$, $\epsilon \in (0,1)$. 
    \begin{enumerate}[(i)]
        \item If $\delta=0$, then
        \begin{align}
      \label{sumirf0}      &\sum_{i=0}^{\infty}\|\Psi_{i}\| =O_p(1),\\
       \label{sumirf02}     &\sum_{i=0}^{\infty}\|\Psi_{i}\|^2 =O_p(1).
        \end{align}
        \item If $\delta=1$, then
                        \begin{align}
       \label{sumirf1}     &\sum_{i=0}^{h-1}\|\Psi_{i}\| =O_p(h),\\
       \label{sumirf12}     &\sum_{i=0}^{h-1}\|\Psi_{i}\|^2 =O_p(h).
        \end{align}
        \item If $\delta=2$, then
                        \begin{align}
       \label{sumirf2}     &\sum_{i=0}^{h-1}\|\Psi_{i}\| =O_p(h^2),\\
       \label{sumirf22}     &\sum_{i=0}^{h-1}\|\Psi_{i}\|^2 =O_p(h^3) .
        \end{align}
    \end{enumerate}
\end{lemma}
See the proof in Subsection \ref{prooflemmairf}.

\begin{lemma}[Boundedness of the LRV]
\label{lemmavarscore}
    Suppose Assumption \ref{assimeanind} and \ref{assmoment}(i) hold. Let $w \in \mathbb R^{pK}$, $\|w\|=1$, $\underline c_{\xi}, \overline  c_{\xi}$ are positive constant, and $0< \underline c_{\xi}\leq \overline  c_{\xi}< \infty$. Then, the long run variance $\Omega_{s,h}$  defined in \eqref{omegash} is bounded:
\begin{align}
  0<\underline c_{\xi} \sum_{j=1}^{h+p-1}\psi_{w,j}
  \leq 
  w'\Omega_{s,h} w 
  \leq 
  \overline  c_{\xi} \sum_{j=1}^{h+p-1}\psi_{w,j},
\end{align}
where $\psi_{w,j}=\|
             \sum_{i=\max(1,j-h+1)}^{\min(p,j)}  w_i \Psi_{1\bullet,h-j+i-1} 
             \|^2$, $w_i$ is the $K$-dimensional sub-vector of $w$, $w=(w_1',w_2',\cdots,w_p')'$.
\end{lemma}
See the proof in Subsection \ref{prooflemmaa4}.

\begin{lemma}[Variance of variance estimates of regression score function]
    \label{lemmavarvarbound}
    Let the conditions of Proposition \eqref{propclt} hold, then 
    \begin{align}
        \text{Var} \left( (\bar T-p+1)^{-1} \sum_{t=p}^{\bar T} (w's_{t,h}^*)^2/ w'\Omega_{s,h} w \right) \xrightarrow{p} 0.
    \end{align}    
\end{lemma}
See the proof in Subsection \ref{prooflemmavar}.

\begin{lemma}[Bounded fourth moments of regression score function]
\label{lemma42}
Suppose Assumption \ref{assimeanind} and \ref{assmoment}(i) hold. Then
\begin{align}
    \mathbb E [ (w's_{t,h}^*)^4/ (w'\Omega_{s,h} w)^2 ] < c_0 <\infty
\end{align}
where $w \in \mathbb R^{pK}$, $\|w\|=1$, for some constant $c_0$ almost surely.    
\end{lemma}
See the proof in Subsection \ref{prooflemma4th}.

\begin{lemma}[Boundedness of the second moments]
\label{lemma2moments}
    Suppose Assumption \ref{assimeanind} and \ref{assmoment} hold. Then
    \begin{align}
  \label{b174}   & \lim_{T\rightarrow \infty}  T^{-1}\sum_{t=1}^{T} u_t u_t' \xrightarrow{p}  \Sigma_u.
  \end{align}
  Moreover, let the autoregressive coefficients be in the parameter space $\mathcal{B}(\delta,c,\epsilon)$. Let $c=(c_0,c_1)$, $c_0,c_1>0$, $\epsilon \in (0,1)$. 
\begin{enumerate}[(i)]
        \item If $\delta=0$ and $0< h/T\leq \alpha$ for $\alpha\in [0,1)$, then
        \begin{align}
        \label{conszx0}           &\lim_{T\rightarrow \infty} T^{-1} \sum_{t=p}^T z_t x_t' \xrightarrow{p}\mathbb E[z_t x_t'],\\
        \label{xx0}    &\lim_{T\rightarrow \infty}\|T^{-1} \sum_{t=1}^{T} x_t x_t'  \| =O_p(1),\\
        \label{ux0}   &\lim_{T\rightarrow \infty}\|T^{-1/2} \sum_{t=1}^T x_t u_{t+k}'\| =O_p(1),\\
         \label{xe0}   &\lim_{T\rightarrow \infty}\|\bar T^{-1/2} \sum_{t=p}^{T-h} x_t e_{t+k,h} \| =O_p(1),
        \end{align}
    \item If $\delta=1$ and $\frac{h}{T}\max(1,\frac{h}{w'\Omega_{\beta,h} w})\xrightarrow{p}0$, then
        \begin{align}
       \label{conszx1} &\lim_{T\rightarrow \infty} T^{-1} \sum_{t=p}^T z_t \Delta_{10} \tilde x_t' \xrightarrow{p}\mathbb E[z_t \Delta_{10} \tilde x_t'],\\
       \label{xx1}     &\lim_{T\rightarrow \infty}\|T^{-1} \sum_{t=1}^{T} G_{1,T} x_t x_t'G_{1,T}'  \| =O_p(1).\\
       \label{ux1} &\lim_{T\rightarrow \infty}\|T^{-1/2} \sum_{t=1}^T   G_{1,T}x_t  u_{t+k}' \| =O_p(1),\\
       \label{uy1} &\lim_{T\rightarrow \infty}\|T^{-1/2} \sum_{t=1}^T  \Upsilon_1 \tilde y_t  u_{t+k} \| =O_p(1),\\
             \label{xe1}   &\lim_{T\rightarrow \infty}\|\bar T^{-1/2} \sum_{t=p}^{T-h}  G_{1,T} x_t e_{t+k,h} \| = O_p(h),
        \end{align}
    \item If $\delta=2$ and $\frac{h^3}{T}\max(1,\frac{h}{w'\Omega_{\beta,h} w})\xrightarrow{p}0$, then
        \begin{align}
     \label{conszx2}  
                    &\lim_{T\rightarrow \infty} T^{-1} \sum_{t=p}^T z_t \Delta_{20} \tilde x_t' \xrightarrow{p}\mathbb E[z_t \Delta_{20} \tilde x_t'],\\
       \label{xx2}   
                    &\lim_{T\rightarrow \infty}\|T^{-1} \sum_{t=1}^{T} G_{2,T} x_t x_t'G_{2,T}'  \| =O_p(1),\\
       \label{ux2}  
                    &\lim_{T\rightarrow \infty}\|T^{-1/2} \sum_{t=1}^T  G_{2,T}x_t u_{t+k}' \| =O_p(1),\\
        \label{uy2} &\lim_{T\rightarrow \infty}\|T^{-1/2} \sum_{t=1}^T u_{t+k} 
                     (\tilde y_t' \Upsilon_2,\Delta_{21}\tilde y_t' \Upsilon_1)   \| =O_p(1),\\
         \label{xe2}   
                     &\lim_{T\rightarrow \infty}\|\bar T^{-1/2} \sum_{t=p}^{T-h} G_{2,T} x_t e_{t+k,h} \| = O_p(h^2).
        \end{align}
    \end{enumerate}
    for any non-negative integer $k$.
\end{lemma}
See the proof in Subsection \ref{prooflemmaa5}.

\begin{lemma}[Asymptotically negligible estimation errors]
\label{lemmanegerror}
        Suppose Assumption \ref{assimeanind}, \ref{assmoment}, and \ref{ass4} hold. Let the autoregressive coefficients be in the parameter space $\mathcal{B}(\delta,c,\epsilon)$. Let $c=(c_0,c_1)$, $c_0,c_1>0$, $\epsilon \in (0,1)$, and $w\in\mathbb R^{pK}$, and $\|w\|=1$. If one of the following conditions hold,
    \begin{enumerate}[(i)]
        \item $\delta=0$ and $0< h/T\leq \alpha$ for $\alpha\in [0,1)$,
         \item $\delta=1$ and $\frac{h}{T}\max(1,\frac{h}{w'\Omega_{\beta,h} w})\xrightarrow{p}0$,
    \item $\delta=2$ and $\frac{h^3}{T}\max(1,\frac{h}{w'\Omega_{\beta,h} w})\xrightarrow{p}0$,
    \end{enumerate}
    then
      \begin{align}
       \label{zzx0} &\lim_{T\rightarrow \infty}\|\bar T^{-1/2}\sum_{t=p}^{T-h} (\hat{z}_t -z_t) x_{t-k}'G_{\delta,T}' \| =O_p(1),\\
       \label{ze0} &\lim_{T\rightarrow \infty}\left\|  \frac{1}{\left(\bar T w'\Omega_{\beta,h} w\right)^{1/2}}\sum_{t=p}^{T-h}  (\hat z_t -z_{t}) e_{t,h} \right\|\xrightarrow{p} 0,
        \end{align}
        for finite non-negative integer $k$.
\end{lemma}
See the proof in Subsection \ref{prooflemmaa6}.

\begin{lemma}[Boundedness of the fourth moment]
\label{lemmacov}
        Suppose Assumption \ref{assimeanind}, \ref{assmoment}, and \ref{ass4} hold. Let the autoregressive coefficients be in the parameter space $\mathcal{B}(\delta,c,\epsilon)$. Let $c=(c_0,c_1)$, $c_0,c_1>0$, $\epsilon \in (0,1)$, and $w\in\mathbb R^{pK}$, and $\|w\|=1$. Denote $\tilde y_t =\Pi y_t$.
    \begin{enumerate}[(i)]
        \item If $\delta=0$, then
        \begin{align}
   \label{y0}     \max_{1\leq t\leq T }\mathbb E[\|\tilde y_{t}\|^4] = O_p(1).
    \end{align}
    \item If $\delta=1$, then
        \begin{align}
    \label{y1}   \max_{1\leq t\leq T }\mathbb E[\|\Upsilon_1 \tilde y_{t}\|^4] =O_p(1).
    \end{align}
    \item If $\delta=2$, then
        \begin{align}
    \label{y2}   \max_{1\leq t\leq T }\mathbb E[\|\Upsilon_2 \tilde y_{t}\|^4] =O_p(1).
    \end{align}
    \end{enumerate}
\end{lemma}
See the proof in Subsection \ref{prooflemmaa7}.

\newpage
\section{Appendix: Proofs}

\subsection{Lemma 4.1 (LRV matrix)}
\label{prooflemma4.1}
\begin{proof}
    First we show $ \text{LRV}(s_{t,h}^*)=\Omega_{s,h}$. Recall the explicit form of $s_{t,h}$ in \eqref{3.13}. We partition $s_{t,h}$ as $s_{t,h}=(s_{1,t,h}',s_{2,t,h}',\cdots,s_{p,t,h}')'$ where $s_{i,t,h}=u_{t+i-1} e_{t,h}$. Thus, given the explicit form of $s_{t,h}^*$ in \eqref{4.19}, $s_{t,h}^* =(s_{1,t,h}',s_{2,t+1,h}',\cdots,s_{p,t+p-1,h}')' $.   
    As $\Omega_{s,h}$ is the LRV of $s_{t,h}$, we partition $\Omega_{s,h}$ into $p\times p$ block matrices of dimension $K$,
    \begin{align}
        \Omega_{s,h} = \left[\Omega_{s,h}^{(ij)} \right]_{i,j=1,2,\cdots,p},
    \end{align}
    where $\Omega_{s,h}^{(ij)}$ is a $K$-dimensional square matrix, $\Omega_{s,h}^{(ij)} = \sum_{k=-\infty}^{\infty}\mathbb{E}[s_{i,t,h} s_{j,t+k,h}']$. Under Assumption \eqref{assimeanind}, $\Omega_{s,h}^{(ij)} = \sum_{k=-h+1}^{h-1}\mathbb{E}[s_{i,t,h} s_{j,t+k,h}']$, as $\mathbb{E}[s_{i,t,h} s_{j,t+k,h}']=0$ for all $|k|\geq h$.

    Since $s_{t,h}^* =(s_{1,t,h}',s_{2,t+1,h}',\cdots,s_{p,t+p-1,h}')' $, to prove $ \text{LRV}(s_{t,h}^*)=\Omega_{s,h}$, it is equivalent to show $\sum_{k=-\infty}^\infty\mathbb{E}[ s_{i,t+i-1,h}s_{j,t+k+j-1,h}']=\Omega_{s,h}^{(ij)} $.
    \begin{align}
        \begin{split}
            \sum_{k=-\infty}^\infty\mathbb{E}[ s_{i,t+i-1,h}s_{j,t+k+j-1,h}'] &= \sum_{k=-\infty}^\infty\mathbb{E}[ s_{i,t,h}s_{j,t+k+j-i,h}']\\
            & \quad (\text{ because of covariance stationary of $s_{t,h}$})\\
            &= \sum_{k=-h+1+i-j}^{h-1+i-j}\mathbb{E}[ s_{i,t,h}s_{j,t+k+j-i,h}']\\
            & \quad (\text{ since $\mathbb{E}[s_{i,t,h} s_{j,t+k,h}']=0$ for all $|k|\geq h$})\\
            &= \sum_{k=-h+1}^{h-1}\mathbb{E}[ s_{i,t,h}s_{j,t+k,h}']=\Omega_{s,h}^{(ij)} \\
            & \quad (\text{ replacing $k+j-i$ by $k$})
        \end{split}
    \end{align}
Thus, we have completed the proof of $ \text{LRV}(s_{t,h}^*)=\Omega_{s,h}$.
    
Next, we show $\text{LRV}(s_{t,h}^*) = \text{Var}(s_{t,h}^*)$. To prove the identity between LRV and variance, it is equivalent to show the serial uncorrelation of $s_{t,h}^*$. Without loss of generality, suppose $k>0$
\begin{align}
\begin{split}
    \mathbb E [s_{t+ k,h}^* s_{t,h}^{*'} ]
    &= \mathbb E \left[ \mathbb E [ s_{t+ k,h}^* s_{t,h}^{*'}  \mid \{u_\tau\}_{\tau>t} ]  \right] \\
    &\quad \text{ (Law of Iterated Expectations)}\\
    &= \mathbb E \left[s_{t+ k,h}^*  \mathbb E [  s_{t,h}^{*'} \mid \{u_\tau\}_{\tau>t} ]  \right]\\
    &\quad \text{ (Since $s_{t+ k,h}^*$ is measureable by the information set $\{u_\tau\}_{\tau>t}$ for $k>0$)}\\
    &=0 \\
    &\quad \text{ (Since $\mathbb E [  s_{t,h}^{*} \mid \{u_\tau\}_{\tau>t} ] = (e_{t,h},e_{t+1,h},\cdots,e_{t+p-1,h})'\otimes \underbrace{\mathbb E [  u_t \mid \{u_\tau\}_{\tau>t} ]}_{=0 \text{ by Assumption \ref{assimeanind}}}=0$ )}
\end{split}
\end{align}
Thus, the proof of $\text{LRV}(s_{t,h}^*) = \text{Var}(s_{t,h}^*)$ is complete.
\end{proof}

\newpage
\subsection{Lemma 4.2 (Asymptotic equivalence)}
\label{prooflemma4.2}
\begin{proof}
Recall $\sum_{t=p}^{T-h} s_{t,h} = \overline{s}_{1,h} + \overline{s}_{2,h} + \sum_{t=p}^{T-h-p+1} s_{t,h}^*$ in \eqref{4.32}. Thus, to prove \eqref{4.39}, it is equivalent to show
\begin{align}
    \bar T^{-1/2} w'(\overline{s}_{1,h} + \overline{s}_{2,h})\xrightarrow{p} 0.
\end{align}
where $\overline{s}_{1,h}$ and $\overline{s}_{2,h}$ are two summations of order ($p-1$), defined as $\overline{s}_{1,h}=\sum_{i=1}^{p-1} s_{i,h}^{**}$ and $\overline{s}_{2,h}=\sum_{i=1}^{p-1} s_{i,h}^{***}$, such that $s_{i,h}^{**}=J_{(-i)} s_{i,h}^{*}$,
$s_{i,h}^{***}=J_{(p-i)} s_{\bar T+i,h}^{*}$, $\bar T=T-h-p+1$, and $J_{(i)}$ and $J_{(-i)}$ respectively denote the first $iK$ rows and the last $(p-i)K$ rows of identity matrix of dimension $pK$, $[J_{(i)}',J_{(-i)}]'=I_{pK}$.

The convergence will be proved by showing $\bar T^{-1/2}s_{i,h}^{**} , \bar T^{-1/2}s_{i,h}^{***} $ have zero mean and zero variance in the limit, and apply Chebyshev's inequality. The mean of $\bar T^{-1/2}s_{i,h}^{**} , \bar T^{-1/2}s_{i,h}^{***} $ are readily seen as zero since $s_{i,h}^{**}=J_{(-i)} s_{i,h}^{*}$,
$s_{i,h}^{***}=J_{(p-i)} s_{\bar T+i,h}^{*}$, and $\mathbb E[s_{t,h}^*]=\mathbb E[(e_{t,h},e_{t+1,h},\cdots,e_{t+p-1,h})'\otimes u_t]=0$ by the serial uncorrelation of $u_t$ nested in Assumption \ref{assimeanind}.

Next, we check the variance of $s_{i,h}^{**} , s_{i,h}^{***} $. Given $s_{i,h}^{**}=J_{(-i)} s_{i,h}^{*}$,
$s_{i,h}^{***}=J_{(p-i)} s_{\bar T+i,h}^{*}$, it is sufficient to check the Frobenius norm of the variance matrix of $s_{i,h}^*$, $\Omega_{s,h}$,
\begin{align}
\label{4.52}
    \|\Omega_{s,h}\|/\bar T\xrightarrow{p} 0 .
\end{align}
It is equivalently to show that each element in $\Omega_{s,h}/\bar T$ converges to zero. In other words, the variance of elements in $s_{t,h}^*$ is asymptotic negligible compared to $\bar T$. Thus, without loss of generality, we check the variance of the $i$-th element in the term $u_t e_{t+k,h}$ for some $0\leq k\leq p-1$,
\begin{align}
\label{omegaspi}
\begin{split}
    \mathbb{E} [ u_{i,t}^2 e_{t+k,h}^2] 
    =& \mathbb{E} [ u_{i,t}^2 \sum_{j=0}^{h-1}(\Psi_{1\bullet ,j} u_{t+k+h-j})^2]\\
    & \quad (\text{since $e_{t,h}$ is the first element in $u_t^{(h)}$, and $u_t^{(h)}=\sum_{j=0}^{h-1}\Psi_j u_{t+h-j}$})\\
     & \quad (\text{recall $\Psi_{1\bullet,j}$ indicates the first row of matrix $\Psi_j$})\\
    \leq &\sum_{j=0}^{h-1} \|\Psi_{1\bullet ,j}\|^2\mathbb{E} [ u_{i,t}^2 
     \| u_{t+k+h-j}\|^2]\\
    \leq  &\mathbb{E} [  
     \| u_{t}\|^4]\sum_{j=0}^{h-1}\|\Psi_{1\bullet ,j}\|^2 
\end{split}
\end{align}


$\mathbb{E} [  
     \| u_{t}\|^4]$ is bounded above by a constant due to Assumption \eqref{assmoment}. By Lemma \eqref{lemmabound}, $\|\Psi_{1\bullet ,j}\|^2$ is a constant when $\delta=0$; $\pi_h \propto h$  when $\delta=1$; $\pi_h \propto h^2$  when $\delta=2$. Thus, under the conditions specified in this lemma, $\|\Psi_{1\bullet ,j}\|^2/\bar T \xrightarrow{p} 0$ for all three cases; and in turn $\|\Omega_{s,h}\|/\bar T$ converges to zero as well. By Chebyshev's inequality, $\bar T^{-1/2}s_{i,h}^{**}$ and $\bar T^{-1/2}s_{i,h}^{***}$ converge to zero. Given the finite order $p$, $\bar T^{-1/2} w'(\overline{s}_{1,h} + \overline{s}_{2,h})\xrightarrow{p} 0$. Thus, the proof is complete.
\end{proof}

\newpage
\newpage
\subsection{Proposition 4.3 (Asymptotic normality)}
\label{proofprop4.3}
\begin{proof}
    Lemma \ref{lemmaneg} shows that $ \bar T^{-1/2} w'(\sum_{t=p}^{T-h} s_{t,h} -  \sum_{t=p}^{\bar T} s_{t,h}^*) \xrightarrow{p} 0$. Therefore, to prove $\bar T^{-1/2}\sum_{t=p}^{T-h} w's_{t,h} / (w'\Omega_{s,h} w)^{1/2} \xrightarrow{d} N(0,1)$, it is sufficient to show the following conditions hold,
    \begin{enumerate}
        \item $\left(\frac{\bar T}{\bar T -p+1}\right)^{-1/2}\xrightarrow{p} 1 $,
        \item $\lambda_{\min}(\Omega_{s,h})$ is bounded below by a non-zero constant.
        \item $(\bar T -p+1)^{-1/2}\sum_{t=p}^{\bar T } w's_{t,h}^* / (w'\Omega_{s,h} w)^{1/2} \xrightarrow{d} N(0,1)$,
    \end{enumerate}
    The first condition is satisfied trivially since $p$ is finite and $\bar T\rightarrow\infty$. The second condition is proved in Lemma \ref{lemmavarscore}. The third condition will be proved by using martingale Central Limit Theorem, \cite[Theorem 24.3]{davidson1994stochastic}.
    
    To apply martingale Central Limit Theorem, we need first to show the underlying process is a martingale difference sequence. Define
    \begin{align}
        \chi_{t,h}: = (\bar T -p+1)^{-1/2} w's_{\bar T+1-t,h}^*/ (w'\Omega_{s,h} w)^{1/2}, 
    \end{align}
    for $t=1,2,\cdots,\bar T-p+1$. Notice $\sum_{t=1}^{\bar T-p+1} \chi_{t,h} = (\bar T -p+1)^{-1/2}\sum_{t=p}^{\bar T } w's_{t,h}^* / (w'\Omega_{s,h} w)^{1/2}$. We also construct an information set but with a reversed time ordering,\footnote{Note: the method of reversed time ordering follows the procedure in \cite[Lemma A.1]{montiel2021local}.} 
    \begin{align}
       &\mathcal{F}_{T,t}=\sigma(u_{\bar T+1-t}, u_{\bar T+2-t},\cdots)
    \end{align}
    for $t=1,2,\cdots,\bar T-p+1$. The sequence of sigma-algebra is a filtration, $\mathcal{F}_{T,1}\subseteq \mathcal{F}_{T,2}\subseteq \cdots$.  
    By the mean-independence assumption (Assumption \ref{assimeanind}), it is readily to obtain that
    \begin{align}
        \mathbb E [\chi_{t,h} \mid \mathcal{F}_{T,t-1}]=0.
    \end{align}
    The zero conditional expectation is because $\chi_{t,h}$ is the linear transformation of $s_{\bar T+1-t,h}^*$, such that $s_{\bar T+1-t,h}^*=(e_{\bar T+1-t,h},e_{\bar T+2-t,h},\cdots,e_{\bar T+p-t,h})'\otimes u_{\bar T+1-t}$. Notice $(e_{\bar T+1-t,h},e_{\bar T+2-t,h},\cdots,e_{\bar T+p-t,h})$ is measureable by $\mathcal{F}_{T,t-1}$ but $u_{\bar T+1-t}$ has zero conditional expectation on $\mathcal{F}_{T,t-1}$ by Assumption \ref{assimeanind}. Therefore, $\chi_{t,h}$ is a martingale difference sequence.
    
    We now verify the conditions in \cite[Theorem 24.3]{davidson1994stochastic} to prove the convergence $\sum_{t=1}^{\bar T-p+1 }\chi_{t,h} \xrightarrow{d} N(0,1)$.  The convergence holds if the following conditions are satisfied:
    \begin{enumerate}
        \item $\sum_{t=1}^{\bar T-p+1 }\mathbb E [\chi_{t,h}^2] = 1$,
        \item $\sum_{t=1}^{\bar T-p+1 } \chi_{t,h}^2 \xrightarrow{p}1$, as $\bar T\rightarrow \infty$.
        \item $\max_{1\leq t\leq \bar T-p+1 } |\chi_{t,h}| \xrightarrow{p}0$, as $\bar T\rightarrow \infty$.
    \end{enumerate}
    The first condition holds due to the definition of $\Omega_{s,h}$, $\Omega_{s,h} = \mathbb E[  s_{t,h}^* s_{t,h}^{*'} ]$. It yields $\mathbb E [\chi_{t,h}^2] = (\bar T-p+1)^{-1}\mathbb E[ w' s_{t,h}^* s_{t,h}^{*'} w]/ w'\Omega_{s,h}w = (\bar T-p+1)^{-1}$. Thus, $\sum_{t=1}^{\bar T-p+1 }\mathbb E [\chi_{t,h}^2] = 1$.
    
    The second condition will be proved by using Chebyshev's inequality. By the first condition, we obtain the mean as  $\sum_{t=1}^{\bar T-p+1 } \mathbb E[\chi_{t,h}^2] = 1$. Since $\chi_{t,h}: = (\bar T -p+1)^{-1/2} w's_{\bar T+1-t,h}^*/ (w'\Omega_{s,h} w)^{1/2}$, $\text{Var}\left(\sum_{t=1}^{\bar T-p+1 }\chi_{t,h}^2\right) = \text{Var}\left( (\bar T-p+1)^{-1}\sum_{t=p}^{\bar T } (w's_{t,h}^*)^2/ (w'\Omega_{s,h} w)\right)$. Lemma \eqref{lemmavarvarbound} shows that $\text{Var}\left( (\bar T-p+1)^{-1}\sum_{t=p}^{\bar T } (w's_{t,h}^*)^2/ (w'\Omega_{s,h} w)\right) \xrightarrow{p}0$. Thus, the term $\sum_{t=1}^{\bar T-p+1 } \chi_{t,h}^2$ has zero variance in the limit. By Chebyshev's inequality, we obtain $\sum_{t=1}^{\bar T-p+1 } \chi_{t,h}^2 \xrightarrow{p}1$.
    
    To prove the third condition, by Theorem 23.16 in \cite{davidson1994stochastic}, it is sufficient to prove that, for arbitrary $c>0$, 
    \begin{align}
       (\bar T-p+1) \mathbb E [ \chi_{t,h}^2 \mathbb{1}_{ |\chi_{t,h}|>c}]\xrightarrow{p}0
    \end{align}
    where $\mathbb{1}$ is an indicator function. We show the above expression:
    \begin{align}
    \begin{split}
    & (\bar T-p+1) \mathbb E [ \chi_{t,h}^2 \mathbb{1}_{ |\chi_{t,h}|>c}]\\
    \leq  &  (\bar T-p+1) \mathbb E [ \chi_{t,h}^2 \mathbb{1}_{ |\chi_{t,h}|>c} \frac{\chi_{t,h}^2}{c^2}]\\
    \leq  &  (\bar T-p+1) \mathbb E [  \frac{\chi_{t,h}^4}{c^2}]\\
    =&
       (\bar T -p+1)^{-1} \mathbb E [ (w's_{t,h}^*)^4/ (w'\Omega_{s,h} w)^2 ] /c^2\\
       &(\text{by the definition of $\chi_{t,h}$})\\
       \xrightarrow{p}&0
    \end{split}
    \end{align}
    The convergence is due to the fact that  $\mathbb E [ (w's_{t,h}^*)^4/ (w'\Omega_{s,h} w)^2 ]$ is bounded above by a constant shown in Lemma \eqref{lemma42}. 
\end{proof}

\newpage
\subsection{Proposition 5.1 (Infeasible estimates)}
\label{proofpropinfea}
\begin{proof}

The proof is conducted across three cases, where $\delta \in \{0, 1, 2\}$. Since $\delta$ controls the persistence of the data process, when $\delta$ equals 1 or 2, the data process is allowed to contain unit root(s). Consequently, the transformation matrices $\bar G_{1,T}$ and $\bar G_{2,T}$, defined in \eqref{barG}, are necessary for inference derivation. This step adheres to the standard procedure in inference derivation for non-stationary time series processes, see \cite{hamilton1994time}.

\textbf{Case 1:} $\delta=0$, $\Phi\in \mathcal{B}(0,c,\epsilon)$, and $0<h\leq \alpha T$ for $\alpha \in (0,1)$. We show
\begin{align}
\label{infea0}
\begin{split}
     &\frac{1}{(w'\Omega_{\beta,h} w)^{1/2}}\bar T^{1/2} (\tilde{\beta}_h^{2S} -\beta_h)\\
        =& \frac{1}{(w'\Omega_{\beta,h} w)^{1/2}}
(\bar T^{-1}\sum_{t=p}^{T-h} 
 z_{t} x_{t}' )^{-1}(\bar T^{-1/2}\sum_{t=p}^{T-h}  z_{t} e_{t,h}) \\
 \xrightarrow{d}&N(0,1).
\end{split}
\end{align}
First, according to \ref{conszx0}, Lemma \ref{lemma2moments}, $\bar T^{-1}\sum_{t=p}^{T-h} 
 z_{t} x_{t}'\xrightarrow{p}\mathbb E[z_t x_t']$. Next, the term $\bar T^{-1/2}\sum_{t=p}^{T-h}  z_{t} e_{t,h}$ converges in law by Proposition \ref{propclt}; and thereby Slutsky's theorem entails that $\bar T^{1/2}(\tilde{\beta}_h^{2S} -\beta_h) \xrightarrow{d} N(0, \Omega_{\beta,h}  )$, where $\Omega_{\beta,h}$ is defined as $\Omega_{\beta,h}=\Sigma_{zx}^{-1}\Omega_{\beta,h} \Sigma_{zx}^{'-1}$ in \eqref{omegabeta} and $\Sigma_{zx}:=\mathbb E[z_t x_t']$. Thus, the proof of \eqref{infea0} is complete.

\textbf{Case 2:} $\delta=1$, $\Phi\in \mathcal{B}(1,c,\epsilon)$, and $0<h\leq  \bar h$, such that $\frac{\bar h}{T}\max (1, \frac{\bar h}{w'\Omega_{\beta,h} w })\xrightarrow{p}0$. We show

\begin{align}
\label{infea1}
\begin{split}
    & \frac{1}{(w'\Omega_{\beta,h} w)^{1/2}}\bar T^{1/2}w' 
        (\tilde{\beta}_h^{LA(1)-2S} -\beta_h)\\
        =&
         \frac{1}{(w'\Omega_{\beta,h} w)^{1/2}}\bar T^{1/2}w'         
        H_1 
(\sum_{t=p}^{T-h} 
 z_{t,1} x_{t,1}' )^{-1}(\sum_{t=p}^{T-h}  z_{t,1} e_{t,h}) \\
 =&
\frac{1}{(w'\Omega_{\beta,h} w)^{1/2}} w'         
        H_1 \bar G_{1,T}'
\left(\bar T^{-1}\sum_{t=p}^{T-h} \left[\begin{array}{cc}
  I_{pK}   & 0 \\
   0  & \Upsilon_1\Pi
\end{array}\right]
 z_{t,1} x_{t,1}' \bar G_{1,T}' \right)^{-1} \\
 &\left(\bar T^{1/2}\sum_{t=p}^{T-h} \left[\begin{array}{cc}
  I_{pK}   & 0 \\
   0  & \Upsilon_1\Pi
\end{array}\right] z_{t,1} e_{t,h} \right)  \\
=& \frac{1}{(w'\Omega_{\beta,h} w)^{1/2}} w'         
        H_1 \bar G_{1,T}'
\left(\bar T^{-1}\sum_{t=p}^{T-h} \left[\begin{array}{cc}
  z_t \Delta_{10}\tilde x_t'   & z_t \tilde y_{t-p}'\Upsilon_1 \\
 \Upsilon_1 \tilde y_{t-p}\Delta_{10}\tilde x_t'  & \Upsilon_1\tilde y_{t-p}\tilde y_{t-p}'\Upsilon_1
\end{array}\right]
\right)^{-1} \\
 & \left(\bar T^{1/2}\sum_{t=p}^{T-h} \left[\begin{array}{c}
  z_t  e_{t,h} \\
 \Upsilon_1 \tilde y_{t-p}e_{t,h}
\end{array}\right]  \right) \\
 \xrightarrow{d}&N(0,1).
\end{split}
\end{align}
The convergence of the above expression will be verified by checking the value of $H_1 \bar G_{1,T}'$, the inverse matrix, and two $\bar T^{1/2}$ terms. First, the definition of $\bar G_{1,T}$ in \eqref{barG} entails that $H_1 \bar G_{1,T}' = [ (I_p\otimes \Pi') \bar P_1', 0_{pK\times K}]$. 

Then we check the invertibility of the block covariance matrix by verifying the convergence of each block matrices. The entire covariance matrix is non-singular if the following conditions hold:
\begin{enumerate}[(1)]
    \item $\bar T^{-1}\sum_{t=p}^{T-h}z_t \Delta_{10}\tilde x_t' \xrightarrow{p} \mathbb E[z_t \Delta_{10}\tilde x_t']$,
    \item $\|\bar T^{-1}\sum_{t=p}^{T-h}z_t \tilde y_{t-p}'\Upsilon_1 \|\xrightarrow{p}0$, 
    \item $\|\bar T^{-1}\sum_{t=p}^{T-h}\Upsilon_1 \tilde y_{t-p}(\Delta_{10}\tilde x_t)'  \| <c<\infty$,
    \item $\lambda_{\min}(\bar T^{-1}\sum_{t=p}^{T-h} \Upsilon_1\tilde y_{t-p}\tilde y_{t-p}'\Upsilon_1)>0 $ a.s.,
\end{enumerate}
Condition (1) is shown in \eqref{conszx1}, Lemma \ref{lemma2moments}. Condition (2) is the immediate result of \eqref{uy1}, Lemma \ref{lemma2moments}, as $z_t$ is a vector of $u_t,\cdots,u_{t-p+1}$. Condition (3) will hold by Cauchy–Schwarz inequality and the boundedness of $\| \bar T^{-1}\sum_{t=p}^{T-h}\Upsilon_1 \tilde y_{t-p} \tilde y_{t-p}' \Upsilon_1\|$ and $\| \bar T^{-1}\sum_{t=p}^{T-h} (\Delta_{10}\tilde x_t)(\Delta_{10}\tilde x_t)'\|$. The term $\| \bar T^{-1}\sum_{t=p}^{T-h}\Upsilon_1 \tilde y_{t-p} \tilde y_{t-p}' \Upsilon_1\|$ is bounded above by a constant, shown in \eqref{xx1}, Lemma \ref{lemma2moments}, as $\Upsilon_1\tilde y_{t-p}  $ is the sub-vector of $G_{1,T} x_{t-1}$; the term $\| \bar T^{-1}\sum_{t=p}^{T-h} (\Delta_{10}\tilde x_t)(\Delta_{10}\tilde x_t)'\|$  is bounded above by a constant, as argued in the proof of \eqref{xx1}, Lemma \ref{lemma2moments}. It is because $\Delta_{10}\tilde x_t$ is a stacked $\Delta_{10}\tilde y_t$'s and $\Delta_{10}\tilde y_t$ is a stationary process. Lastly, condition (4) is imposed in Assumption \ref{ass4}. Thus, the entire block covariance matrix is invertible. 

Moreover, condition (1)-(4) induces that the upper-left block in the inverse of the covariance matrix converges to $\mathbb E[z_t \Delta_{10}\tilde x_t']^{-1}$; and the upper-right block in the inverse of the covariance matrix is a $O_p(\bar T^{-1/2})$ term because of $\|\bar T^{-1/2}\sum_{t=p}^{T-h}z_t \tilde y_{t-p}'\Upsilon_1 \| =O_p(1)$ as an immediate result of \eqref{ux1}, Lemma \ref{lemma2moments} ($z_t$ is a stacked $u_t$'s). Note that the value of lower-left(right) block in the inverse matrix does not matter since the last $K$ columns of matrix $H_1 \bar G_{1,T}'$ are zeros.

In the limit, $\bar T^{1/2}\sum_{t=p}^{T-h} z_t  e_{t,h}$ converges in law by Proposition \eqref{clt}. The term $\| \bar T^{1/2}\sum_{t=p}^{T-h} \Upsilon_1 \tilde y_{t-p}e_{t,h} \|$ is bounded above by an $O_p(h)$, shown in \eqref{xe1} in Lemma \ref{lemma2moments}, as $\Upsilon_1 \tilde y_{t-p}$ is a sub-vector of $G_{1,T}x_{t-1}$.

Thus, combining all limiting results, \eqref{infea1} in the limit can be written as a summation:
\begin{align}
  \frac{1}{(w'\Omega_{\beta,h} w)^{1/2}}  w'H_1 \bar G_{1,T}' \left(\mathbb E[z_t \Delta_{10}\tilde x_t']^{-1} N(0,\Omega_{s,h}) +
    \underbrace{ O_p(\bar T^{-1/2}) O_p(h)}_{\text{bias due to lag-augmentation}} \right)
\end{align}
Since $H_1 \bar G_{1,T}' \mathbb E[z_t \Delta_{10}\tilde x_t']^{-1} = E[z_t  x_t']^{-1}$ and $\Sigma_{zx}=E[z_t  x_t']$, we obtain the first term has a Gaussian distribution with unity variance because of $\Omega_{\beta,h}=\Sigma_{zx}^{-1}\Omega_{s,h}\Sigma_{zx}^{'-1}$. Moreover, to make sure the bias term converges to zero, it is sufficient to impose restriction such that
\begin{align}
    \frac{h}{\bar T^{1/2}(w'\Omega_{\beta,h} w)^{1/2}} \xrightarrow{p}0.
\end{align}
It is satisfied under the condition $h\leq \bar h$, such that $\frac{\bar h}{T}\max (1, \frac{\bar h}{w'\Omega_{\beta,h} w })\xrightarrow{p}0$. Thus, the proof of \eqref{infea1} is complete.

\textbf{Case 3:} $\delta=2$, $\Phi\in \mathcal{B}(2,c,\epsilon)$ and $0<h\leq \bar h$, such that $\frac{\bar h^3}{T}\max(1,\frac{\bar h}{w'\Omega_{\beta,h} w})\xrightarrow{p}0$. We show

\begin{align}
\label{infea2}
\begin{split}
    & \frac{1}{(w'\Omega_{\beta,h} w)^{1/2}}\bar T^{1/2}w' 
        (\tilde{\beta}_h^{LA(2)-2S} -\beta_h)\\
        =&
       \frac{1}{(w'\Omega_{\beta,h} w)^{1/2}} \bar T^{1/2}w'         
        H_2 
(\sum_{t=p}^{T-h} 
 z_{t,2} x_{t,2}' )^{-1}(\sum_{t=p}^{T-h}  z_{t,2} e_{t,h}) \\
       = & 
       \frac{1}{(w'\Omega_{\beta,h} w)^{1/2}}  \bar T^{1/2}w'         
        H_2 \bar G_{2,T}'
       \left(\sum_{t=p}^{T-h} 
\left[\begin{array}{ccc}
    I_{pK}  & & \\
      & \Pi & -P_2\Pi \\
      && \Pi
  \end{array} \right] z_{t,2} x_{t,2}'\bar G_{2,T}' \right)^{-1}\\
  &
  \left(\sum_{t=p}^{T-h} \left[\begin{array}{ccc}
    I_{pK}  & & \\
      & \Pi & -P_2\Pi \\
      && \Pi
  \end{array} \right] z_{t,2} e_{t,h}\right) \\
   = & 
     \frac{1}{(w'\Omega_{\beta,h} w)^{1/2}}   w'         
        H_2 \bar G_{2,T}'\\
       &\left(\bar T^{-1}\sum_{t=p}^{T-h} 
\left[\begin{array}{ccc}
    z_t \Delta_{20}\tilde x_t'  &z_t \Delta_{21}\tilde y_{t-p}'\Upsilon_1 & z_t \tilde y_{t-p-1}'\Upsilon_2\\
   \Upsilon_1 \Delta_{21}\tilde y_{t-p} \Delta_{20}\tilde x_t'  & \Upsilon_1\Delta_{21}\tilde y_{t-p} \Delta_{21}\tilde y_{t-p}'&
   \Upsilon_1 \Delta_{21}\tilde y_{t-p} \tilde y_{t-p-1}'\Upsilon_2\\
    \Upsilon_2\tilde y_{t-p-1}\Delta_{20}\tilde x_t'  & \Upsilon_2\tilde y_{t-p-1}\Delta_{21}\tilde y_{t-p}'\Upsilon_1 & \Upsilon_2\tilde y_{t-p-1} \tilde y_{t-p-1}'\Upsilon_2
  \end{array} \right] \right)^{-1}\\
  &
  \left(\bar T^{-1/2}\sum_{t=p}^{T-h} \left[\begin{array}{c}
   z_{t } e_{t,h}\\
   \Upsilon_1\Delta_{21}\tilde y_{t-p} e_{t,h}  \\
   \Upsilon_2\tilde y_{t-p-1} e_{t,h}  
  \end{array} \right] \right) 
  \\
 \xrightarrow{d}&N(0,1).
    \end{split}
\end{align}
The convergence of the above expression will be verified by checking the value of $H_2 \bar G_{2,T}'$, the inverse matrix, and three $\bar T^{1/2}$ terms. The definition of $\bar G_{2,T}$ in \eqref{barG} entails that $H_2 \bar G_{2,T}' = [ (I_p\otimes \Pi') \bar P_2', 0_{pK\times 2K}]$. 

Then we check the invertibility of the block covariance matrix by verifying the convergence of each block matrices. The entire covariance matrix is non-singular if the following conditions hold:
\begin{enumerate}[(1)]
    \item $\bar T^{-1}\sum_{t=p}^{T-h}z_t \Delta_{20}\tilde x_t' \xrightarrow{p} \mathbb E[z_t \Delta_{20}\tilde x_t']$,
    \item $\|\bar T^{-1}\sum_{t=p}^{T-h}z_t \Delta_{21}\tilde y_{t-p}'\Upsilon_1 \| + \|\bar T^{-1}\sum_{t=p}^{T-h}z_t \tilde y_{t-p-1}'\Upsilon_2 \|\xrightarrow{p}0$,
    \item $\|\bar T^{-1}\sum_{t=p}^{T-h}\Upsilon_1 \Delta_{21} \tilde y_{t-p}\Delta_{20}\tilde x_t'  \|, \|\bar T^{-1}\sum_{t=p}^{T-h}\Upsilon_2  \tilde y_{t-p-1}\Delta_{20}\tilde x_t'  \| <c<\infty$,
    \item the $2\times 2$ lower-right block in the covariance matrix is non-singular almost surely
\end{enumerate}
Condition (1) is shown in \eqref{conszx2}, Lemma \ref{lemma2moments}. Condition (2) is the immediate result of \eqref{uy2}, Lemma \ref{lemma2moments}, as $z_t$ is a vector of $u_t,\cdots,u_{t-p+1}$. Condition (3) will hold by Cauchy-Schwarz inequality and boundedness of $\|\bar T^{-1}\sum_{t=p}^{T-h}\Upsilon_1 (\Delta_{21} \tilde y_{t-p})(\Delta_{21} \tilde y_{t-p})' \Upsilon_1 \|$, $\|\bar T^{-1}\sum_{t=p}^{T-h} (\Delta_{20}\tilde x_t) (\Delta_{20}\tilde x_t)'\|$, and $\|\bar T^{-1}\sum_{t=p}^{T-h} \Upsilon_2 \tilde y_{t-p-1}\tilde y_{t-p-1} '\Upsilon_2 \|$. The term $\|\bar T^{-1}\sum_{t=p}^{T-h}\Upsilon_1 (\Delta_{21} \tilde y_{t-p})(\Delta_{21} \tilde y_{t-p})' \Upsilon_1 \|$ and $\|\bar T^{-1}\sum_{t=p}^{T-h} \Upsilon_2 \tilde y_{t-p-1}\tilde y_{t-p-1} '\Upsilon_2 \|$ are bounded by a constant, shown in \eqref{xx2}, Lemma \ref{lemma2moments}, as $\Delta_{21} \tilde y_{t-p}, \tilde y_{t-p-1}$ are sub-vector of $G_{2,t}x_{t-2}$. The term $\|\bar T^{-1}\sum_{t=p}^{T-h} (\Delta_{20}\tilde x_t) (\Delta_{20}\tilde x_t)'\|$ is bounded above by a constant, as argued in the proof of \eqref{xx2}. It is because $\Delta_{20}\tilde x_t$ by definition is a stacked $\Delta_{20}\tilde y_t$'s and $\Delta_{20}\tilde y_t$ is a stationary process. Lastly, condition (4) is imposed in Assumption \ref{ass4}. Thus, the entire block covariance matrix is invertible. 

As the covariance matrix is partitioned into $3\times 3$ block matrix, condition (1)-(4) induces that the (11)-th block in the inverse matrix converges to $\mathbb E[z_t \Delta_{20}\tilde x_t']^{-1}$; and (12)-th and (13)-th block in the inverse of the covariance matrix is a $O_p(\bar T^{-1/2})$ term because $\|\bar T^{-1/2}\sum_{t=p}^{T-h}z_t \Delta_{21}\tilde y_{t-p}'\Upsilon_1 \| , \|\bar T^{-1/2}\sum_{t=p}^{T-h}z_t \tilde y_{t-p-1}'\Upsilon_2 \| $ are $O_p(1)$ terms by \eqref{ux2}, Lemma \ref{lemma2moments} ($z_t$ is a vector of $u_t,\cdots,u_{t-p+1}$). Note that the value of the rest of blocks in the inverse matrix does not matter since the last $2K$ columns of matrix $H_2 \bar G_{2,T}'$ are zeros.

In the limit, $\bar T^{1/2}\sum_{t=p}^{T-h} z_t  e_{t,h}$ converges in law by Proposition \eqref{clt}; and $\| \bar T^{1/2}\sum_{t=p}^{T-h}\Upsilon_1\Delta_{21}\tilde y_{t-p} e_{t,h} \|,\| \bar T^{1/2}\sum_{t=p}^{T-h}\Upsilon_2\tilde y_{t-p-1} e_{t,h} \|$ are $O_p(h^2)$ terms, shown in \eqref{xe2}, Lemma \ref{lemma2moments}, as $\Upsilon_1\Delta_{21}\tilde y_{t-p}, \Upsilon_2 y_{t-p-1}$ are sub-vectors of $G_{2,T} x_{t-2}$.

In summary, combining all limiting results, \eqref{infea2} in the limit can be written as
\begin{align}
    \frac{1}{(w'\Omega_{\beta,h} w)^{1/2}}w'H_2 \bar G_{2,T}' \left( \mathbb E[z_t \Delta_{20}\tilde x_t']^{-1} N(0,\Omega_{s,h}) +
    \underbrace{ O_p(\bar T^{-1/2}) O_p(h^2)}_{\text{bias due to lag-augmentation}} \right)
\end{align}
Since $H_2 \bar G_{2,T}' \mathbb E[z_t \Delta_{20}\tilde x_t']^{-1} = E[z_t  x_t']^{-1}= E[z_t  x_t']^{-1}$ and $\Sigma_{zx}=E[z_t  x_t']$, we obtain the first term has a Gaussian distribution with unity variance because of $\Omega_{\beta,h}=\Sigma_{zx}^{-1}\Omega_{s,h}\Sigma_{zx}^{'-1}$. Moreover, to make sure the bias term converges to zero, it is sufficient to impose restriction such that
\begin{align}
    \frac{h^2}{\bar T^{1/2}(w'\Omega_{\beta,h} w)^{1/2}} \xrightarrow{p}0.
\end{align}
It is satisfied under the condition $h\leq \bar h$ such that $\frac{\bar h^3}{T}\max (1, \frac{\bar h}{w'\Omega_{\beta,h} w })\xrightarrow{p}0$. Thus, the proof of \eqref{infea2} is complete.

\end{proof}

\newpage
\subsection{Proposition 5.2 (Estimates equivalence)}
\label{proofpropiden}
\begin{proof}
The proof is conducted across three cases, where $\delta \in \{0, 1, 2\}$.

\textbf{Case 1:} $\delta=0$, $\Phi\in \mathcal{B}(0,c,\epsilon)$ and $0<h\leq \alpha T$ for $\alpha \in (0,1)$. We show
\begin{align}
\label{a80}
\begin{split}
     &|\bar T^{1/2}w' 
        (\hat{\beta}_h^{2S}-\tilde{\beta}_h^{2S})|\\
        =&|  w'\left(  (\bar T^{-1}\sum_{t=p}^{T-h} 
\hat z_{t} x_{t}' )^{-1}(\bar T^{-1/2}\sum_{t=p}^{T-h} \hat z_{t} e_{t,h})
-
(\bar T^{-1}\sum_{t=p}^{T-h} 
 z_{t} x_{t}' )^{-1}(\bar T^{-1/2}\sum_{t=p}^{T-h}  z_{t} e_{t,h}) \right)|\\
 \leq &\| w\|\| (\bar T^{-1}\sum_{t=p}^{T-h} 
\hat z_{t} x_{t}' )^{-1} -(\bar T^{-1}\sum_{t=p}^{T-h} 
 z_{t} x_{t}' )^{-1}\|\|\bar T^{-1/2}\sum_{t=p}^{T-h}  z_{t} e_{t,h}\|\\
 &+\|w\|\| (\bar T^{-1}\sum_{t=p}^{T-h} 
\hat z_{t} x_{t}' )^{-1}\|\| \bar T^{-1/2}\sum_{t=p}^{T-h}  (\hat z_t -z_{t}) e_{t,h} \|\\
 \xrightarrow{p}& 0.
\end{split}
\end{align}
First, we check the convergence of $\| (\bar T^{-1}\sum_{t=p}^{T-h} 
\hat z_{t} x_{t}' )^{-1} -(\bar T^{-1}\sum_{t=p}^{T-h} 
 z_{t} x_{t}' )^{-1}\| \|\bar T^{-1/2}\sum_{t=p}^{T-h}  z_{t} e_{t,h}\|$.
The term $\| (\bar T^{-1}\sum_{t=p}^{T-h} 
\hat z_{t} x_{t}' )^{-1} -(\bar T^{-1}\sum_{t=p}^{T-h} 
 z_{t} x_{t}' )^{-1}\| \xrightarrow{p}0$ because of (1) $\bar T^{-1}\sum_{t=p}^{T-h} 
 z_{t} x_{t}' \xrightarrow{p}\mathbb E[z_t x_t']$ (by \eqref{ux0}, Lemma \ref{lemma2moments}), (2) $\bar T^{-1}\sum_{t=p}^{T-h} 
 \hat z_{t} x_{t}' \xrightarrow{p}\mathbb E[z_t x_t']$ (by \eqref{ux0} in Lemma \ref{lemma2moments} and \eqref{zzx0} in Lemma \ref{lemmanegerror}), and (3) $\mathbb E[z_tx_t']$ is a non-singular matrix. The term $\|\bar T^{-1/2}\sum_{t=p}^{T-h}  z_{t} e_{t,h}\|$ is $O_p(1)$ in the limit, as $\bar T^{-1/2}\sum_{t=p}^{T-h}  z_{t} e_{t,h} \xrightarrow{d}N(0,\Omega_{s,h})$ shown by Proposition \ref{propclt} and the maximum eigenvalue of $\Omega_{s,h}$ is bounded above by a constant when $\delta=0$ (see illustration in the discussion following \eqref{4.52} and \eqref{omegaspi}).

Next, we check the convergence of $\| (\bar T^{-1}\sum_{t=p}^{T-h} 
\hat z_{t} x_{t}' )^{-1}\|\| \bar T^{-1/2}\sum_{t=p}^{T-h}  (\hat z_t -z_{t}) e_{t,h} \|$. The term  $\|(\bar T^{-1}\sum_{t=p}^{T-h} 
\hat z_{t} x_{t}' )^{-1}\|$ is bounded above by a constant since  $\bar T^{-1}\sum_{t=p}^{T-h} 
\hat z_{t} x_{t}' \xrightarrow{p}\mathbb E[z_tx_t']$  due to \eqref{ux0}, Lemma \ref{lemma2moments} and \eqref{zzx0}, Lemma \ref{lemmanegerror}. The term $\| \bar T^{-1/2}\sum_{t=p}^{T-h} (\hat{z}_t - z_{t} )e_{t,h}\| \xrightarrow{p}0$ by \eqref{ze0}, Lemma \ref{lemmanegerror} and the non-singularity of $\Omega_{\beta,h}$ (implied by the non-singularity of $\Sigma_{zx}$ and $\Omega_{s,h}$). 

In summary, combining all the boundedness results yields the expression of \eqref{a80} converges to zero. Thus, \eqref{a80} is proved.

\textbf{Case 2:} $\delta=1$, $\Phi\in \mathcal{B}(1,c,\epsilon)$ and $0<h\leq  \bar h$, such that $\frac{\bar h}{T}\max (1, \frac{\bar h}{w'\Omega_{\beta,h} w })\xrightarrow{p}0$. We show

\begin{align}
\label{a83}
\begin{split}
    &\frac{1}{(w'\Omega_{\beta,h} w)^{1/2}} \bar T^{1/2}w' 
        (\hat{\beta}_h^{LA(1)-2S}-\tilde{\beta}_h^{LA(1)-2S})\\
        =&
        \frac{1}{(w'\Omega_{\beta,h} w)^{1/2}} \bar T^{1/2}w'         
        H_1 \left( (\sum_{t=p}^{T-h} 
\hat z_{t,1} x_{t,1}' )^{-1}(\sum_{t=p}^{T-h} \hat z_{t,1} e_{t,h}) - 
(\sum_{t=p}^{T-h} 
 z_{t,1} x_{t,1}' )^{-1}(\sum_{t=p}^{T-h}  z_{t,1} e_{t,h}) \right)\\
 =&\frac{1}{(w'\Omega_{\beta,h} w)^{1/2}} \bar T^{1/2} w'  H_1 
 \left( (\sum_{t=p}^{T-h} \hat z_{t,1} x_{t,1}' )^{-1}
 - (\sum_{t=p}^{T-h} 
 z_{t,1} x_{t,1}' )^{-1} \right) (\sum_{t=p}^{T-h}  z_{t,1} e_{t,h})\\
 & +\frac{1}{(w'\Omega_{\beta,h} w)^{1/2}}\bar T^{1/2} w'  H_1 (\sum_{t=p}^{T-h} \hat z_{t,1} x_{t,1}' )^{-1} (\sum_{t=p}^{T-h}  (\hat z_{t,1} - z_{t,1}) e_{t,h}).
\end{split}
\end{align}
The convergence of the first term in the last expression of \eqref{a83} can be shown as
\begin{align}
\label{a89}
    \begin{split}
        &\frac{1}{(w'\Omega_{\beta,h} w)^{1/2}} \left| \bar T^{1/2} w'  H_1 
 \left( (\sum_{t=p}^{T-h} \hat z_{t,1} x_{t,1}' )^{-1}
 - (\sum_{t=p}^{T-h} 
 z_{t,1} x_{t,1}' )^{-1} \right) (\sum_{t=p}^{T-h}  z_{t,1} e_{t,h})\right|
 \\
 = & 
\frac{1}{(w'\Omega_{\beta,h} w)^{1/2}} \left| w' H_1 
\bar G_{1,T}' \left( \left(\bar T^{-1}
\left[\begin{array}{cc}
  I_{pK}   & 0 \\
   0  & \Upsilon_1\Pi
\end{array}\right]
\sum_{t=p}^{T-h} \hat z_{t,1} x_{t,1}'\bar G_{1,T}' \right)^{-1} \right.\right.\\
 & 
\left.\left. - \left(\bar T^{-1}\left[\begin{array}{cc}
  I_{pK}   & 0 \\
   0  & \Upsilon_1\Pi
\end{array}\right]\sum_{t=p}^{T-h} 
 z_{t,1} x_{t,1}' \bar G_{1,T}'\right)^{-1} \right)
 \left(\bar T^{-1/2}\sum_{t=p}^{T-h}\left[\begin{array}{cc}
  I_{pK}   & 0 \\
   0  & \Upsilon_1\Pi
\end{array}\right]  z_{t,1} e_{t,h}\right) \right|
\\
 \leq  & \frac{1}{(w'\Omega_{\beta,h} w)^{1/2}}
 \| w\|\| H_1  \bar G_{1,T}'\|\\
 & \left\| 
\left(\bar T^{-1}\sum_{t=p}^{T-h} \left[\begin{array}{cc}
 \hat z_t \Delta_{10}\tilde x_t'   &\hat z_t \tilde y_{t-p}'\Upsilon_1 \\
 \Upsilon_1 \tilde y_{t-p}\Delta_{10}\tilde x_t'  & \Upsilon_1\tilde y_{t-p}\tilde y_{t-p}'\Upsilon_1
\end{array}\right]
\right)^{-1} \right.\\
&\left. - \left(\bar T^{-1}\sum_{t=p}^{T-h} \left[\begin{array}{cc}
  z_t \Delta_{10}\tilde x_t'   & z_t \tilde y_{t-p}'\Upsilon_1 \\
 \Upsilon_1 \tilde y_{t-p}\Delta_{10}\tilde x_t'  & \Upsilon_1\tilde y_{t-p}\tilde y_{t-p}'\Upsilon_1
\end{array}\right]
\right)^{-1} \right\| \\
 & \left\|
 \left[\begin{array}{c}
      \bar T^{-1/2}\sum_{t=p}^{T-h}  z_{t} e_{t,h}  \\
      \bar T^{-1/2}\sum_{t=p}^{T-h} \Upsilon_1 \tilde y_{t-p} e_{t,h} 
 \end{array} \right]\right\| \\
 \xrightarrow{p}&  0.
    \end{split}
\end{align}

To show the convergence, we need to first verify the invertibility of two covariance matrices. The first covariance matrix is non-singular because it is a consistent estimate of the second covariance matrix, and the second covariance matrix has been shown as invertible in the case of $\delta=1$ in the proof of Proposition \ref{propinfea}, see Appendix \ref{proofpropinfea}. The consistency is verified by the results of $\|\bar T^{-1}\sum_{t=p}^{T-h}(\hat {z}_t -z_t) \Delta_{10} \tilde x_t'\|\xrightarrow{p}0$,  $\|\bar T^{-1}\sum_{t=p}^{T-h}(\hat {z}_t -z_t) \Delta_{10} \tilde y_{t-p}'\Upsilon_1 \|\xrightarrow{p}0$. The two convergence results hold due to \eqref{zzx0}, Lemma \ref{lemmanegerror}, as $\Delta_{10} \tilde x_t$ is a vector of stacked $ \Delta_{10} \tilde y_{t-k}$'s (by the definition of $\Delta_{10} \tilde x_{t}$) and $\Delta_{10} \tilde y_{t-k}$ is a sub-vector of $G_{1,T}x_{t-k}$ (Similar for $\Delta_{10} \tilde y_{t-p}$).

As both matrices are invertibility, we obtain
\begin{align}
\label{a922}
    \begin{split}
        & 
\left\| 
\left(\bar T^{-1}\sum_{t=p}^{T-h} \left[\begin{array}{cc}
 \hat z_t \Delta_{10}\tilde x_t'   &\hat z_t \tilde y_{t-p}'\Upsilon_1 \\
 \Upsilon_1 \tilde y_{t-p}\Delta_{10}\tilde x_t'  & \Upsilon_1\tilde y_{t-p}\tilde y_{t-p}'\Upsilon_1
\end{array}\right]
\right)^{-1} \right.\\
& \left. - \left(\bar T^{-1}\sum_{t=p}^{T-h} \left[\begin{array}{cc}
  z_t \Delta_{10}\tilde x_t'   & z_t \tilde y_{t-p}'\Upsilon_1 \\
 \Upsilon_1 \tilde y_{t-p}\Delta_{10}\tilde x_t'  & \Upsilon_1\tilde y_{t-p}\tilde y_{t-p}'\Upsilon_1
\end{array}\right]
\right)^{-1} \right\|\\
\leq & \text{constant}\left( \| \bar T^{-1}\sum_{t=p}^{T-h}(\hat{z}_t-z_t) \Delta_{10} \tilde x_t' \| +\|(\bar T^{-1}\sum_{t=p}^{T-h}(\hat{z}_t-z_t) \tilde y_{t-p}'\Upsilon_1) \| \right)\\
= & O_p(T^{-1/2})
    \end{split}
\end{align}
The inequality is due to the general formula, $\|\hat{A}^{-1} -A^{-1}\|\leq \|\hat{A}^{-1}(\hat{A}-A)A^{-1}\|$. In this case, both inverse matrices are non-singular and thereby the norm of the inverse matrix is bounded above by a constant. The last equality is because \eqref{zzx0} in Lemma \ref{lemmanegerror} shows that the above two norms are $O_p(T^{-1/2})$ terms as component in $\Delta_{10}\tilde x_t$ and $\Upsilon_1 \tilde y_{t-p}$ are sub-vectors of $G_{1,T}x_{t-k}$ for some positive integer $k$. 

In addition, \eqref{xe1}, Lemma \ref{lemma2moments} show that $\|
      \bar T^{-1/2}\sum_{t=p}^{T-h} \Upsilon_1  \tilde y_{t-p} e_{t,h} \|$ is $O_p(h)$ as $ \Upsilon_1  \tilde y_{t-p}$ is a sub-vector of $G_{1,T} x_{t-1}$ . The term $\| \bar T^{-1/2}\sum_{t=p}^{T-h}  z_{t} e_{t,h} \|$ is bounded by a $O_p(h)$ in the case of $\delta=1$ because $\bar T^{-1/2}\sum_{t=p}^{T-h}  z_{t} e_{t,h} \xrightarrow{d}N(0,\Omega_{s,h})$ shown by Proposition \ref{propclt} and the maximum eigenvalue of $\Omega_{s,h}$ bounded above by a $O_p(h)$ when $\delta=1$ (see illustration in the discussion following \eqref{4.52} and \eqref{omegaspi}). Since $0<h\leq  \bar h$, such that $\frac{\bar h}{T}\max (1, \frac{\bar h}{w'\Omega_{\beta,h} w })\xrightarrow{p}0$, it yields $\frac{h}{(T w'\Omega_{\beta,h} w)^{1/2}} \rightarrow 0$ and therefore the convergence of \eqref{a89} holds.

Next we check the second term in the last expression of \eqref{a83}.
\begin{align}
\label{a107}
    \begin{split}
        &\frac{1}{(w'\Omega_{\beta,h} w)^{1/2}}|\bar T^{1/2}  w' H_1 (\sum_{t=p}^{T-h} \hat z_{t,1} x_{t,1}' )^{-1} (\sum_{t=p}^{T-h}  (\hat z_{t,1} - z_{t,1}) e_{t,h})|\\
        =&\frac{1}{(w'\Omega_{\beta,h} w)^{1/2}}|w' H_1 
\bar G_{1,T}'\left(\bar T^{-1}
\left[\begin{array}{cc}
  I_{pK}   & 0 \\
   0  & \Upsilon_1\Pi
\end{array}\right]
\sum_{t=p}^{T-h} \hat z_{t,1} x_{t,1}'\bar G_{1,T}' \right)^{-1}\\
& \left[\begin{array}{cc}
  I_{pK}   & 0 \\
   0  & \Upsilon_1\Pi
\end{array}\right] (\sum_{t=p}^{T-h}  (\hat z_{t,1} - z_{t,1}) e_{t,h})|\\
\leq &
\frac{1}{(w'\Omega_{\beta,h} w)^{1/2}}\|w\|\|H_1 
\bar G_{1,T}'\|\\
&\| \left(\bar T^{-1}\sum_{t=p}^{T-h} \left[\begin{array}{cc}
 \hat z_t \Delta_{10}\tilde x_t'   &\hat z_t \tilde y_{t-p}'\Upsilon_1 \\
 \Upsilon_1 \tilde y_{t-p}\Delta_{10}\tilde x_t'  & \Upsilon_1\tilde y_{t-p}\tilde y_{t-p}'\Upsilon_1
\end{array}\right]
\right)^{-1}
\|\\
 &\| \left[\begin{array}{c}
      \bar T^{-1/2}\sum_{t=p}^{T-h}  (\hat z_{t} - z_{t}) e_{t,h}  \\
      0
 \end{array} \right]\|\\
 \xrightarrow{p}& 0
    \end{split}
\end{align}
The convergence will be verified through the boundedness of each norm. $\|H_1 
\bar G_{1,T}'\|=O_p(1)$ as it is a constant matrix. The norm of the second term is bounded above by a constant almost surely because the argument for \eqref{a89} entails that it is a consistent estimate for a non-singular matrix. Lastly, $\frac{1}{(w'\Omega_{\beta,h} w)^{1/2}}\| \bar T^{-1/2}\sum_{t=p}^{T-h}  (\hat z_{t} - z_{t}) e_{t,h} \|\xrightarrow{p}0$ shown in \eqref{ze0}, Lemma \ref{lemmanegerror}. Therefore the convergence of \eqref{a107} holds. 

Thus, the convergence of \eqref{a89} and \eqref{a107} imply that \eqref{a83} holds.

\textbf{Case 3:} $\delta=2$, $\Phi\in \mathcal{B}(2,c,\epsilon)$ and $0<h\leq  \bar h$, such that  $\frac{\bar h^3}{T}\max(1,\frac{\bar h}{w'\Omega_{\beta,h} w})\xrightarrow{p}0$. We show
\begin{align}
\label{a96}
\begin{split}
    &\frac{1}{(w'\Omega_{\beta,h} w)^{1/2}} \bar T^{1/2}w' 
        (\hat{\beta}_h^{LA(2)-2S}-\tilde{\beta}_h^{LA(2)-2S})\\
        =&
        \frac{1}{(w'\Omega_{\beta,h} w)^{1/2}}\bar T^{1/2}w'         
        H_2 \left( (\sum_{t=p}^{T-h} 
\hat z_{t,2} x_{t,2}' )^{-1}(\sum_{t=p}^{T-h} \hat z_{t,2} e_{t,h}) - 
(\sum_{t=p}^{T-h} 
 z_{t,2} x_{t,2}' )^{-1}(\sum_{t=p}^{T-h}  z_{t,2} e_{t,h}) \right)\\
 =&\frac{1}{(w'\Omega_{\beta,h} w)^{1/2}} \bar T^{1/2} w'  H_2 
 \left( (\sum_{t=p}^{T-h} \hat z_{t,2} x_{t,2}' )^{-1}
 - (\sum_{t=p}^{T-h} 
 z_{t,2} x_{t,2}' )^{-1} \right) (\sum_{t=p}^{T-h}  z_{t,2} e_{t,h})\\
 & +\frac{1}{(w'\Omega_{\beta,h} w)^{1/2}}\bar T^{1/2} w'  H_2 (\sum_{t=p}^{T-h} \hat z_{t,2} x_{t,2}' )^{-1} (\sum_{t=p}^{T-h}  (\hat z_{t,2} - z_{t,2}) e_{t,h}).
\end{split}
\end{align}

The first term in the last expression of \eqref{a96} can be shown as
\begin{align}
\label{a110}
    \begin{split}
        & \frac{1}{(w'\Omega_{\beta,h} w)^{1/2}}
        | \bar T^{1/2} w'  H_2 
 \left( (\sum_{t=p}^{T-h} \hat z_{t,2} x_{t,2}' )^{-1}
 - (\sum_{t=p}^{T-h} 
 z_{t,2} x_{t,2}' )^{-1} \right) (\sum_{t=p}^{T-h}  z_{t,2} e_{t,h})|\\
 =&\frac{1}{(w'\Omega_{\beta,h} w)^{1/2}}
 \left|w' H_2 G_{2,T}' 
 \left( \left(\bar T^{-1}\sum_{t=p}^{T-h}
  \left[\begin{array}{ccc}
    I_{pK}  & & \\
      & \Upsilon_1 \Pi & -\Upsilon_1 P_2\Pi \\
      && \Upsilon_2\Pi
  \end{array} \right]\hat z_{t,2} x_{t,2}' G_{2,T}'\right)^{-1} \right. \right.
 \\
 & \left. - \left(\bar T^{-1}\sum_{t=p}^{T-h} 
 \left[\begin{array}{ccc}
    I_{pK}  & & \\
      & \Upsilon_1 \Pi & -\Upsilon_1 P_2\Pi \\
      && \Upsilon_2\Pi
  \end{array} \right]
 z_{t,2} x_{t,2}' G_{2,T}'\right)^{-1} \right) \\
 &\left.  \left(\bar T^{-1/2}\sum_{t=p}^{T-h}  \left[\begin{array}{ccc}
    I_{pK}  & & \\
      & \Upsilon_1 \Pi & -\Upsilon_1 P_2\Pi \\
      && \Upsilon_2\Pi
  \end{array} \right] z_{t,2} e_{t,h}\right) \right|
  \\
  \leq &
  \frac{1}{(w'\Omega_{\beta,h} w)^{1/2}}
 \|w\| \| H_2 G_{2,T}' \|\\
  &
  \| \left(\bar T^{-1}\sum_{t=p}^{T-h} 
\left[\begin{array}{ccc}
  \hat  z_t \Delta_{20}\tilde x_t'  &\hat z_t \Delta_{21}\tilde y_{t-p}'\Upsilon_1 &\hat z_t \tilde y_{t-p-1}'\Upsilon_2\\
   \Upsilon_1 \Delta_{21}\tilde y_{t-p} \Delta_{20}\tilde x_t'  & \Upsilon_1\Delta_{21}\tilde y_{t-p} \Delta_{21}\tilde y_{t-p}'&
   \Upsilon_1 \Delta_{21}\tilde y_{t-p} \tilde y_{t-p-1}'\Upsilon_2\\
    \Upsilon_2\tilde y_{t-p-1}\Delta_{20}\tilde x_t'  & \Upsilon_2\tilde y_{t-p-1}\Delta_{21}\tilde y_{t-p}'\Upsilon_1 & \Upsilon_2\tilde y_{t-p-1} \tilde y_{t-p-1}'\Upsilon_2
  \end{array} \right] \right)^{-1}
 \\
 &- \left(\bar T^{-1}\sum_{t=p}^{T-h} 
\left[\begin{array}{ccc}
    z_t \Delta_{20}\tilde x_t'  &z_t \Delta_{21}\tilde y_{t-p}'\Upsilon_1 & z_t \tilde y_{t-p-1}'\Upsilon_2\\
   \Upsilon_1 \Delta_{21}\tilde y_{t-p} \Delta_{20}\tilde x_t'  & \Upsilon_1\Delta_{21}\tilde y_{t-p} \Delta_{21}\tilde y_{t-p}'&
   \Upsilon_1 \Delta_{21}\tilde y_{t-p} \tilde y_{t-p-1}'\Upsilon_2\\
    \Upsilon_2\tilde y_{t-p-1}\Delta_{20}\tilde x_t'  & \Upsilon_2\tilde y_{t-p-1}\Delta_{21}\tilde y_{t-p}'\Upsilon_1 & \Upsilon_2\tilde y_{t-p-1} \tilde y_{t-p-1}'\Upsilon_2
  \end{array} \right] \right)^{-1}\| \\
 &\|\bar T^{-1/2}\sum_{t=p}^{T-h}  \left[\begin{array}{ccc}
    z_t e_{t,h}\\
    \Upsilon_1 \Delta_{21}\tilde y_{t-p}e_{t,h}  \\
    \Upsilon_2 \tilde y_{t-p-1}e_{t,h}
  \end{array} \right] \| \\
  \xrightarrow{p}& 0
    \end{split}
\end{align}
To show the convergence, we need to first verify the invertibility of two covariance matrices.
The first covariance matrix is non-singular because it is a consistent estimate of the second covariance
matrix, and the second covariance matrix has been shown as invertible in the case of $\delta=2$ in the proof of Proposition \ref{propinfea}, see Appendix \ref{proofpropinfea}. The consistency is verified by the results of $\|\bar T^{-1}\sum_{t=p}^{T-h}(\hat {z}_t -z_t) \Delta_{20} \tilde x_t'\|\xrightarrow{p}0$,  $\|\bar T^{-1}\sum_{t=p}^{T-h}(\hat {z}_t -z_t) \Delta_{21} \tilde y_{t-p}'\Upsilon_1 \|\xrightarrow{p}0$, and  $\|\bar T^{-1}\sum_{t=p}^{T-h}(\hat {z}_t -z_t)  \tilde y_{t-p-1}'\Upsilon_2 \|\xrightarrow{p}0$. The three convergence results hold due to \eqref{zzx0}, Lemma \ref{lemmanegerror}, as $\Delta_{20} \tilde x_t$ is a stacked $\Delta_{20} \tilde y_{t}$'s and $\Delta_{20} \tilde y_{t}$ is sub-vector of $G_{2,T}x_{t}$; also $\Upsilon_1 \Delta_{21} \tilde y_{t-p}$ and $\Upsilon_2\tilde y_{t-p-1}$ are sub-vectors of $G_{2,T}x_{t-2}$.

As both matrices are invertibility, we obtain
\begin{align}
\label{a9222}
    \begin{split}
        & 
 \| \left(\bar T^{-1}\sum_{t=p}^{T-h} 
\left[\begin{array}{ccc}
  \hat  z_t \Delta_{20}\tilde x_t'  &\hat z_t \Delta_{21}\tilde y_{t-p}'\Upsilon_1 &\hat z_t \tilde y_{t-p-1}'\Upsilon_2\\
   \Upsilon_1 \Delta_{21}\tilde y_{t-p} \Delta_{20}\tilde x_t'  & \Upsilon_1\Delta_{21}\tilde y_{t-p} \Delta_{21}\tilde y_{t-p}'&
   \Upsilon_1 \Delta_{21}\tilde y_{t-p} \tilde y_{t-p-1}'\Upsilon_2\\
    \Upsilon_2\tilde y_{t-p-1}\Delta_{20}\tilde x_t'  & \Upsilon_2\tilde y_{t-p-1}\Delta_{21}\tilde y_{t-p}'\Upsilon_1 & \Upsilon_2\tilde y_{t-p-1} \tilde y_{t-p-1}'\Upsilon_2
  \end{array} \right] \right)^{-1}
 \\
 &- \left(\bar T^{-1}\sum_{t=p}^{T-h} 
\left[\begin{array}{ccc}
    z_t \Delta_{20}\tilde x_t'  &z_t \Delta_{21}\tilde y_{t-p}'\Upsilon_1 & z_t \tilde y_{t-p-1}'\Upsilon_2\\
   \Upsilon_1 \Delta_{21}\tilde y_{t-p} \Delta_{20}\tilde x_t'  & \Upsilon_1\Delta_{21}\tilde y_{t-p} \Delta_{21}\tilde y_{t-p}'&
   \Upsilon_1 \Delta_{21}\tilde y_{t-p} \tilde y_{t-p-1}'\Upsilon_2\\
    \Upsilon_2\tilde y_{t-p-1}\Delta_{20}\tilde x_t'  & \Upsilon_2\tilde y_{t-p-1}\Delta_{21}\tilde y_{t-p}'\Upsilon_1 & \Upsilon_2\tilde y_{t-p-1} \tilde y_{t-p-1}'\Upsilon_2
  \end{array} \right] \right)^{-1}\|\\
\leq & \text{constant}
\left( 
\|\bar T^{-1}\sum_{t=p}^{T-h}(\hat{z}_t-z_t) \Delta_{20} \tilde x_t' \|
+
\|(\bar T^{-1}\sum_{t=p}^{T-h}(\hat{z}_t-z_t)\Delta_{21} \tilde y_{t-p}'\Upsilon_1) \| \right.\\
& \left. +
\|(\bar T^{-1}\sum_{t=p}^{T-h}(\hat{z}_t-z_t)\tilde y_{t-p-1}'\Upsilon_1) \|
\right)\\
= & O_p(T^{-1/2})
    \end{split}
\end{align}
The inequality shares the same argument for \eqref{a922}. The last equality is because \eqref{zzx0} in Lemma \ref{lemmanegerror} shows that the above three norms are $O_p(T^{-1/2})$ terms, as components in $\Delta_{20} \tilde x_t$, $\Upsilon_1\Delta_{21} \tilde y_{t-p}$ and $\Upsilon_1 \tilde y_{t-p-1}$ are sub-vectors of $G_{2,T}x_{t-k}$ for some positive integer $k$.

In addition, \eqref{xe2}, Lemma \ref{lemma2moments} show that $\|\bar T^{-1}\sum_{t=p}^{T-h}
z_t e_{t,h}\|$,
$\|\bar T^{-1}\sum_{t=p}^{T-h} \Upsilon_1 \Delta_{21}\tilde y_{t-p}e_{t,h}\|$, and
    $\|\bar T^{-1}\sum_{t=p}^{T-h}
    \Upsilon_2 \tilde y_{t-p-1}e_{t,h}\|$ is $O_p(h^2)$ 
    as 
    $\Upsilon_1 \Delta_{21}\tilde y_{t-p}$ and $\Upsilon_2 \tilde y_{t-p-1}$ are sub-vectors of $G_{2,T} x_{t-2}$ . The term $\| \bar T^{-1/2}\sum_{t=p}^{T-h}  z_{t} e_{t,h} \|$ is bounded by a $O_p(h^2)$ in the case of $\delta=2$ because $\bar T^{-1/2}\sum_{t=p}^{T-h}  z_{t} e_{t,h} \xrightarrow{d}N(0,\Omega_{s,h})$ shown by Proposition \ref{propclt} and the maximum eigenvalue of $\Omega_{s,h}$ bounded above by a $O_p(h)$ when $\delta=1$ (see illustration in the discussion following \eqref{4.52} and \eqref{omegaspi}). Since $0<h\leq  \bar h$, such that $\frac{\bar h}{T}\max (1, \frac{\bar h}{w'\Omega_{\beta,h} w })\xrightarrow{p}0$, it yields $\frac{h}{(T w'\Omega_{\beta,h} w)^{1/2}} \rightarrow 0$ and therefore the convergence of \eqref{a89} holds.

      The last equality is because Lemma \ref{lemmanegerror} shows that the above three norms are $O_p(T^{-1/2})$ terms. In addition, Lemma \ref{lemma2moments} proved that $\|\bar T^{-1}\sum_{t=p}^{T-h}
z_t e_{t,h}\|$,
$\|\bar T^{-1}\sum_{t=p}^{T-h} \Upsilon_1 \Delta_{21}\tilde y_{t-p}e_{t,h}\|$, and
    $\|\bar T^{-1}\sum_{t=p}^{T-h}
    \Upsilon_2 \tilde y_{t-p-1}e_{t,h}\|$    are bounded above by a $O_p(h^2)$ terms because $\bar T^{-1/2}\sum_{t=p}^{T-h}  z_{t} e_{t,h} \xrightarrow{d}N(0,\Omega_{s,h})$ shown by Proposition \ref{propclt} and the maximum eigenvalue of $\Omega_{s,h}$ bounded above by a $O_p(h^2)$ when $\delta=2$ (see illustration in the discussion following \eqref{4.52} and \eqref{omegaspi}). Since $0<h\leq  \bar h$, such that $\frac{\bar h^3}{T}\max (1, \frac{\bar h}{w'\Omega_{\beta,h} w })\xrightarrow{p}0$, it yields $\frac{h^2}{(T w'\Omega_{\beta,h} w)^{1/2}} \rightarrow 0$ and therefore the convergence of \eqref{a110} holds.

Next, the second term in the last expression of \eqref{a96} can be shown as
\begin{align}
\label{a111}
    \begin{split}
        & \frac{1}{(w'\Omega_{\beta,h} w)^{1/2}}|\bar T^{1/2} w'  H_2 (\sum_{t=p}^{T-h} \hat z_{t,2} x_{t,2}' )^{-1} (\sum_{t=p}^{T-h}  (\hat z_{t,2} - z_{t,2}) e_{t,h})|
        \\
        = 
        &
        \frac{1}{(w'\Omega_{\beta,h} w)^{1/2}}| w' 
         H_2 G_{2,T}'
         \left(\sum_{t=p}^{T-h} \left[\begin{array}{ccc}
    I_{pK}  & & \\
      & \Upsilon_1 \Pi & -\Upsilon_1 P_2\Pi \\
      && \Upsilon_2\Pi
  \end{array} \right]\hat z_{t,2} x_{t,2}'G_{2,T}' \right)^{-1} \\
  &
  \sum_{t=p}^{T-h}  \left[\begin{array}{ccc}
    I_{pK}  & & \\
      & \Upsilon_1 \Pi & -\Upsilon_1 P_2\Pi \\
      && \Upsilon_2\Pi
  \end{array} \right] (\hat z_{t,2} - z_{t,2}) e_{t,h})|\\
  \leq &
  \frac{1}{(w'\Omega_{\beta,h} w)^{1/2}}
  \|w\|  \| H_2 G_{2,T}'\|\\
  &
  \| \left(\bar T^{-1}\sum_{t=p}^{T-h} 
\left[\begin{array}{ccc}
  \hat  z_t \Delta_{20}\tilde x_t'  &\hat z_t \Delta_{21}\tilde y_{t-p}'\Upsilon_1 &\hat z_t \tilde y_{t-p-1}'\Upsilon_2\\
   \Upsilon_1 \Delta_{21}\tilde y_{t-p} \Delta_{20}\tilde x_t'  & \Upsilon_1\Delta_{21}\tilde y_{t-p} \Delta_{21}\tilde y_{t-p}'&
   \Upsilon_1 \Delta_{21}\tilde y_{t-p} \tilde y_{t-p-1}'\Upsilon_2\\
    \Upsilon_2\tilde y_{t-p-1}\Delta_{20}\tilde x_t'  & \Upsilon_2\tilde y_{t-p-1}\Delta_{21}\tilde y_{t-p}'\Upsilon_1 & \Upsilon_2\tilde y_{t-p-1} \tilde y_{t-p-1}'\Upsilon_2
  \end{array} \right] \right)^{-1} \| \\
  &\|\bar T^{-1/2}\sum_{t=p}^{T-h} \left[\begin{array}{c}
    (\hat z_t -z_t)e_{t,h}  \\
     0 \\
     0
  \end{array} \right] \| 
  \\
  \xrightarrow{p}& 0.
    \end{split}
\end{align}
The convergence will be verified through the boundedness of each norm. $\|H_2 
\bar G_{2,T}'\|=O_p(1)$ as it is a constant matrix. The norm of the inverse matrix is bounded above by a constant almost surely because the argument for \eqref{a110} entails that it is a consistent estimate for a non-singular matrix. Lastly, $\frac{1}{(w'\Omega_{\beta,h} w)^{1/2}}\| \bar T^{-1/2}\sum_{t=p}^{T-h}  (\hat z_{t} - z_{t}) e_{t,h} \|\xrightarrow{p}0$ shown in \eqref{ze0}, Lemma \ref{lemmanegerror}. Therefore the convergence of \eqref{a107} holds. 

Thus, the convergence of \eqref{a110} and \eqref{a111} imply that \eqref{a96} holds. The proof is complete.

\end{proof}

\newpage
\subsection{Proposition 5.3 (Uniform convergence)}
\label{proofunitheo}
\begin{proof}
We decompose the following term as
\begin{align}
\label{a100}
        \begin{split}
            & \bar T^{1/2} \frac{ w' ( 
        \hat {\beta}_h^{LA(\delta)-\textup{2S}}         - 
       \beta_h ) }{ (w'\hat \Omega_{\beta,h}w)^{1/2}}\\
       = & \left( \bar T^{1/2} \frac{ w' ( 
        \hat {\beta}_h^{LA(\delta)-\textup{2S}}  - \tilde{\beta}_h^{LA(\delta)-\textup{2S}}  ) }{(w'\Omega_{\beta}w)^{1/2}} 
       +
       \bar T^{1/2} \frac{w'(\tilde{\beta}_h^{LA(\delta)-\textup{2S}}   - 
       \beta_h)}{(w'\Omega_{\beta}w)^{1/2}}
       \right)
       \frac{(w'\Omega_{\beta}w)^{1/2}}{ (w'\hat \Omega_{\beta,h}w)^{1/2}}.
        \end{split}
    \end{align}
    for $\delta=0,1,2$. Proposition 5.1 and 5.2 respectively show that $\bar T^{1/2} \frac{ w' ( 
        \hat {\beta}_h^{LA(\delta)-\textup{2S}}  - \tilde{\beta}_h^{LA(\delta)-\textup{2S}}  ) }{(w'\Omega_{\beta}w)^{1/2}}  \xrightarrow{p}0$ and $\bar T^{1/2} \frac{w'(\tilde{\beta}_h^{LA(\delta)-\textup{2S}}   - 
       \beta_h)}{(w'\Omega_{\beta}w)^{1/2}}\xrightarrow{d}N(0,1)$ for all cases of $\delta=0,1,2$. Thus, to show the convergence of \eqref{a100}, given the explicit formula of $\Omega_{\beta,h}=\Sigma_{zx}^{-1} {\Omega}_{s,h} \Sigma_{zx}^{'-1}$ and $\Sigma_{zx}$ is a constant matrix, $\Sigma_{zx}=(I_p\otimes \Sigma_u) {\overline \Psi}_p'$, it is sufficient to prove
       \begin{align}
      \label{convergencesigu}     & \hat{\Sigma}_u \xrightarrow{p} {\Sigma}_u ,\\
       \label{convergencepsip}     & \hat{\overline \Psi}_p \xrightarrow{p} {\overline \Psi}_p,\\
       \label{convergenceomegash}     &w'\hat{\Omega}_{s,h} w / w'{\Omega}_{s,h}w \xrightarrow{p} 1
       \end{align}
       for all cases of $\delta=0,1,2$, $w\in \mathbb R^{pK}$, and $\|w\|=1$.
       
First, we prove \eqref{convergencesigu}. Given the definition of $\hat{\Sigma}_u =\frac{1}{\bar T}\sum_{t=p}^{T} \hat{u}_t\hat{u}_t' $, $\hat{u}_t = y_t - \hat{\Phi}(p) x_t $, and $\hat{\Phi}(p) = \sum_{t=1}^{T} y_t x_{t-1}' (\sum_{t=1}^{T} x_{t-1} x_{t-1}' )^{-1} $, then
        \begin{align}
        \label{b125}
            \begin{split}
                & \frac{1}{ T}\sum_{t=1}^{T} \hat{u}_t\hat{u}_t' \\
                = & \frac{1}{ T} \sum_{t=1}^{T}\left({u}_t+ (\Phi(p)-\hat{\Phi}(p))x_{t-1} \right)\left({u}_t+ (\Phi(p)-\hat{\Phi}(p))x_{t-1} \right)' \\
                = & \frac{1}{ T}\sum_{t=1}^{T} u_t u_t'+2(\frac{1}{T}\sum_{t=1}^{T} u_tx_{t-1}')(\frac{1}{T}\sum_{t=1}^{T}x_{t-1}x_{t-1}')^{-1}(\frac{1}{ T}\sum_{t=1}^{T}x_{t-1} {u}_t')  \\
                &+(\frac{1}{T}\sum_{t=1}^{T} u_tx_{t-1}')(\frac{1}{T}\sum_{t=1}^{T}x_{t-1}x_{t-1}')^{-1}
                (\frac{1}{T}\sum_{t=1}^{T}x_{t-1}x_{t-1}')
                (\frac{1}{T}\sum_{t=1}^{T}x_{t-1}x_{t-1}')^{-1}(\frac{1}{ T}\sum_{t=1}^{T}x_{t-1} {u}_t')  \\         
                = & \frac{1}{ T}\sum_{t=1}^{T} u_t u_t'+3(\frac{1}{T}\sum_{t=1}^{T} u_tx_{t-1}'G_{\delta,T}')(\frac{1}{T}\sum_{t=1}^{T}G_{\delta,T}x_{t-1}x_{t-1}'G_{\delta,T}')^{-1}(\frac{1}{ T}\sum_{t=1}^{T}G_{\delta,T}x_{t-1} {u}_t')  \\       
                \xrightarrow{p}&\Sigma_u 
            \end{split}
        \end{align}
        The convergence of $\frac{1}{ T}\sum_{t=1}^{T} u_t u_t'$ is shown in \eqref{b174}. In addition, the second term converges in probability to zero because $\|\frac{1}{T}\sum_{t=1}^{T} u_tx_{t-1}'G_{\delta,T}'\|\xrightarrow{p}0$ by \eqref{ux0}, \eqref{ux1}, and \eqref{ux2} in Lemma \ref{lemma2moments} for $\delta=0,1,2$, respectively; and $\|(\frac{1}{T}\sum_{t=1}^{T}G_{\delta,T}x_{t-1}x_{t-1}'G_{\delta,T}')^{-1}\|$ is bounded above by a constant almost surely by Assumption \ref{ass4}.  

        Next, we prove \eqref{convergencepsip}. As $\hat{\overline \Psi}_p$ is constructed through recursive VAR-based impulse response estimates and $p$ is a finite integer, it is sufficient to show the LS estimator of VAR slope coefficients is consistent. It is equivalently to show the VAR slope coefficients are consistent for $\delta=0,1,2$,
        \begin{align}
            \begin{split}
                & (\frac{1}{T}\sum_{t=1}^{T} u_tx_{t-1}')(\frac{1}{T}\sum_{t=1}^{T}x_{t-1}x_{t-1}')^{-1}\\
                =& 
                (\frac{1}{T}\sum_{t=1}^{T} u_tx_{t-1}'G_{\delta,T}')(\frac{1}{T}\sum_{t=1}^{T}G_{\delta,T}x_{t-1}x_{t-1}'G_{\delta,T}')^{-1}G_{\delta,T}\\
                 \xrightarrow{p}&0.
            \end{split}
        \end{align}
        The convergence is because $\|\frac{1}{T}\sum_{t=1}^{T} u_tx_{t-1}'G_{\delta,T}'\|\xrightarrow{p}0$ and  $\|(\frac{1}{T}\sum_{t=1}^{T}G_{\delta,T}x_{t-1}x_{t-1}'G_{\delta,T}')^{-1}\|$ is bounded above by a constant almost surely. The boundedness and convergence are argued in the paragraph following \eqref{b125}. The matrix  $\|G_{\delta,T}\|$ is a $O_p(1)$ term by its definition.

        Lastly, we prove \eqref{convergenceomegash}. Note that
        in Appendix \eqref{proofprop4.3}, we have shown that $(\bar T -p+1)^{-1} \sum_{t=p}^{\bar T} (w's_{t,h}^*)^2/ w'\Omega_{s,h} w \xrightarrow{p}1$, for any $w\in \mathbb R^{pk\times 1}$, $\|w\|=1$. 
        Given the definition of $\hat{\Omega}_{s,h}=(\bar T-p+1)^{-1}\sum_{t=p}^{\bar T} \hat{s}_{t,h}^{*}\hat{s}_{t,h}^{*'}$ and $\hat{s}_{t,h}^{*}= (\hat{e}_{t,h},\hat{e}_{t+1,h},\cdots,\hat{e}_{t+p-1,h})'\otimes \hat{u}_t$, to prove \eqref{convergenceomegash}, it is equivalent to show that
        \begin{align}
        \label{b.127}
            \begin{split}
                &\frac{1}{(\bar T-p+1)w'\Omega_{s,h}w}
               \left( \sum_{t=p}^{\bar T}  (w' \hat{s}_{t,h}^{*} )^2 -\sum_{t=p}^{\bar T} (w' {s}_{t,h}^{*} )^2 \right)\\
               =&\frac{1}{(\bar T-p+1)w'\Omega_{s,h}w}\sum_{t=1}^{\bar T} (w' \hat{s}_{t,h}^{*}- w' {s}_{t,h}^{*} ) (w' \hat{s}_{t,h}^{*}+ w' {s}_{t,h}^{*} )\\
               =&\frac{1}{(\bar T-p+1)w'\Omega_{s,h}w}\sum_{t=1}^{\bar T} (w' \hat{s}_{t,h}^{*}- w' {s}_{t,h}^{*} ) (w' \hat{s}_{t,h}^{*}- w' {s}_{t,h}^{*}+2w' {s}_{t,h}^{*} )\\
                 =&\frac{1}{(\bar T-p+1)w'\Omega_{s,h}w}\sum_{t=1}^{\bar T} (w' \hat{s}_{t,h}^{*}- w' {s}_{t,h}^{*} )^2 + \frac{2}{(\bar T-p+1)w'\Omega_{s,h}w}\sum_{t=1}^{\bar T} (w' \hat{s}_{t,h}^{*}- w' {s}_{t,h}^{*} )w' {s}_{t,h}^{*} \\
                \leq &
                \frac{\sum_{t=1}^{\bar T} (w' \hat{s}_{t,h}^{*}- w' {s}_{t,h}^{*} )^2}{(\bar T-p+1)w'\Omega_{s,h}w}+
                2\left( \frac{\sum_{t=1}^{\bar T} (w' \hat{s}_{t,h}^{*}- w' {s}_{t,h}^{*} )^2}{(\bar T-p+1)w'\Omega_{s,h}w} \right)^{1/2}\left( \frac{\sum_{t=1}^{\bar T} (w' {s}_{t,h}^{*})^2}{(\bar T-p+1)w'\Omega_{s,h}w} \right)^{1/2}\\
                \xrightarrow{p}&0.
            \end{split}
        \end{align}
        The convergence can be shown if (1) $\frac{\sum_{t=1}^{\bar T} (w' {s}_{t,h}^{*})^2}{(\bar T-p+1)w'\Omega_{s,h}w} \xrightarrow{p} 1$ and (2) $\frac{\sum_{t=1}^{\bar T} (w' \hat{s}_{t,h}^{*}- w' {s}_{t,h}^{*} )^2}{(\bar T-p+1)w'\Omega_{s,h}w} \xrightarrow{p} 0$.    In Appendix \eqref{proofprop4.3}, we proved the first condition by showing $\sum_{t=1}^{\bar T-p+1 } \chi_{t,h}^2 \xrightarrow{p}1$, where $\chi_{t,h} = (\bar T -p+1)^{-1/2} w's_{\bar T +1 -t,h}^* / (w'\Omega_{s,h} w)^{1/2}$. The second condition will be proved as follows,
        \begin{align}
            \begin{split}
                & \frac{1}{(\bar T-p+1)w'\Omega_{s,h}w}\sum_{t=1}^{\bar T} (w' \hat{s}_{t,h}^{*} - w' {s}_{t,h}^{*})^2\\
                =
                & \frac{1}{(\bar T-p+1)w'\Omega_{s,h}w}\sum_{t=1}^{\bar T} \left( \sum_{k=1}^p w_k' \hat{u}_t \hat  e_{t+k-1,h} -w_k'u_t e_{t+k-1,h}\right)^2\\
                &(\text{by partition $w$ into $p$ sub-vectors, $w=(w_1',w_2',\cdots,w_p')'$, and $w_k\in \mathbb R^K$ for $k=1,2,\cdots,p$},\\
                &\quad \text{also the explicit form of ${s}_{t,h}^{*} = (e_{t,h},e_{t+1,h},\cdots,e_{t+p-1,h})'\otimes u_t$})\\
                \leq 
                & \frac{1}{(\bar T-p+1)w'\Omega_{s,h}w}\sum_{t=1}^{\bar T} \left( p\sum_{k=1}^p \left( w_k'  \hat{u}_t \hat  e_{t+k-1,h} -w_k' u_t e_{t+k-1,h} \right)^2 \right)\\
                &(\text{Cauchy-Schwarz inequality})
                       \end{split}
        \end{align}
Since $p$ is finite and the minimum eigenvalue of $\Omega_{s,h}$ is bounded below by a constant, without loss of generality, it is equivalently to show that for all $k=1,2,\cdots,p$,
          \begin{align}
            \begin{split}
            &\frac{1}{\bar T-p+1}\sum_{t=1}^{\bar T} \left( w_k'  \hat{u}_t \hat  e_{t+k-1,h} -w_k' u_t e_{t+k-1,h} \right)^2 \\
            = &\frac{1}{\bar T-p+1}\sum_{t=1}^{\bar T} \left( w_k'  (\hat{u}_t - u_t)e_{t+k-1,h}+ w_k' u_t (
            \hat  e_{t+k-1,h}-e_{t+k-1,h})
            +w_k'( \hat u_t - u_t )(\hat e_{t+k-1,h} - e_{t+k-1,h}) \right)^2 \\
                \leq &
                 \frac{\|w_1\|^2}{\bar T-p+1} \left(
                \|\sum_{t=1}^{\bar T} ( \hat u_t -  u_t ) e_{t+k-1,h}\|
                +
                \|\sum_{t=1}^{\bar T}  u_t (\hat e_{t+k-1,h}- e_{t+k-1,h} )\| \right.\\
                  &\left.+
                \|\sum_{t=1}^{\bar T}  ( \hat u_t -  u_t )  (\hat e_{t+k-1,h}- e_{t+k-1,h} )\|\right)\\
               = &
                 \frac{\|w_1\|^2}{\bar T-p+1} 
                \left(
                \|\sum_{t=1}^{T} u_t x_{t-1}' (\sum_{t=1}^{T} x_{t-1} x_{t-1}' )^{-1}\sum_{t=1}^{\bar T} x_{t-1} e_{t+k-1,h}\| \right. \\
                &                +
                \|\sum_{t=1}^{\bar T}  u_t x_{t}'(\sum_{t=1}^{\bar T}  x_t x_{t}')^{-1}\sum_{t=1}^{\bar T}  x_t e_{t+k-1,h}\|
                 \\
                 & \left. +
                \sum_{t=1}^{T} u_t x_{t-1}' (\sum_{t=1}^{T} x_{t-1} x_{t-1}' )^{-1}\sum_{t=1}^{\bar T} x_{t-1}x_{t}'(\sum_{t=1}^{\bar T}  x_t x_{t}')^{-1}\sum_{t=1}^{\bar T}  x_t e_{t+k-1,h}
                \|\right)\\
                \leq  &
                  \frac{\|w_1\|^2}{\bar T-p+1} \\
                &\left(
                \|T^{-1/2}\sum_{t=1}^{T} u_t x_{t-1}' G_{\delta,T}'\|
                \|(T^{-1}\sum_{t=1}^{T} G_{\delta,T}x_{t-1} x_{t-1}' G_{\delta,T}')^{-1}\|
                \|T^{-1/2}\sum_{t=1}^{\bar T} G_{\delta,T}x_{t-1} e_{t+k-1,h}\| \right. \\
                &               +
                \|T^{-1/2}\sum_{t=1}^{\bar T}  u_t x_{t}'G_{\delta,T}'\|
                \|(T^{-1}\sum_{t=1}^{\bar T}G_{\delta,T}  x_t x_{t}'G_{\delta,T}')^{-1}\|
                \|T^{-1/2}\sum_{t=1}^{\bar T} G_{\delta,T} x_t e_{t+k-1,h}\|\\
               &  +
                \|T^{-1/2}\sum_{t=1}^{T} u_t x_{t-1}' G_{\delta,T}'\|\|(T^{-1}\sum_{t=1}^{T}G_{\delta,T} x_{t-1} x_{t-1}' G_{\delta,T}')^{-1}\|\|T^{-1}\sum_{t=1}^{\bar T}G_{\delta,T} x_{t-1}x_{t}'G_{\delta,T}'\|\\
                & \left. \quad \|(T^{-1}\sum_{t=1}^{\bar T}G_{\delta,T}  x_t x_{t}'G_{\delta,T}')^{-1}\| \|T^{-1/2}\sum_{t=1}^{\bar T} G_{\delta,T} x_t e_{t+k-1,h}
                \|\right)\\
                \xrightarrow{p} & 0 
            \end{split}
        \end{align}
        The convergence holds because (1) $\|T^{-1/2}\sum_{t=1}^{T} u_t x_{t-1}' G_{\delta,T}'\|$ is a $O_p(1)$ term by \eqref{ux0}, \eqref{ux1}, \eqref{ux2} in Lemma \ref{lemma2moments} for cases of $\delta=0,1,2$, respectively; (2) $\|(T^{-1}\sum_{t=1}^{\bar T}G_{\delta,T}  x_t x_{t}'G_{\delta,T}')^{-1}\|$ is bounded above by a constant almost surely by Assumption \ref{ass4}; (3) $\|T^{-1}\sum_{t=1}^{\bar T}G_{\delta,T} x_{t-1}x_{t}'G_{\delta,T}'\|$ is $O_p(1)$  in the limit for all cases of $\delta=0,1,2$ due to Cauchy-Schwarz inequality and $\|T^{-1}\sum_{t=1}^{\bar T}G_{\delta,T} x_{t}x_{t}'G_{\delta,T}'\|=O_p(1)$ shown in  \eqref{xx0}, \eqref{xx1}, \eqref{xx2} in Lemma \ref{lemma2moments} for cases of $\delta=0,1,2$, respectively; (4)  $\|T^{-1/2}\sum_{t=1}^{\bar T} G_{\delta,T} x_t e_{t+k,h}
                \|$ is a $O_p(1)$, $O_p(h)$, and $O_p(h^2)$ term in the limit for the cases of $\delta=0,1,2$, respectively, by \eqref{xe0}, \eqref{xe1}, and \eqref{xe2} in Lemma \ref{lemma2moments}. Therefore, given the constraints imposed on horizon $h$ for cases of $\delta=1,2$, the big summation in the last expression above divided by $\bar T-p+1$ converges to zero. Thus, \eqref{b.127} is proved.

                In summary, we have proved \eqref{convergencesigu}, \eqref{convergencepsip}, and \eqref{convergenceomegash} and in turn, the term of \eqref{a100} converges to a standard Gaussian distribution. The proof is complete.
\end{proof}

\newpage
\subsection{Lemma A.1}
\label{prooflemmairf}
\begin{proof}
    
    The proof will be conducted in three cases, $\delta\in \{0,1,2\}$. Define $J_1=(I_K,0_{K\times (p-1)K})$.
    
    \textbf{Case 1: } $\delta=0$, $\Phi(L)=B_0(L)\Pi$. 
    
    Rewrite $\Phi(L)y_t = u_t$ as $B_0(L)\Pi y_t = u_t$. It yields $\Pi B_0(L)\Pi y_t =\Pi u_t$, and the impulse response of $\Pi y_t$ to $\Pi u_t$ is gauged by the polynomial $\Pi B_0(L)$, which has explicit expression of $J_1\mathbf B_0^h J_1'$ where $\mathbf B_0$ is the companion matrix defined in Definition \ref{defps} (recall $B_{\delta}(L)=\sum_{i=0}^{p-\delta} B_iL^i$ and $B_0=\Pi^{-1}$). Therefore, we obtain $\Psi_h=\Pi^{-1} J_1\mathbf B_0^h J_1' \Pi$. The norm of impulse response matrix is bounded as
    \begin{align}
        \| \Psi_{h}\|=\| \Pi^{-1}J_1\mathbf B_0^h J_1'  \Pi \| \leq \|\Pi^{-1} \| \|\Pi\| \|J_1\|^2  \|\mathbf B_0^h \| \leq \|\Pi^{-1}\|\|\Pi\| c_1(1-\epsilon)^h,
    \end{align}
    where the last inequality is due to the definition of parameter space, $\| \mathbf B_0^h \| \leq c_1(1-\epsilon)^h$ for some constant $c_1$ and $\epsilon \in (0,1)$, and $\Pi$ is a non-singular constant matrix.  Thus, apply the formula for the sum of geometric terms and it yields 
    \begin{align}
    \label{pi0}
        &\sum_{i=0}^{\infty}\|\Psi_{i}\| 
        \leq \|\Pi^{-1}\| c_1  \frac{1}{\epsilon}\|\Pi\|=O_p(1)\\
        & \sum_{i=0}^{\infty}\|\Psi_{i}\|^2 
        \leq \|\Pi^{-1}\| c_1  \frac{1}{1-(1-\epsilon)^2}\|\Pi\|=O_p(1)
    \end{align}

    \textbf{Case 2: } $\delta=1$, $\Phi(L)=B_1(L)U_1(L) \Pi$, where $U_1(L)=I -P_1 L$. 
    
    Rewrite $\Phi(L)y_t = u_t$ as $B_1(L)U_1(L) \Pi y_t = u_t$. It yields  $\Pi B_1(L)U_1(L) \Pi y_t =\Pi u_t$. First, follow Case 1 procedure, we obtain the impulse response of $ U_1(L)\Pi y_t$ to $\Pi u_t$ as $J_1\mathbf B_1^h J_1'$. Then, we derive the impulse response of $\Pi y_t$ to $\Pi u_t$ as $\sum_{i=0}^{h}P_1^i J_1\mathbf B_1^i J_1'$ due to the form of $U_1(L)=I -P_1 L$. Lastly, we obtain $\Psi_h=\Pi^{-1} ( \sum_{i=0}^{h}P_1^i J_1\mathbf B_1^i J_1') \Pi$. The norm of impulse response matrix is bounded as
    \begin{align}
    \begin{split}
        \|\Psi_{h}\| \leq &
        \|\Pi^{-1}\|
        \sum_{i=0}^{h}
        \|P_1^i\| \| J_1\mathbf B_1^{h-i} J_1'\| \|\Pi\|\\
        \leq & 
        \|\Pi^{-1}\|
        \sum_{i=0}^{h} c_1 (1-\epsilon)^{h-i} \|\Pi\| \\
        & \text{ (since $P_1=\text{diag}[p_{k,1}]_{k=1,\cdots,K}$ and $|p_{k,1}|\leq 1$)}\\
        \leq & 
        \|\Pi^{-1}\|c_1 
        \sum_{i=0}^{\infty} (1-\epsilon)^{i} \|\Pi\|\\
        \leq & 
        \|\Pi^{-1}\|c_1  \frac{1}{\epsilon}\|\Pi\|
        \end{split}
    \end{align}
    Thus, apply the formula for the sum of geometric terms and it yields 
    \begin{align}
        & \sum_{i=0}^{h-1}\|\Psi_{i}\|
        \leq h (\|\Pi^{-1}\| c_1  \frac{1}{\epsilon}) =O_p(h)\\
        &\sum_{i=0}^{h-1}\|\Psi_{i}\|^2 
        \leq h (\|\Pi^{-1}\| c_1  \frac{1}{\epsilon})^2=O_p(h)
    \end{align}

    \textbf{Case 3: } $\delta=2$, $\Phi(L)=B_2(L)U_1(L)U_2(L) \Pi$, where $U_1(L)=I -\Pi_1 L$  and $U_2(L)=I -\Pi_2 L$.

    Following Case 2 procedure, we obtain the impulse response of $ U_1(L)U_2(L)\Pi y_t$ to $\Pi u_t$ as $J_1\mathbf B_2^h J_1'$. Then, we derive the impulse response of $U_2(L)\Pi y_t$ to $\Pi u_t$ as $\sum_{i=0}^{h}P_1^i J_1\mathbf B_1^i J_1'$ due to the form of $U_1(L)=I -P_1 L$; and we derive the impulse response of $\Pi y_t$ to $\Pi u_t$ as $\sum_{i=0}^{h}P_2^i\sum_{j=0}^{h-i}P_1^i J_1\mathbf B_1^{h-i-j} J_1'$ due to the form of $U_2(L)=I -P_2 L$.   Lastly, we obtain the impulse response as the following form,
   \begin{align}
   \label{b104}
       \begin{split}
          \Psi_h = \Pi^{-1} \left\{\sum_{i=0}^{h}P_2^i\left( \sum_{j=0}^{h-i}P_1^i J_1\mathbf B_1^{h-i-j} J_1'\right) \right\}\Pi.
       \end{split}
   \end{align}
    The norm of impulse response matrix is bounded as
    \begin{align}
    \label{b103}
    \begin{split}
        \|\Psi_{h}\| 
        \leq & 
        \|\Pi^{-1}\|
        \sum_{i=0}^{h}
        \|P_2^i\| \sum_{j=0}^{h-i} \|P_1^j\|
        \| J_1\mathbf B_2^{h-i-j} J_1' \| \|\Pi\|\\
        \leq & 
        \|\Pi^{-1}\|
        \sum_{i=0}^{h}
         \sum_{j=0}^{h-i}  \| J_1\mathbf B_2^{h-i-j} J_1' \| \|\Pi\|\\
        \leq & 
        \|\Pi^{-1}\|
        \sum_{i=0}^{h}
         \sum_{j=0}^{\infty}  \| J_1\mathbf B_2^{j} J_1'\| \|\Pi\|\\
         = & 
        \|\Pi^{-1}\|
        (h+1) c_1 \frac{1}{\epsilon}\|\Pi\|\\
        \end{split}
    \end{align}
    Thus, apply the formula for the sum of geometric terms and it yields 
    \begin{align}
    & \sum_{i=0}^{h-1}\|\Psi_{i}\|
        \leq   (\|\Pi^{-1}\|
         c_1 \frac{1}{\epsilon})\sum_{i=0}^{h-1} (i+1)\|\Pi\|
         =\text{constant}* h(h+1)/2 =O_p(h^2) \\
     &   \sum_{i=0}^{h-1}\|\Psi_{i}\|^2
        \leq   (\|\Pi^{-1}\|
         c_1 \frac{1}{\epsilon})^2\sum_{i=0}^{h-1} (i+1)^2 \|\Pi\|^2
         =\text{constant}* h(h+1)(2h+1)/6 =O_p(h^3)
    \end{align}
Thus, the proof is complete.
\end{proof}

\newpage
\subsection{Lemma A.2}
\label{prooflemmaa4}

\begin{proof}
        Since $\Omega_{s,h}$ is the covariance matrix for $s_{t,h}^*$, $w'\Omega_{s,h} w=\mathbb E[w's_{t,h}^* s_{t,h}^{*'}w]$. We therefore partition $w$ into $p$ sub-vectors of dimension $K$, $w=(w_1',w_2',\cdots,w_K')'$ and rewrite $w's_{t,h}^*$ as
    \begin{align}
    \label{b140}
    \begin{split}
        w' s_{t,h}^* =& w'(  (e_{t,h},e_{t+1,h},\cdots,e_{t+p-1,h})'\otimes u_t) \\
                      =& \sum_{i=1}^p w_i' u_t e_{t+i-1,h} \\
                      =& \sum_{i=1}^p w_i' u_t (\sum_{j=1}^h \Psi_{1\bullet,j-1}u_{t+h+i-j})\\
                      &(\text{by definition, $e_{t+i-1,h} = \sum_{j=1}^h \Psi_{1\bullet,j-1}u_{t+h+i-j}$ })\\
                      =&u_t' \sum_{i=1}^p  w_i \sum_{j=1}^h \Psi_{1\bullet,j-1}u_{t+h+i-j}\\
            =& u_t' \sum_{j=1}^{h+p-1} \left(\sum_{i=\max(1,j-h+1)}^{\min(p,j)} w_i \Psi_{1\bullet,h-j+i-1}\right) u_{t+j}\\
            &(\text{re-organize the summation}) \\
             =&\sum_{j=1}^{h+p-1} \xi_{j,t}
        \end{split}
    \end{align}
    where we define
    \begin{align}
        \xi_{j,t} := u_t'  \left(\sum_{i=\max(1,j-h+1)}^{\min(p,j)} w_i \Psi_{1\bullet,h-j+i-1}\right) u_{t+j}.
    \end{align}
    
    Because of the mean-independence assumption, it is readily to check that $\xi_{j,t}$ is serially uncorrelated, $\mathbb E[\xi_{i,t} \xi_{j,t}]=0$, $\forall i\neq j$. Thus,
    \begin{align}
    \label{b149}
        \begin{split}
        &w'\Omega_{s,h} w\\
        =&\mathbb E[ (w' s_{t,h}^*)^2] \\
            =&\mathbb E[ (\sum_{j=1}^{h+p-1} \xi_{j,t})^2] \\
            =&\sum_{j=1}^{h+p-1}\mathbb E[\xi_{j,t}^2].
        \end{split}
    \end{align}
    Now, we argue that the variance of $\xi_{j,t}$ is bounded,
    \begin{align}
    \label{b150}
        \underline c_{\xi} \psi_{w,j}\leq E[\xi_{j,t}^2] \leq \overline  c_{\xi} \psi_{w,j},
    \end{align}
    where $\psi_{w,j}:=\|
             \sum_{i=\max(1,j-h+1)}^{\min(p,j)}  w_i \Psi_{1\bullet,h-j+i-1} 
             \|^2$,
    for some constant $0< \underline c_{\xi}\leq \overline  c_{\xi}< \infty$. 

    First, the upper bound of $E[\xi_{j,t}^2]$ is shown as follows
    \begin{align}
    \label{upbound}
    \begin{split}
        &\mathbb E[ \| u_t'(\sum_{i=\max(1,j-h+1)}^{\min(p,j)} w_i \Psi_{1\bullet,h-j+i-1}) u_{t+j}\|^2 ]\\
           \leq &
           \mathbb E[ \| u_t\|^2\| \sum_{i=\max(1,j-h+1)}^{\min(p,j)} w_i' u_t \Psi_{1\bullet,h-j+i-1}\|^2 
           \| u_{t+j}\|^2]\\
            \leq & 
            \mathbb E[\| u_{t}\|^4] 
            \|
             \sum_{i=\max(1,j-h+1)}^{\min(p,j)}  w_i \Psi_{1\bullet,h-j+i-1} 
             \|^2\\
             =& \overline c_\xi
           \psi_{w,j}
    \end{split}
    \end{align}
    where we define $\overline c_\xi=\mathbb E[\| u_{t}\|^4]   $ and
    \begin{align}
             & \psi_{w,j} =\|
             \sum_{i=\max(1,j-h+1)}^{\min(p,j)}  w_i \Psi_{1\bullet,h-j+i-1} \|^2 .
    \end{align}
   
    The lower bound of $E[\xi_{j,t}^2]$ is verified as follows:
    \begin{align}
    \begin{split}
        &\mathbb E[ (\sum_{i=\max(1,j-h+1)}^{\min(p,j)} w_i' u_t \Psi_{1\bullet,h-j+i-1}) u_{t+j}
            u_{t+j}' (\sum_{i=\max(1,j-h+1)}^{\min(p,j)}\Psi_{1\bullet,h-j+i-1}' w_i' u_t )]\\
            =&
            \mathbb E[ \mathbb E[ (\sum_{i=\max(1,j-h+1)}^{\min(p,j)} w_i' u_t \Psi_{1\bullet,h-j+i-1}) u_{t+j}
            u_{t+j}' (\sum_{i=\max(1,j-h+1)}^{\min(p,j)}\Psi_{1\bullet,h-j+i-1}' w_i' u_t ) \mid \mathcal{F}_{t+j-1} ] ]  \\
            &(\text{By Law of Iterated Expectations})\\
            =&
            \mathbb E[ (\sum_{i=\max(1,j-h+1)}^{\min(p,j)} w_i' u_t \Psi_{1\bullet,h-j+i-1})
            \mathbb E[  u_{t+j}
            u_{t+j}'\mid \mathcal{F}_{t+j-1} ]  (\sum_{i=\max(1,j-h+1)}^{\min(p,j)}\Psi_{1\bullet,h-j+i-1}' w_i' u_t ) ]\\
            \geq& 
            c
            \mathbb E[ \|\sum_{i=\max(1,j-h+1)}^{\min(p,j)} w_i' u_t \Psi_{1\bullet,h-j+i-1}\|^2 ]\\
             &(\text{By Assumption \ref{assmoment}, } \lambda_{\min}(\mathbb E[  u_{t+j}
            u_{t+j}'\mid \mathcal{F}_{t+j-1} ])>c>0,a.s. \text{ and quadratic inequality}) \\
            =& 
            c
            \mathbb E[ \| u_t'
            \sum_{i=\max(1,j-h+1)}^{\min(p,j)} w_i \Psi_{1\bullet,h-j+i-1} \|^2 ]\\
            =& 
            c
            \mathbb E[ \text{trace}\left(u_t'
            (\sum_{i=\max(1,j-h+1)}^{\min(p,j)}  w_i \Psi_{1\bullet,h-j+i-1} )
            (\sum_{i=\max(1,j-h+1)}^{\min(p,j)} \Psi_{1\bullet,h-j+i-1}' w_i') u_t \right) ]\\
            =& 
           c*
            \text{trace}\left(\mathbb E[ u_t u_t']
            (\sum_{i=\max(1,j-h+1)}^{\min(p,j)}  w_i \Psi_{1\bullet,h-j+i-1} )
            (\sum_{i=\max(1,j-h+1)}^{\min(p,j)} \Psi_{1\bullet,h-j+i-1}' w_i')  \right) ]\\ 
              &(\text{since $\mathbb  E[\text{trace}(A)]=\text{trace}(\mathbb 
 E[A])$})\\
 =& 
           c*
            \text{trace}\left((\sum_{i=\max(1,j-h+1)}^{\min(p,j)} \Psi_{1\bullet,h-j+i-1}' w_i') \Sigma_u
            (\sum_{i=\max(1,j-h+1)}^{\min(p,j)}  w_i \Psi_{1\bullet,h-j+i-1} )
             \right) ]\\
             &(\text{since  the trace is invariant under cyclic permutation})\\ 
             \geq & 
           c  
             \lambda_{\min}( \Sigma_u)
            \text{trace}\left(
             (\sum_{i=\max(1,j-h+1)}^{\min(p,j)}  w_i \Psi_{1\bullet,h-j+i-1} )
            (\sum_{i=\max(1,j-h+1)}^{\min(p,j)} \Psi_{1\bullet,h-j+i-1}' w_i')\right)\\
            &(\text{By quadratic inequality})\\
            =&
             c \lambda_{\min}(\Sigma_u)
           \|
             \sum_{i=\max(1,j-h+1)}^{\min(p,j)}  w_i \Psi_{1\bullet,h-j+i-1} \|^2\\
              =&
             \underline c_\xi
           \psi_{w,j}
    \end{split}
    \end{align}
    where we define $
             \underline c_\xi =  c \lambda_{\min}(\Sigma_u)$, and $\psi_{w,j}$ is defined in \eqref{upbound}. 
            
    By Assumption \ref{assmoment}(i), $\underline c_\xi$ is some positive constant. Moreover, to show $\underline c_{\xi} \sum_{j=1}^{h+p-1}\psi_{w,j}>0$, it is sufficient to show $\sum_{j=1}^{h+p-1}\psi_{w,j}>0$. We verify this result by explicitly writing out the form of $\sum_{j=1}^{h+p-1}\psi_{w,j}$:
    \begin{align}
        \begin{split}
            &\sum_{j=1}^{h+p-1}\psi_{w,j}\\
            \geq& \sum_{j=h}^{h+p-1}\psi_{w,j}\\
            =& \underbrace{\|w_p \Psi_{1\bullet,0}\|^2}_{=\psi_{w,h+p-1}} 
            + \underbrace{\|
            w_{p-1} \Psi_{1\bullet,0}+w_{p} \Psi_{1\bullet,1}\|^2\mathbb 1_{(p>1)}}_{=\psi_{w,h+p-2}}\\
            & + \underbrace{\|
            w_{p-2} \Psi_{1\bullet,0}+w_{p-1} \Psi_{1\bullet,1}+w_{p-3} \Psi_{1\bullet,3}\|^2\mathbb 1_{(p>2)}}_{=\psi_{w,h+p-3}} + \cdots  + \underbrace{\| \sum_{i=0}^{p-1}
            w_{p-i} \Psi_{1\bullet,i}\|^2}_{=\psi_{w,h}}
        \end{split}
    \end{align}
     Then, we check the last expression term by term. Given any constant vector $w\in \mathbb R^{Kp}\backslash$, $\|w\|=1$, and its partition $w=(w_1',w_2',\cdots,w_p')'$, suppose the last nonzero sub-vector is denoted as $w_q$, such that $\|w_q\|^2 \neq 0$ and $\|w_{j}\|^2=0$ for all $j>q$. Thus, we could state that $\psi_{w,h+j-1}=0$ for all $j=q+1,\cdots,p$ and $\psi_{w,h+q-1}=\|w_q \Psi_{1\bullet,0}\|^2 = \|w_q\|^2>0$ because $\Psi_0=I_K$ by the definition of impulse response function. Therefore, we have shown the last expression above is a positive constant and it induces $
         \sum_{j=1}^{h+p-1}\psi_{w,j} >0, \text{ given }w\in \mathbb R^{Kp}$, $\|w\|=1$, and thereby the variance matrix $\Omega_{s,h}$ is positive definite,
     \begin{align}
           w'\Omega_{s,h} w \geq c_{\xi} \sum_{j=1}^{h+p-1} \psi_{w,h} >0.
     \end{align}

Lastly, we need to check $\overline  c_{\xi}\geq \underline c_{\xi} $,
\begin{align}
\begin{split}
    \overline  c_{\xi}&=\mathbb E[\| u_{t}\|^4]\\
    &\geq \mathbb E[ \| u_{t+j}\|^2 \| u_{t}\|^2] = \mathbb E[ \mathbb E[\| u_{t+j}\|^2 \| u_{t}\|^2|\mathcal{F}_{t+j-1}]]=\mathbb E[  \| u_{t}\|^2 \mathbb E[\| u_{t+j}\|^2 |\mathcal{F}_{t+j-1}]]\\
    &\geq c\mathbb E[  \| u_{t}\|^2]\\
    &\geq c \lambda_{\min}(\Sigma_u)=\underline c_{\xi}
\end{split}
\end{align}
   In summary,  we obtain
   \begin{align}
       0<\underline c_{\xi} \sum_{j=1}^{h+p-1}\psi_{w,j}
  \leq 
  w'\Omega_{s,h} w 
  \leq 
  \overline  c_{\xi} \sum_{j=1}^{h+p-1}\psi_{w,j}.
   \end{align}
\end{proof}

\newpage

\subsection{Lemma A.3}
\label{prooflemmavar}
\begin{proof}
    As shown in \eqref{b140}, $
        w's_{t,h}^* = \sum_{j=1}^{h+p-1} \xi_{j,t}$. Thus, we could write
    \begin{align}
        \begin{split}
            &\text{Var} \left( (\bar T-p+1)^{-1} \sum_{t=p}^{\bar T} (w's_{t,h}^*)^2/ w'\Omega_{s,h} w \right)\\
            =&(\bar T-p+1)^{-2} (w'\Omega_{s,h} w)^{-2}
            \text{Var} \left(\sum_{t=p}^{\bar T} (\sum_{j=1}^{h+p-1} \xi_{j,t})^2 \right)\\
            =&(\bar T-p+1)^{-2} (w'\Omega_{s,h} w)^{-2} 
            \sum_{|m| \leq  \bar T-p}(\bar T-p+1-|m|) \text{Cov}(
            (\sum_{j=1}^{h+p-1} \xi_{j,0})^2,
            (\sum_{j=1}^{h+p-1} \xi_{j,m})^2 )\\
            = &(\bar T -p+1)^{-1} (w'\Omega_{s,h} w)^{-2} 
            \sum_{|m| \leq  \bar T-p}(1-|m|/\bar T)  |\Gamma_m|\\
            \leq &2 \bar T^{-1} (w'\Omega_{s,h} w)^{-2} 
             \sum_{m=0}^{\bar T-p} |\Gamma_m|,
        \end{split}
    \end{align}
    where we define
    \begin{align}
        \Gamma_m :=\text{Cov}(
            (\sum_{j=1}^{h+p-1} \xi_{j,0})^2,
            (\sum_{j=1}^{h+p-1} \xi_{j,m})^2 ). 
    \end{align}
    By expanding the summation of $(\sum_{j=1}^{h+p-1}\xi_{j,0})^2,(\sum_{j=1}^{h+p-1}\xi_{j,m})^2$, we obtain
    \begin{align}
        \Gamma_m = \sum_{i=1}^{h+p-1}\sum_{j=1}^{h+p-1}\sum_{k=1}^{h+p-1}\sum_{l=1}^{h+p-1}\text{Cov}(
            \xi_{i,0}\xi_{j,0},\quad 
            \xi_{k,m}\xi_{l,m}). 
    \end{align}
    
    Given the definition of $\xi_{j,t}$, $
        \xi_{j,t} := u_t'  (\sum_{i=\max(1,j-h+1)}^{\min(p,j)} w_i \Psi_{1\bullet,h-j+i-1}) u_{t+j}$, and mean-independence assumption (Assumption \ref{assimeanind}), if $i=j$, then the covariance equals to zero for all $k\neq l$. It is because $k+m\neq l+m$ and only at most one value can be equal to $i$ (or $j$). If $i\neq j$, then the covariance has non-zero value unless $\{i,j\}=\{m+k,m+l\}$. Thus,
        \begin{align}
            \Gamma_m = 
            \sum_{j=1}^{h+p-1}\sum_{l=1}^{h+p-1}
            \text{Cov}(
            \xi_{j,0}^2, 
            \xi_{l,m}^2) 
            +
            \sum_{j=1}^{h+p-1}\sum_{i\neq j}
            \text{Cov}(
            \xi_{i,0}\xi_{j,0}, \quad
            \xi_{i-m,m}\xi_{j-m,m})
        \end{align}
        and
        \begin{align}
        \begin{split}
            &\sum_{m=0}^{\bar T-p} |\Gamma_m|\\
         \leq  
            &\sum_{m=0}^{\bar T-p} \sum_{j=1}^{h+p-1}\sum_{l=1}^{h+p-1}
           | \text{Cov}(
            \xi_{j,0}^2, 
            \xi_{l,m}^2) | 
            +
            \sum_{m=0}^{\bar T-p}
            \sum_{j=1}^{h+p-1}\sum_{i\neq j}
           | \text{Cov}(
            \xi_{i,0}\xi_{j,0}, \quad
            \xi_{i-m,m}\xi_{j-m,m})|
        \end{split}
        \end{align}
Notice that for the second term, it is implicitly restricted that $m<\min(i,j)$, since both $i-m$ and $j-m$ shall be strict positive. 

We will now show the boundedness of above two terms.

\textbf{(1) Boundedness of }$ \sum_{m=0}^{\bar T-p}\sum_{j=1}^{h+p-1}\sum_{l=1}^{h+p-1}
            \text{Cov}(
            \xi_{j,0}^2, 
            \xi_{l,m}^2) $: Since $\xi_{j,t}=u_t'  (\sum_{i=\max(1,j-h+1)}^{\min(p,j)} w_i \Psi_{1\bullet,h-j+i-1}) u_{t+j})$, we rewrite $\xi_{j,t}^2$ as
            \begin{align}
            \begin{split}
                \xi_{j,t}^2
                =
                \psi_{w,j} (u_t' M_j u_{t+j})^2,
            \end{split}
            \end{align}
            where $M_j$ is a $K\times K$ constant matrix, $M_j = (\sum_{i=\max(1,j-h+1)}^{\min(p,j)} w_i \Psi_{1\bullet,h-j+i-1}) / \psi_{w,j}^{1/2}$, and $\|M_j\|=1$.
            It yields the first term is bounded above by
            \begin{align}
            \label{b108}
                \begin{split}
                    &         \sum_{m=0}^{\bar T-p} \sum_{j=1}^{h+p-1}\sum_{l=1}^{h+p-1}
           | \text{Cov}(
            \xi_{j,0}^2, 
            \xi_{l,m}^2) |\\
            \leq & 
            \sum_{j=1}^{h+p-1}\sum_{l=1}^{h+p-1} \psi_{w,j}\psi_{w,l}
            \underset{\|M_j\|,\|M_l\|=1}{\text{sup}}
            \left(\sum_{m=0}^{\bar T-p} 
            | \text{Cov}(
            (u_0' M_j u_{j})^2, 
            (u_m' M_l u_{m+l})^2 )|  \right)
                \end{split}
            \end{align}
            Let's expand two terms, $(u_0' M_j u_{j})^2$ and $(u_m' M_l u_{m+l})^2 $. We obtain
            \begin{align}
                \begin{split}
                    &\underset{\|M_j\|,\|M_l\|=1}{\text{sup}}
                    \sum_{m=0}^{\bar T-p}
            | \text{Cov}(
            (u_0' M_j u_{j})^2, 
            (u_m' M_l u_{m+l})^2 )|\\
            \leq &
            (\sum_{l_1=1}^K\sum_{l_2=1}^K |M_{l_1 l_2,j}|)^2
            (\sum_{l_1=1}^K\sum_{l_2=1}^K |M_{l_1 l_2,l}|)^2
            \underset{j_1,j_2,j_3,j_4}{\text{sup}}
             \sum_{m=0}^{\bar T-p}
            | \text{Cov}(
            \tilde u_{j_1,0}\tilde u_{j_2,j}, 
            \tilde u_{j_3,m}\tilde u_{j_4,m+l} )| \\
            &(\text{$M_{l_1l_2,j}$ denotes the $l_1l_2$-th element in $M_j$})\\
            \leq &
            K^4
            \underset{j_1,j_2,j_3,j_4}{\text{sup}}
             \sum_{m=0}^{\bar T-p}
            | \text{Cov}(
            \tilde u_{j_1,0}\tilde u_{j_2,j}, 
            \tilde u_{j_3,m}\tilde u_{j_4,m+l} )| \\
            \leq & \text{constant}
                \end{split}
            \end{align}
            where $\tilde u_t =u_t\otimes u_t$, $j_1,j_2,j_3,j_4 =1,2,\cdots, K^2$. The second last inequality is due to $(\sum_{l_1=1}^K\sum_{l_2=1}^K|M_{l_1 l_2,j}|)\leq K \|M_j\|^2=K$ by Cauchy–Schwarz inequality. The last inequality is because the dimension $K$ is finite and $\underset{j_1,j_2,j_3,j_4}{\text{sup}}
             \sum_{m=0}^{\bar T-p}
            | \text{Cov}(
            \tilde u_{j_1,0}\tilde u_{j_2,j}, 
            \tilde u_{j_3,m}\tilde u_{j_4,m+l} )| $ is bounded by a constant. The boundedness of $\underset{j_1,j_2,j_3,j_4}{\text{sup}}
             \sum_{m=0}^{\bar T-p}
            | \text{Cov}(
            \tilde u_{j_1,0}\tilde u_{j_2,j}, 
            \tilde u_{j_3,m}\tilde u_{j_4,m+l} )| $ is implied by the condition of finite fourth order of cumulant of series $u_t\otimes u_t$ given in Assumption \ref{assmoment}(ii), which is argued in the Footnote 3 in the Supplement of \cite{montiel2021local}. Consequently, the term in \eqref{b108} is bounded from above  by 
            \begin{align}
                \text{constant}* (\sum_{j=1}^{h-p+1}\psi_{w,j})^2,
            \end{align}
            since $\sum_{j=1}^{h+p-1}\sum_{l=1}^{h+p-1} \psi_{w,j}\psi_{w,l} = (\sum_{j=1}^{h-p+1}\psi_{w,j})^2$ .
            
             Given $(w'\Omega_{s,h} w)^2\geq \underline c_\xi (\sum_{j=1}^{h-p+1}\psi_{w,j})^2$ by Lemma \ref{lemmavarscore}, it yields
            \begin{align}
            \label{b160}
                \begin{split}
                    2 (\bar T -p+1)^{-1} (w'\Omega_{s,h} w)^{-2} 
             \sum_{m=0}^{\bar T-p} \sum_{j=1}^{h+p-1}\sum_{l=1}^{h+p-1}
           | \text{Cov}(
            \xi_{j,0}^2, 
            \xi_{l,m}^2) | \xrightarrow{p} 0
                \end{split}
            \end{align}
            as long as $(\bar T -p+1) \rightarrow \infty$.

            \textbf{(2) Boundedness of } $\sum_{m=0}^{\bar T-p}
            \sum_{j=1}^{h+p-1}\sum_{i\neq j}
           | \text{Cov}(
            \xi_{i,0}\xi_{j,0}, \quad
            \xi_{i-m,m}\xi_{j-m,m})|$.

            Since $\mathbb E[ \xi_{i,0}\xi_{j,0}] =0$ for all $i\neq j$ under mean-independence assumption, it yields
            \begin{align}
            \label{b126}
                \begin{split}
                    &
                    \sum_{m=0}^{\bar T-p}
            \sum_{j=1}^{h+p-1}\sum_{i\neq j}
           | \text{Cov}(
            \xi_{i,0}\xi_{j,0}, \quad
            \xi_{i-m,m}\xi_{j-m,m})| \\
            = & \sum_{m=0}^{\bar T-p}
            \sum_{j=1}^{h+p-1}\sum_{i\neq j}
           | \mathbb E[ \xi_{i,0}\xi_{j,0}
            \xi_{i-m,m}\xi_{j-m,m})]\\
            \leq 
            & 
            \sum_{m=0}^{\bar T-p}
            \sum_{j=1}^{h+p-1}\sum_{i\neq j}
            \psi_{w,i}^{1/2}\psi_{w,j}^{1/2}\psi_{w,i-m}^{1/2}\psi_{w,j-m}^{1/2}\mathbb E[ \|u_0\|^2 \|u_m\|^2 \|u_i\|^2\|u_j\|^2]\\
            & (\text{ we use $\xi_{j,t}=u_t'  (\sum_{i=\max(1,j-h+1)}^{\min(p,j)} w_i \Psi_{1\bullet,h-j+i-1}) u_{t+j})$, $\psi_{w,j}^{1/2}=\|\sum_{i=\max(1,j-h+1)}^{\min(p,j)} w_i \Psi_{1\bullet,h-j+i-1}\|$})\\
            \leq 
            &
            \mathbb  E[\|u_0\|^8] \sum_{m=0}^{\bar T-p}
            \sum_{j=1}^{h+p-1}\sum_{i\neq j}
            \psi_{w,i}^{1/2}\psi_{w,j}^{1/2}\psi_{w,i-m}^{1/2}\psi_{w,j-m}^{1/2}\\
            =
            &
            \mathbb  E[\|u_0\|^8] 
            \sum_{m=0}^{h+p-1}\sum_{j=1}^{h+p-1}\sum_{i\neq j}
            \psi_{w,i}^{1/2}\psi_{w,j}^{1/2}\psi_{w,i-m}^{1/2}\psi_{w,j-m}^{1/2}\\
            &(\text{ since $i-m,j-m$ shall be positive integers})
                \end{split}
            \end{align}
            By Assumption \ref{assmoment} that the eighth moment of $u_t$ is bounded, the boundedness of the last term in \eqref{b126} will be checked under three scenarios $\delta\in \{0,1,2\}$ with the respective condition imposed on the projection horizon $h$.

            For the cases of $\delta=1,2$, the boundedness of the last term in \eqref{b126} can be obtained by Cauchy–Schwarz inequality:
            \begin{align}
            \label{b127}
                \begin{split}
                    &\sum_{m=0}^{h+p-1}\sum_{j=1}^{h+p-1}\sum_{i\neq j}
            \psi_{w,i}^{1/2}\psi_{w,j}^{1/2}\psi_{w,i-m}^{1/2}\psi_{w,j-m}^{1/2}
            \\
            \leq 
            &
            (h+p) (\sum_{j=1}^{h+p-1}\sum_{i=1}^{h+p-1}
            \psi_{w,i}\psi_{w,j})^{1/2}
            (\sum_{j=1}^{h+p-1}\sum_{i=1}^{h+p-1}
            \psi_{w,i}\psi_{w,j})  ^{1/2}\\
            =
            &
            (h+p) (\sum_{j=1}^{h+p-1}
            \psi_{w,j})^2.
                \end{split}
            \end{align}
            When $h/T\xrightarrow{p} 0 $ holds, $
                \frac{h}{\bar T -p+1} =  
                \frac{h}{T}
                \frac{T}{T-h}
                \frac{T-h}{\bar T-p+1} 
                \xrightarrow{p}0$, 
            because $\frac{T}{T-h}=\frac{1}{1-h/T}\rightarrow{1}$ and $\frac{T-h}{\bar T-p+1} \rightarrow{1}$.

            Combining the result of \eqref{denomenat} that $(w'\Omega_{s,h} w)^2\geq \underline c_\xi (\sum_{j=1}^{h-p+1}\psi_{w,j})^2$, the term of $\sum_{m=0}^{h+p-1}\sum_{j=1}^{h+p-1}\sum_{i\neq j}
            \psi_{w,i}^{1/2}\psi_{w,j}^{1/2}\psi_{w,i-m}^{1/2}\psi_{w,j-m}^{1/2}$ is bounded:
\begin{align}
\label{b163}
    (\bar T -p+1)^{-1} (w'\Omega_{s,h} w)^{-2} \sum_{m=0}^{h+p-1}
    \sum_{j=1}^{h+p-1}\sum_{i\neq j}
            \psi_{w,i}^{1/2}\psi_{w,j}^{1/2}\psi_{w,i-m}^{1/2}\psi_{w,j-m}^{1/2} \xrightarrow{p}0. 
\end{align}

Notice that the above boundedness holds under the condition $h/T\xrightarrow{p} 0 $, which is satisfied for the case $\delta=1$ or $\delta=2$. Thus, we have shown that the boundedness for \eqref{b126} when $\delta=1,2$.

On the other hand, for the stationary case that $\delta=0$, the projection horizon is allowed to be a non-trivial fraction of the sample size $T$, the boundedness for \eqref{b126} can not be shown by the above method. The proof will be conducted by showing the implicit relationship between $\psi_{w,j}$ and $\psi_{w,j-m}$. Recall the form of $\psi_{w,j}$,
\begin{align}
    \psi_{w,j}^{1/2} =\|\sum_{i=\max(1,j-h+1)}^{\min(p,j)} w_i \Psi_{1\bullet,h-j+i-1}\|.              
\end{align}
The number of items in the summation, $\sum_{i=\max(1,j-h+1)}^{\min(p,j)} w_i \Psi_{1\bullet,h-j+i-1}$, depends on the value of $j$:
\begin{enumerate}
    \item $1\leq j \leq p-1$, there are $j$ terms in the summation.
    \item $p\leq j \leq h$, there are $p$ terms in the summation.
    \item $h+1 \leq j \leq  h+p-1$, there are $(h+p-j)$ terms in the summation.
\end{enumerate}
Thus, we use indicator functions to partition $\sum_{m=0}^{h+p-1}\sum_{j=1}^{h+p-1}\sum_{i\neq j}
            \psi_{w,i}^{1/2}\psi_{w,j}^{1/2}\psi_{w,i-m}^{1/2}\psi_{w,j-m}^{1/2}$ into three parts,
\begin{align}
\begin{split}
     &\sum_{m=0}^{h+p-1}\sum_{j=1}^{h+p-1}\sum_{i\neq j}
            \psi_{w,i}^{1/2}\psi_{w,j}^{1/2}\psi_{w,i-m}^{1/2}\psi_{w,j-m}^{1/2}\\
=&2 \sum_{j=1}^{h+p-1}\sum_{i=1}^{j-1}\sum_{m=0}^{i-1}
            \psi_{w,i}^{1/2}\psi_{w,j}^{1/2}\psi_{w,i-m}^{1/2}\psi_{w,j-m}^{1/2}\\
            &(\text{ since $i,j$ are exchangeable and $j>i>m$})\\
            \leq &
2(\mathbb 1_{\{p\leq i,j,i-m,j-m\leq h\}}
\sum_{j=1}^{h+p-1}\sum_{i=1}^{j-1}\sum_{m=0}^{i-1}
            \psi_{w,i}^{1/2}\psi_{w,j}^{1/2}\psi_{w,i-m}^{1/2}\psi_{w,j-m}^{1/2}\\
             &+
            \mathbb 1_{\{j>h\}}
\sum_{j=1}^{h+p-1}\sum_{i=1}^{j-1}\sum_{m=0}^{i-1}
            \psi_{w,i}^{1/2}\psi_{w,j}^{1/2}\psi_{w,i-m}^{1/2}\psi_{w,j-m}^{1/2}\\
            &+
            \mathbb 1_{\{i-p<m<i\}}
\sum_{j=1}^{h+p-1}\sum_{i=1}^{j-1}\sum_{m=0}^{i-1}
            \psi_{w,i}^{1/2}\psi_{w,j}^{1/2}\psi_{w,i-m}^{1/2}\psi_{w,j-m}^{1/2} )
\end{split}
\end{align}
The rule of partition:
\begin{align}
\begin{cases}
     j>h \quad \quad \quad \quad \quad \quad \quad \quad \quad \quad \quad \quad \quad \quad \quad \quad \quad  ( \text{2nd indicator }\mathbb 1_{\{j>h\}})\\
     p\leq j \leq h \begin{cases}
         i\geq p \begin{cases}
             j-m\geq p \begin{cases}
                 i-m\geq p \quad \quad( \text{1st indicator }\mathbb 1_{\{p\leq i,j,i-m,j-m\leq h\}})\\
                 i-m<p\quad \quad( \text{3rd indicator }\mathbb 1_{\{i-p<m<i\}})
             \end{cases}\\
             j-m<p \Rightarrow i-m<p\quad ( \text{3rd indicator }\mathbb 1_{\{i-p<m<i\}})
         \end{cases}\\
         i<p \Rightarrow i-m<p \quad \quad \quad \quad \quad \quad ( \text{3rd indicator }\mathbb 1_{\{i-p<m<i\}})
     \end{cases}\\
     j<p \Rightarrow i<p \Rightarrow i-m<p\quad \quad \quad \quad \quad \quad \quad \quad ( \text{3rd indicator }\mathbb 1_{\{i-p<m<i\}})
    \end{cases} 
\end{align}
If we could show the boundedness of the above three terms with indicator functions for the case of $\delta=0$, then the boundedness of \eqref{b126} can be proved. 

\textbf{1. Boundedness for the term with $\mathbb 1_{\{p\leq i,j,i-m,j-m\leq h\}}$}  The proof relies on the explicit form of $\psi_{w,j-m}^{1/2}$ for $p\leq j, j-m\leq h$,
    \begin{align}
    \label{b124}
        \begin{split}
            &\psi_{w,j-m}^{1/2}=
            \|\sum_{i=1}^{p} w_i \Psi_{1\bullet,h-j+m+i-1}\|=
            \|\sum_{i=1}^{p} W_i \Psi_{h-j+m+i-1}\|,
            \end{split}
            \end{align}
            $\text{by defining } W_i =w_i [1,0_{1\times (K-1)}]$. Then, we use the result of Lemma E.6 from Supplement of \cite{montiel2021local}:
\begin{align}
    \Psi_{h+m} = \Psi_h \Psi_m + \sum_{l=2}^p (( \sum_{k=0}^{p-l}\Psi_{h-1-k} \Phi_{l+k}) J_{l-1}' \mathrm \Phi^{m-1} J_1')
\end{align}
for all $h,m>0$ and $\Psi_i=0$ when $i<0$, where $\mathrm \Phi$ is the $pK\times pK$ companion matrix containing the coefficients in $\Phi(L)$, and $J_l$ is the $K\times pK$ matrix obtained from the partition of a $pK$-dimension identity matrix, $I_{pK}=[J_1',J_2',\cdots,J_p']$. We reorder the double summation above and obtain
\begin{align}
\label{b1266}
    \Psi_{h+m} = \Psi_h \Psi_m +\sum_{l=2}^p
            \Psi_{h+1-l} (\sum_{k=l}^p  \Phi_k J_{k-1}\mathrm \Phi^{m-1} J_1').
\end{align}
Then, we combine the results of \eqref{b1266} and \eqref{b124},
\begin{align}
\label{b169}
\begin{split}           
\psi_{w,j-m}^{1/2}=& \|\sum_{i=1}^{p} W_i \Psi_{h-j+m+i-1}\|\\
 =&
            \|\sum_{i=1}^{p} W_i 
            \left(\Psi_{h-j+i-1} \Psi_m
            + \sum_{l=2}^p
            \Psi_{h-j+i-l} (\sum_{k=l}^p  \Phi_k J_{k-1}\mathrm \Phi^{m-1} J_1')
            \right)\|
            \\ 
            =
            &
           \|\sum_{i=1}^{p} W_i 
            \Psi_{h-j+i-1} \Psi_m
            + \sum_{l=2}^p \left( \sum_{i=1}^{p} W_i
            \Psi_{h-j+i-l} \sum_{k=l}^p  \Phi_k J_{k-1} \right)\mathrm \Phi^{m-1} J_1'
            \|
            \\
            \leq
            &
            \|\sum_{i=1}^{p}   W_i  \Psi_{h-j+i-1} \| \| \Psi_m \|
            + 
            \sum_{l=2}^p  \| \sum_{i=1}^p W_i \Psi_{h-j+i-l} \| 
            \| \sum_{k=l}^p \Phi_k J_{k-1}\| \| \mathrm \Phi^{m-1} J_1' \| 
            \\ 
            =
            &
            \psi_{w,j}^{1/2} \| \Psi_m \|+ 
            \sum_{l=2}^p  \psi_{w,j-1+l}^{1/2}
            \| \sum_{k=l}^p  \Phi_k J_{k-1}\| \| \mathrm \Phi^{m-1} J_1' \| 
            \\
            \leq &
            \psi_{w,j}^{1/2} \| \Psi_m \|+ 
            \sum_{l=2}^p  \psi_{w,j-1+l}^{1/2}
            \|\mathrm \Phi\| \| \mathrm \Phi^{m-1}  \| 
            \\
&(\| \sum_{k=l}^p  \Phi_k J_{k-1}\| \| = \| [\Phi_l,\Phi_{l+1},\cdots,\Phi_p ]\| = \| J_1 \mathrm \Phi[J_{l}',J_{l+1}',\cdots,J_{p}']\| \leq \|\mathrm  \Phi\|)
            \\
            = &
            \psi_{w,j}^{1/2} \| J_1 \mathrm \Phi^m J_1' \|+ 
            \sum_{l=2}^p  \psi_{w,j-1+l}^{1/2}
            \|\mathrm \Phi\| \| \mathrm \Phi^{m-1}  \| 
            \\
            &(\text{by definition, $\Psi_m=J_1 \mathrm \Phi^m J_1'$ })\\
            \leq  &
            \psi_{w,j}^{1/2} \|  \mathrm \Phi^m  \|+ 
            \sum_{l=2}^p  \psi_{w,j-1+l}^{1/2}
            \|\mathrm \Phi\| \| \mathrm \Phi^{m-1}  \| 
            \\
            &(\text{since $\| J_1 \mathrm \Phi^m J_1' \|\leq \| \mathrm \Phi^m \| $ }) \\
            \leq 
            &
            \text{constant}*(1-\epsilon)^m \sum_{l=1}^{p}   \psi_{w,j-1+l}^{1/2} 
        \end{split}
    \end{align}
    where the last inequality is due to $\|\mathrm \Phi^h \| \leq \text{constant}* (1-\epsilon)^h$. The boundedness,  $\|\mathrm \Phi^h \| \leq \text{constant}* (1-\epsilon)^h$, is because $\mathrm \Phi^h=(I_p\otimes \Pi^{-1}) \mathbf B_0^h (I_p\otimes \Pi)$ when the data process is stationary ($\delta=0$), such that $ \| \mathbf B_0^h \| \leq c_1 (1-\epsilon)^h,$ and constant rotation matrix $\Pi$ satisfies $\lambda_{\min}(\Pi)>c_0$.
It yields,
\begin{align}
    \begin{split}
        &
        \sum_{j=1}^{h+p-1}\sum_{i\neq j}
            \psi_{w,i}^{1/2}\psi_{w,j}^{1/2}\psi_{w,i-m}^{1/2}\psi_{w,j-m}^{1/2}\\
            \leq 
            &
          \text{constant}* (1-\epsilon)^{2m} \sum_{j=1}^{h+p-1}\sum_{i\neq j}
            \psi_{w,i}^{1/2}\psi_{w,j}^{1/2}
             (\sum_{k=1}^p \psi_{w,i-1+k}^{1/2})(\sum_{k=1}^p \psi_{w,j-1+k}^{1/2})\\
             &(\text{use the result of \eqref{b169}, $\psi_{w,j-m}^{1/2} \leq \text{constant}*(1-\epsilon)^m \sum_{l=1}^{p}   \psi_{w,j-1+l}^{1/2} $ }) \\
             \leq 
            &
           \text{constant}*  (1-\epsilon)^{2m} p^2(\sum_{j=1}^{h+p-1}\sum_{i\neq j}
            \psi_{w,i}\psi_{w,j})^{1/2}
            (\sum_{j=1}^{h+p-1}\sum_{i\neq j}
            \psi_{w,i}\psi_{w,j})^{1/2}\\
            &(\text{note that $(\sum_{k=1}^p \psi_{w,i-1+k}^{1/2})(\sum_{k=1}^p \psi_{w,j-1+k}^{1/2})$ contains $p^2$ number of terms, and obtain} \\
            &\quad \text{$\sum_{j=1}^{h+p-1}\sum_{i\neq j}
            \psi_{w,i}^{1/2}\psi_{w,j}^{1/2} \psi_{w,i-1+k_1}^{1/2}\psi_{w,j-1+k_2}^{1/2} \leq (\sum_{j=1}^{h+p-1}\sum_{i\neq j}
            \psi_{w,i}\psi_{w,j})^{1/2}
            (\sum_{j=1}^{h+p-1}\sum_{i\neq j}
            \psi_{w,i}\psi_{w,j})^{1/2} $} \\
            &\quad \text{for all $k_1,k_2=1,2,\cdots,p$ }) \\
             \leq 
            &
           \text{constant}*  (1-\epsilon)^{2m} p^2 (\sum_{j=1}^{h+p-1}
            \psi_{w,j})^2\\
            &(\text{since $\sum_{j=1}^{h+p-1}\sum_{i\neq j}
            \psi_{w,i}\psi_{w,j} \leq 
            (\sum_{j=1}^{h+p-1} \psi_{w,j})(\sum_{i=1}^{h+p-1} \psi_{w,i})= 
            (\sum_{j=1}^{h+p-1}
            \psi_{w,j})^2$})\\
    \end{split}
\end{align}
Thus, when $i,j,i-m,j-m \in [p,h]$,
\begin{align}
\label{b133}
\begin{split}
    &\mathbb 1_{\{p\leq i,j,i-m,j-m\leq h\}}
\sum_{j=1}^{h+p-1}\sum_{i=1}^{j-1}\sum_{m=0}^{i-1}
            \psi_{w,i}^{1/2}\psi_{w,j}^{1/2}\psi_{w,i-m}^{1/2}\psi_{w,j-m}^{1/2}\\
    = &\mathbb 1_{\{p\leq i,j,i-m,j-m\leq h\}} \sum_{m=0}^{h+p-1}
    \sum_{j=1}^{h+p-1}\sum_{i\neq j}\psi_{w,i}^{1/2}\psi_{w,j}^{1/2}\psi_{w,i-m}^{1/2}\psi_{w,j-m}^{1/2}\\
    \leq  &\mathbb 1_{\{p\leq i,j,i-m,j-m\leq h\}} \sum_{m=0}^{h+p-1}
    \text{constant}*  (1-\epsilon)^{2m} p^2 (\sum_{j=1}^{h+p-1}
            \psi_{w,j})^2\\
            \leq  & \sum_{m=0}^{\infty}
    \text{constant}*  (1-\epsilon)^{2m} p^2 (\sum_{j=1}^{h+p-1}
            \psi_{w,j})^2\\
    \leq& 
          \text{constant}*    \frac{p^2}{1-(1-\epsilon)^{2}} (\sum_{j=1}^{h+p-1}
            \psi_{w,j})^2
            \end{split}
\end{align}

\textbf{2. Boundedness for the term with $\mathbb 1_{\{j\geq  h\}}$}

\begin{align}
\label{b131}
    \begin{split}
        &\mathbb 1_{\{j>h\}}
\sum_{j=1}^{h+p-1}\sum_{i=1}^{j-1}\sum_{m=0}^{i-1}
            \psi_{w,i}^{1/2}\psi_{w,j}^{1/2}\psi_{w,i-m}^{1/2}\psi_{w,j-m}^{1/2}\\
            \leq  &(p-1)\underset{j\in (h,h+p)}{\sup}\psi_{w,j}^{1/2} \sum_{i=1}^{h+p-1}\sum_{m=0}^{i-1}
            \psi_{w,i}^{1/2}\psi_{w,i-m}^{1/2}\psi_{w,j-m}^{1/2}\\
            \leq  &(p-1)\underset{j\in (h,h+p)}{\sup}\psi_{w,j}^{1/2} \sum_{m=0}^{h+p-1}\sum_{i=m+1}^{h+p-1}
            \psi_{w,i}^{1/2}\psi_{w,i-m}^{1/2}\psi_{w,j-m}^{1/2}\\
         \leq  &  
        (p-1)\underset{j\in (h,h+p)}{\sup}\psi_{w,j}^{1/2} 
        \sum_{m=0}^{h+p-1}\psi_{w,j-m}^{1/2}
        \sum_{i=m+1}^{h+p-1}\psi_{w,i}^{1/2}  \psi_{w,i-m}^{1/2}\\ 
        \leq  &  
        (p-1)\underset{j\in (h,h+p)}{\sup}\psi_{w,j}^{1/2} 
        \sum_{m=0}^{h+p-1}\psi_{w,j-m}^{1/2}
        (\sum_{i=0}^{h+p-1}\psi_{w,i})\\ 
        & (\text{Cauchy–Schwarz inequality: }  \sum_{i=m+1}^{h+p-1}\psi_{w,i}^{1/2}  \psi_{w,i-m}^{1/2} \leq  \sum_{i=0}^{h+p-1}\psi_{w,i})\\
             \leq & \text{constant}*(\sum_{i=0}^{h+p-1}\psi_{w,i})^2
    \end{split}
\end{align}
The last inequality is because $p$ is finite order, $\underset{j\in (h,h+p)}{\sup}\psi_{w,j}^{1/2} \leq \sum_{j=0}^{h+p-1}\psi_{w,j}^{1/2}$, and $(
        \sum_{j=0}^{h+p-1}\psi_{w,j}^{1/2} )$ is finite when $\delta=0$, such that
        \begin{align}
        \label{b.174}
            \begin{split}
            \sum_{j=0}^{h+p-1}\psi_{w,j}^{1/2}
            =& \sum_{j=0}^{h+p-1} \|\sum_{i=\max(1,j-h+1)}^{\min(p,j)} w_i \Psi_{1\bullet,h-j+i-1} \|\\
            \leq & \sum_{j=0}^{h+p-1} \|\sum_{i=1}^{p} w_i \Psi_{1\bullet,h-j+i-1} \|\\
            = & \sum_{i=1}^{p} \sum_{j=0}^{h+p-1} \| \Psi_{1\bullet,h-j+i-1} \|\\
            \leq  & \sum_{i=1}^{p} \sum_{h=0}^{\infty} \| \Psi_{1\bullet,h} \| \leq  \text{constant}.
            \end{split}
        \end{align}
The last inequality is shown in \eqref{sumirf0}, Lemma \ref{lemmabound} when $\delta=0$.

\textbf{3. Boundedness for the term with $\mathbb 1_{\{i-p<m<i\}}$}
\begin{align}
\label{b132}
    \begin{split}
        &\mathbb 1_{\{i-p<m<i\}}
\sum_{j=1}^{h+p-1}\sum_{i=1}^{j-1}\sum_{m=0}^{i-1}
            \psi_{w,i}^{1/2}\psi_{w,j}^{1/2}\psi_{w,i-m}^{1/2}\psi_{w,j-m}^{1/2}\\
            = & \sum_{j=1}^{h+p-1}\sum_{i=1}^{j-1}\sum_{k=1}^{p-1}            \psi_{w,i}^{1/2}\psi_{w,j}^{1/2}\psi_{w,k}^{1/2}\psi_{w,k+j-i}^{1/2}\\
            &(\text{replace $i-m$ by $k$. Since $i-p<m<i$, then $0<i-m<p$}) \\
         = & \sum_{j=1}^{h+p-1}\sum_{n=1}^{j-1}\sum_{k=1}^{p-1}
            \psi_{w,j-n}^{1/2}\psi_{w,j}^{1/2}\psi_{w,k}^{1/2}\psi_{w,k+n}^{1/2}\\
               & ( \text{denote } j-i=n )\\
       \leq 
        & (p-1)\underset{k\in [1,p)}{\sup}
         \psi_{w,k}^{1/2}
         \sum_{n=1}^{h+p-1}
         \psi_{w,k+n}^{1/2} 
         \sum_{j=1}^{h+p-1} \psi_{w,j}^{1/2}\psi_{w,j-n}^{1/2} \\
     \leq & 
        \text{constant}*(\sum_{i=0}^{h+p-1}\psi_{w,i})^2
    \end{split}
\end{align}
The last inequality uses some intermediate results from \eqref{b131}:
\begin{enumerate}
\item $\underset{m\in [1,p)}{\sup}
         \psi_{w,m}^{1/2} \leq \sum_{i=0}^{h+p-1}\psi_{w,i}$. 
    \item $         \sum_{n=1}^{h+p-1}
         \psi_{w,k+n}^{1/2}\leq \text{constant} $ when $\delta=0$, similarly as \eqref{b.174}.
         \item $\sum_{j=1}^{h+p-1} \psi_{w,j}^{1/2}\psi_{w,j-n}^{1/2} \leq (\sum_{i=0}^{h+p-1}\psi_{w,i})$ by Cauchy–Schwarz inequality.
\end{enumerate}

In summary, combining the result of \eqref{b133}, \eqref{b131}, \eqref{b132}, when $\delta=0$, $T\rightarrow \infty$, $h\leq \bar h = \alpha T$ and $\alpha\in(0,1)$,
\begin{align}
    (\bar T -p+1)^{-1} (w'\Omega_{s,h} w)^{-2} \sum_{m=0}^{h+p-1}
    \sum_{j=1}^{h+p-1}\sum_{i\neq j}
            \psi_{w,i}^{1/2}\psi_{w,j}^{1/2}\psi_{w,i-m}^{1/2}\psi_{w,j-m}^{1/2} \xrightarrow{p}0, 
\end{align}
The convergence is because Lemma \ref{lemmavarscore} shows $0<\underline c_{\xi} \sum_{j=1}^{h+p-1}\psi_{w,j}
  \leq 
  w'\Omega_{s,h} w 
  \leq 
  \overline  c_{\xi} \sum_{j=1}^{h+p-1}\psi_{w,j}$, and $\underline c_{\xi}$ and $\overline  c_{\xi}$ are positive constant. 

Combining the convergence shown in \eqref{b163} for cases of $\delta=1,2$, we have shown the boundedness for the term in \eqref{b126} such that
\begin{align}
  (\bar T -p+1)^{-1} (w'\Omega_{s,h} w)^{-2}  \sum_{m=0}^{\bar T-p}
            \sum_{j=1}^{h+p-1}\sum_{i\neq j}
           | \text{Cov}(
            \xi_{i,0}\xi_{j,0}, \quad
            \xi_{i-m,m}\xi_{j-m,m})|\xrightarrow{p}0.
\end{align}
for any $\delta=0,1,2$. Combining the convergence of \eqref{b160}, we show the convergence,
\begin{align}
     \text{Var} \left( (\bar T-p+1)^{-1} \sum_{t=p}^{\bar T} (w's_{t,h}^*)^2/ w'\Omega_{s,h} w \right) \xrightarrow{p} 0.
\end{align}
Thus, the proof is complete.

\end{proof}

\newpage

\subsection{Lemma A.4}
\label{prooflemma4th}
\begin{proof}
First we write $\mathbb E[ (w' s_{t,h}^*)^4]= \mathbb E[ (\sum_{j=1}^{h+p-1} \xi_{j,t})^4] $, as shown in \eqref{b140} that $w' s_{t,h}^* = \sum_{j=1}^{h+p-1} \xi_{j,t}$, where $\xi_{j,t} = u_t'  (\sum_{i=\max(1,j-h+1)}^{\min(p,j)} w_i \Psi_{1\bullet,h-j+i-1}) u_{t+j}.$ Because of the mean-independence assumption, it is readily to check that $\mathbb E[\xi_{i,t}^3\xi_{j,t}]=0$, $\forall i\neq j$:
    \begin{align}
        \begin{split}
            &\mathbb E[\xi_{i,t}^3\xi_{j,t}] \\
            =& \mathbb E[ E[\xi_{i,t}^3\xi_{j,t} \mid u_t,u_{t+i}] ]\\
            =&\mathbb E[ \xi_{i,t}^3 (\sum_{i=\max(1,j-h+1)}^{\min(p,j)} w_i' u_t \Psi_{1\bullet,h-j+i-1}) \mathbb E[u_{t+j} \mid u_t,u_{t+i}] ]\\
            =&0 \text{ (by mean-independence, $\mathbb E[u_{t+j} \mid u_t,u_{t+i}]=0$ )}
        \end{split}
    \end{align}
    Similarly, $\mathbb E[\xi_k^2\xi_{i,t}\xi_{j,t}]=0$, $\forall i\neq j$.
    Thus, 
    \begin{align}
        \begin{split}
            &\mathbb E[ (w' s_{t,h}^*)^4] \\
            =& \mathbb E[(\sum_{j=1}^{h+p-1} \xi_{j,t})^4]\\
            =&\sum_{j=1}^{h+p-1} \mathbb E[ \xi_{j,t}^4] + 6 
            \sum_{j=1}^{h+p-1}\sum_{ l=j+1}^{h+p-1} \mathbb E[ \xi_{j,t}^2 \xi_l^2] .
        \end{split}
    \end{align}
    Then, we show the boundedness of $\mathbb E[ \xi_{j,t}^4]$ and $\mathbb E[ \xi_{j,t}^2 \xi_l^2] .$
    \begin{align}
        \begin{split}
            &\mathbb E[ \xi_{j,t}^4]\\
            =&\mathbb E[(\sum_{i=\max(1,j-h+1)}^{\min(p,j)} w_i' u_t \Psi_{1\bullet,h-j+i-1} u_{t+j})^4 ]\\
            \leq &\mathbb E[\|\sum_{i=\max(1,j-h+1)}^{\min(p,j)} w_i' u_t \Psi_{1\bullet,h-j+i-1}\|^4 \| u_{t+j}\|^4 ]\\    
            = &\mathbb E[\|u_t\|^4 \|\sum_{i=\max(1,j-h+1)}^{\min(p,j)} w_i \Psi_{1\bullet,h-j+i-1}\|^4 \| u_{t+j}\|^4 ]\\
            \leq & \|\sum_{i=\max(1,j-h+1)}^{\min(p,j)} w_i \Psi_{1\bullet,h-j+i-1}\|^4\mathbb E[\|u_t\|^8]\\
            = & \psi_{w,j}^2 \mathbb E[\|u_t\|^8].
        \end{split}
    \end{align}
    Also,
    \begin{align}
        \begin{split}
            &\mathbb E[ \xi_{j,t}^2 \xi_l^2] \\
           =&\mathbb E[(\sum_{i=\max(1,j-h+1)}^{\min(p,j)} w_i' u_t \Psi_{1\bullet,h-j+i-1} u_{t+j})^2 (\sum_{i=1}^{\min(p,l)} w_i' u_t \Psi_{1\bullet,h-l+i-1} u_{t+l})^2]\\
           \leq & 
           \mathbb E[
           \|u_t\|^4 \|u_{t+j}\|^2\|u_{t+l}\|^2]
           \|\sum_{i=\max(1,j-h+1)}^{\min(p,j)} w_i'  \Psi_{1\bullet,h-j+i-1} \|^2 
           \|\sum_{i=1}^{\min(p,l)} w_i' u_t \Psi_{1\bullet,h-l+i-1} u_{t+l}\|^2\\
           &(\text{since }(\sum_{i=\max(1,j-h+1)}^{\min(p,j)} w_i' u_t \Psi_{1\bullet,h-j+i-1} u_{t+j})^2 \leq \|u_t\|^2 \|u_{t+j}\|^2
           \|\sum_{i=\max(1,j-h+1)}^{\min(p,j)} w_i'  \Psi_{1\bullet,h-j+i-1} \|^2)\\
           \leq & 
           \mathbb E[
           \|u_t\|^8]
           \|\sum_{i=\max(1,j-h+1)}^{\min(p,j)} w_i'  \Psi_{1\bullet,h-j+i-1} \|^2 
           \|\sum_{i=1}^{\min(p,l)} w_i' u_t \Psi_{1\bullet,h-l+i-1} u_{t+l}\|^2\\
           =& 
           \mathbb E[
           \|u_t\|^8]
           \psi_{w,j} \psi_{w,l}
        \end{split}
    \end{align}
Now, let's wrap up all results for $\mathbb E [ (w's_{t,h}^*)^4/ (w'\Omega_{s,h} w)^2 ]$. The numerator is bounded from above: 
\begin{align}
\label{numerat}
\begin{split}
     \mathbb E [ (w's_{t,h}^*)^4]=&\sum_{j=1}^{h+p-1} \mathbb E[ \xi_{j,t}^4] + 6 
            \sum_{j=1}^{h+p-1}\sum_{ l=j+1}^{h+p-1} \mathbb E[ \xi_{j,t}^2 \xi_l^2] \\
            \leq &
            \mathbb E[ \|u_t\|^8] (
           \sum_{j=1}^{h+p-1} \psi_{w,j}^2
           + 6 
            \sum_{j=1}^{h+p-1}\sum_{ l=j+1}^{h+p-1}\psi_{w,j} \psi_{w,l})\\
           \leq &  
           \mathbb E[ \|u_t\|^8] 3 (
           \sum_{j=1}^{h+p-1} \psi_{w,j})^2.
\end{split}
\end{align}
By \eqref{b149} and \eqref{b150} in the proof of Lemma \ref{lemmavarscore}, $w'\Omega_{s,h} w = \sum_{j=1}^{h+p-1} \mathbb E [ \xi_{j,t}^2 ] $ and 
\begin{align}
\label{denomenat}
   \underline c_\xi^2 (\sum_{j=1}^{h+p-1} \psi_{w,j})^2 
   \leq 
   (w'\Omega_{s,h} w)^2 
   \leq 
   \overline c_\xi^2 (\sum_{j=1}^{h+p-1} \psi_{w,j})^2
\end{align}
Combining \eqref{numerat} and \eqref{denomenat} and bounded eighth moment of $u_t$ (Assumption \ref{assmoment}(i)) yields
\begin{align}
    \mathbb E [ (w's_{t,h}^*)^4/ (w'\Omega_{s,h} w)^2 ] \leq  3\mathbb E[ \|u_t\|^8]/ \underline c_\xi^2 <c_0<\infty
\end{align}
for some constant $c_0$.  It completes the proof.
\end{proof}

\newpage
\subsection{Lemma A.5}
\label{prooflemmaa5}

\begin{proof}

\textbf{Proof of} \eqref{b174}: The convergence in probability will be proved by Cheybeshev's inequality. First, $T^{-1}\mathbb E[\sum_{t=1}^{T} u_t u_t']=\Sigma_u$. Then, we need to show its variance converges to zero. Without loss of generality, we focus on $i$-th element of $u_t$,
\begin{align}
    \begin{split}
        & \textup{Var}(T^{-1}\sum_{t=1}^{T} u_{i,t}^2)\\
        =&T^{-1} \sum_{|m|=0}^{T}(1-\frac{|m|}{T})\textup{Cov}( u_{i,0}^2 ,u_{i,m}^2)\\
        \leq & \frac{2}{T} \sum_{|m|=0}^{T} |\textup{Cov}( u_{i,0}^2 ,u_{i,m}^2)|
    \end{split}
\end{align}
By Assumption \ref{assmoment}(ii), the autocovariance of $u_t\otimes u_t$ is absolutely summable. Thus, $\sum_{|m|=0}^{T} |\textup{Cov}( u_{i,0}^2 ,u_{i,m}^2)|$ is bounded by a constant from above and it induces the last expression above converges to zero in probability. In summary, using Cheybeshev's inequality we could show the convergence \eqref{b174}.

\textbf{Proof of} \eqref{conszx0}  (Similarly for \eqref{conszx1} and \eqref{conszx2}):

Given $x_t$ contains $p$-lag $y_t$'s and $z_t$ contains $p$-lag $u_t$'s, to prove \eqref{conszx0}, it is sufficient to prove
\begin{align}
 \label{conv1}   & \lim_{T\rightarrow\infty }T^{-1}\sum_{t=1}^T  y_{t} u_{t+k}'\xrightarrow{p} 0\\
 \label{conv2}   & \lim_{T\rightarrow\infty }T^{-1}\sum_{t=1}^{T}  y_{t} u_{t-k}'\xrightarrow{p} \Psi_{k}\Sigma_u
\end{align}
 for $k\geq 0$ .  Here, we use the information that when $\delta=0$, $ y_t $ is stationary, such that $y_t = \sum_{i=0}^\infty \Psi_i u_{t-i}$. We first show the convergence of \eqref{conv1},
    \begin{align}
        \begin{split}
            &\| T^{-1}\sum_{t=1}^T  y_{t} u_{t+k}'\|\\
             =&\|T^{-1}\sum_{t=1}^T  \sum_{i=0}^\infty \Psi_i u_{t-i} u_{t+k}'\|  \\
             =&\|\sum_{i=0}^\infty \Psi_i  T^{-1}\sum_{t=1}^T u_{t-i} u_{t+k}'\|\\
    \leq & \sum_{i=0}^\infty \|\Psi_i \| \|T^{-1}\sum_{t=1}^T u_{t-i} u_{t+k}'\|\\
            \xrightarrow{p}& 0 \quad \text{ (as $ \sum_{i=0}^\infty \|\Psi_i \|$ is bounded by a constant shown in \eqref{sumirf0}, Lemma \ref{lemmabound} when $\delta=0$})
        \end{split}
    \end{align}
    The convergence $\|T^{-1}\sum_{t=1}^T u_{t} u_{t+k}'\| \xrightarrow{p}0$ for all $k>0$ can be readily checked element-by-element by Chebyshev's inequality. Due to mean-independence by Assumption \ref{assimeanind}, its mean equals to zero. Its variance is given below
    \begin{align}
        \begin{split}
            &\text{Var}(T^{-1}\sum_{t=1}^T u_{i,t} u_{j,t+k})\\
            =& \mathbb E [ (T^{-1}\sum_{t=1}^T u_{i,t} u_{j,t+k})^2]\\
            =& T^{-2 }\mathbb E [ \sum_{t=1}^T u_{i,t}^2 u_{j,t+k}^2]\\
            &(\text{as mean-independence by Assumption \ref{assimeanind}})\\
            \leq & T^{-1 }\mathbb E [ \| u_{t}\|^2 \|u_{t+k}\|^2]\\
            \leq & T^{-1 }\mathbb E [ \| u_{t}\|^4 ]\\
            \xrightarrow{p}& 0 \quad \text{ (bounded fourth moment implied by Assumption \ref{assmoment}})
        \end{split}
    \end{align}

     Next, we prove \eqref{conv2},
    \begin{align}
        \begin{split}
            &T^{-1}\sum_{t=1}^{T}  y_{t} u_{t-k}'\\
            =&T^{-1}\sum_{t=1}^{T}  \sum_{j=0}^\infty \Psi_j u_{t-j} u_{t-k}'\\
            =&T^{-1}\sum_{t=1}^{T}  \sum_{j\neq k} \Psi_j u_{t-j} u_{t-k}'
            +T^{-1}\sum_{t=1}^{T}   \Psi_j u_{t-k} u_{t-k}'\\
             =&\sum_{j\neq k} \Psi_j  T^{-1}\sum_{t=1}^{T}  u_{t-j} u_{t-k}'
            +\Psi_j T^{-1}\sum_{t=1}^{T}    u_{t-k} u_{t-k}'\\
            \xrightarrow{p}& \Psi_j\Sigma_u.
        \end{split}
    \end{align}
    The convergence is because we have shown $\|T^{-1}\sum_{t=1}^T u_{t} u_{t+k}'\| \xrightarrow{p}0$ for all $k>0$, and \eqref{b174} shows $T^{-1}\sum_{t=1}^{T}  u_{t-k} u_{t-k}' \xrightarrow{p}\Sigma_u$, and $\sum_{j=0}^\infty\|\Psi_j\|$ is bounded above by a constant when the data process is stationary $\delta=0$ (given in \eqref{sumirf0}, Lemma \ref{lemmabound}). In summary, the proof is complete.

    The proofs of convergence of \eqref{conszx1} and \eqref{conszx2} follow the similar procedure. It is due to the fact that when $\delta=1$, $\Delta_{10}\tilde x_t$ contains $p$-lag $\Delta_{10}\tilde y_t $ and $\Delta_{10}\tilde y_t$ is stationary; when $\delta=2$, $\Delta_{20}\tilde x_t$ contains $p$-lag $\Delta_{20}\tilde y_t $ and $\Delta_{20}\tilde y_t$ is stationary. Thus, the convergence of sample means of $\Delta_{10}\tilde y_t u_{t+k}$ /$\Delta_{20}\tilde y_t u_{t-k}$  and $\Delta_{20}\tilde y_t u_{t+k}$ /$\Delta_{20}\tilde y_t u_{t-k}$ can be shown in the same manner as the proofs for \eqref{conv1} /\eqref{conv2}.


        \textbf{Proof of} \eqref{xx0} (Similarly for \eqref{xx1} and \eqref{xx2}): First, the boundedness of \eqref{xx0} will be proved by Cheybeshev's inequality. As $\Pi y_t =\tilde y_t$, Without loss of generality, we consider $i$-th element in $\tilde y_{t}$. First, $\mathbb E[T^{-1}\sum_{t=1}^{T}\tilde y_{i,t}^2]=\mathbb E[\tilde y_{i,t}^2]$ which is a constant term since $\tilde y_t=B_0(L)u_t$ is a covariance stationary process. Then, we need to show its variance is bounded above by a constant. 
    \begin{align}
    \label{b184}
        \begin{split}
            &\textup{Var}(T^{-1}\sum_{t=1}^{T}\tilde y_{i,t}^2)\\
            =&T^{-2}\textup{Var}(\sum_{t=1}^{T}\tilde y_{i,t}^2)\\
            \leq &\max_{1\leq t\leq T} \textup{Var}(\tilde y_{i,t}^2)\\
            \leq &\max_{1\leq t\leq T}\mathbb E[y_{i,t}^4]
        \end{split}
    \end{align}
The last expression is a $O_p(1)$ term as shown in \eqref{y0}, Lemma \ref{lemmacov}.

    Next, we prove \eqref{xx1}. Similar to the proof of \eqref{xx0}, we need to check that the term has a bounded mean and variance. Note that the first $(p-1)K$ elements in $G_{1,T}x_t$ are lagged $\Delta_{10}\tilde y_t$'s, which are covariance stationary. Thus, its sample variance has a finite mean and variance, as can be proved in a similar manner to the proof for \eqref{xx0}. The last $K$ elements are potentially non-stationary processes scaled by a probability scaling matrix, $\Upsilon_1 \tilde y_{t-p}$. Without loss of generality, we focus on its $i$-th element. Recall the $i$-th element on the main diagonal of $\Upsilon_1$ is $g_{i,1}^{-1/2}$. The mean is bounded as follows,
    \begin{align}
        \begin{split}
            & \mathbb E [T^{-1}\sum_{t=1}^T g_{i,1}^{-1} \tilde y_{i,t}^2] \leq  \max_{1\leq t\leq T } \mathbb E [g_{i,1}^{-1}\tilde y_{i,t}^2]\leq  \max_{1\leq t\leq T }\mathbb E [\| \Upsilon_1 \tilde y_{t}\|^4 ] =O_p(1),
        \end{split}
    \end{align}
    by \eqref{y1}, Lemma \ref{lemmacov}. Next, its variance is shown to be bounded as follows, 
        \begin{align}
        \begin{split}
            \textup{Var}(T^{-1}\sum_{t=1}^T g_{i,1}^{-1} \tilde y_{i,t}^2)
            \leq   \max_{1\leq t\leq T }\textup{Var}(g_{i,1}^{-1} y_{i,t}^2)
            \leq    \max_{1\leq t\leq T }\mathbb E [g_{i,1}^{-2}\tilde y_{i,t}^4] \leq  \max_{1\leq t\leq T }\mathbb E [\| \Upsilon_1 \tilde y_{t}\|^4 ] =O_p(1),
        \end{split}
    \end{align}
    by \eqref{y1}, Lemma \ref{lemmacov}. Thus, for each element in $G_{1,T}x_t$, its sample variance has finite mean and variance. In turn, \eqref{xx1} is proved.

    Lastly, we prove \eqref{xx2}. Similar to the proof of \eqref{xx0}. We need to check the term has bounded mean and variance. In particular, the first $(p-2)K$ elements in $G_{2,T}x_t$ are lagged $\Delta_{20}\tilde y_t = B_2(L) u_t$, which are covariance stationary. Thus, its sample variance has a finite mean and variance, as can be proved in a similar manner to the proof for \eqref{xx0}. The second last $K$ elements in $G_{2,T}x_t$ are potentially I(1) processes scaled by a probability scaling matrix $\Upsilon_1 \Delta_{21}\tilde y_{t-p} $. Its sample variance has finite mean and variance, as can be proved in a similar manner to the proof for \eqref{xx1}. Lastly, the last $K$ elements in $G_{2,T}x_t$ are also potentially I(2) processes scaled by a probability scaling matrix  $\Upsilon_2 \tilde y_{t-p-1}$. To show its sample variance is bounded, without loss of generality, we focus on its $i$-th element. Its mean is shown bounded as follows,
    \begin{align}
        \begin{split}
            \mathbb E [T^{-1}\sum_{t=1}^T g_{i,2}^{-1} \tilde y_{i,t}^2]
            \leq   \max_{1\leq t\leq T } \mathbb E [g_{i,2}^{-1}\tilde y_{i,t}^2] \leq  \max_{1\leq t\leq T }\mathbb E [\| \Upsilon_2 \tilde y_{t}\|^4 ] =O_p(1),
        \end{split}
    \end{align}
    by \eqref{y2}, Lemma \ref{lemmacov}. Next, its variance is shown to be bounded as follows,
        \begin{align}
        \begin{split}
             \textup{Var}(T^{-1}\sum_{t=1}^T g_{i,2}^{-1} \tilde y_{i,t}^2)
            \leq   \max_{1\leq t\leq T }\textup{Var}(g_{i,2}^{-1} y_{i,t}^2)
            \leq    \max_{1\leq t\leq T }\mathbb E [g_{i,2}^{-2}\tilde y_{i,t}^4]\leq  \max_{1\leq t\leq T }\mathbb E [\| \Upsilon_2 \tilde y_{t}\|^4 ] =O_p(1),
        \end{split}
    \end{align}
    by \eqref{y2}, Lemma \ref{lemmacov}. Thus, for each element in $G_{2,T}x_t$, its sample variance has finite mean and variance. In turn, \eqref{xx2} is proved.

    \textbf{Proof of} \eqref{ux0} (Similarly for \eqref{ux1}, \eqref{uy1}; and \eqref{ux2}, \eqref{uy2}): the boundedness will be proved by Cheybeshev's inequality (finite mean and finite variance). As $x_t$ contains $p$-lag $y_t$'s, without loss of generality, it is equivalent to show $T^{-1/2}\sum_{t=1}^T u_{i,t+k} y_{j,t} =O_p(1)$ for all $k>0$ and any $i=1,2,\cdots,K$. By mean-independence assumption (Assumption \ref{assimeanind}, the mean $\mathbb E[T^{-1/2}\sum_{t=1}^T u_{i,t+k} y_{j,t}]=0$. Then, we need to show its variance is bounded above by a constant.  
    \begin{align}
    \label{b183}
        \begin{split}
            &\textup{Var}(T^{-1/2}\sum_{t=1}^T u_{i,t+k} y_{j,t})\\
            =&\mathbb E [ (T^{-1/2}\sum_{t=1}^T u_{i,t+k} y_{j,t})^2]\\
            =&T^{-1}\sum_{t=1}^{T}\mathbb E [ y_{j,t}^2 u_{i,t-k}^2]\\
            & (\text{due to mean independence by Assumption \ref{assimeanind} })\\
            = &\mathbb E [ y_{j,t}^2 u_{i,t-k}^2]\\
            & (\text{as $y_t$ is covariance stationary when $\delta=0$})\\ 
            \leq &\mathbb E [ y_{j,t}^4]^{1/2} \mathbb E [ u_{i,t}^4]^{1/2}    
        \end{split}
    \end{align}
The last expression is a $O_p(1)$ term because $ y_t =\Pi^{-1}\tilde y_t$, $\Pi$ is a non-singular rotation matrix,  $\tilde y_t$ has boundedness of fourth moments shown in \eqref{y0} of Lemma \ref{lemmacov} when $\delta=0$, and $u_t$ has bounded fourth moment by Assumption \ref{assmoment}. 

  Then we prove \eqref{ux1} and \eqref{uy1}. The boundedness will be proved in the similar manner as \eqref{ux0}. We will show both expressions have finite mean and finite variable. The mean of two expressions are equal to zero because of mean independence of $u_t$. The boundedness of the variance is shown through the fourth moment of $u_t$ and $G_{1,T}x_t$.  When $\delta=1$, the first $(p-1)K$ elements in $G_{1,T}x_t$ is stationary and thereby its fourth moment can be shown as bounded in a similar manner as the proof for \eqref{y0}; its last $K$ elements in $G_{1,T}x_t$ is lagged $\Upsilon_1 \tilde y_{t}$ whose fourth moment is shown as bounded in \eqref{y1} in Lemma \ref{lemmacov}. 
  
  Lastly we prove \eqref{ux2} and \eqref{uy2}. The boundedness will be proved in the similar manner as \eqref{ux0}. We will show both expressions have finite mean and finite variable. The mean of two expressions are equal to zero because of mean independence of $u_t$. The boundedness of the variance is shown through the fourth moment of $u_t$ and $G_{2,T}x_t$. When $\delta=2$, the first $(p-2)K$ elements in $G_{2,T}x_t$ are stationary whose fourth moment can be shown as bounded in a similar manner as the proof for \eqref{y0}; its second last $K$ elements $(\Upsilon_1 \Delta_{21}\tilde y_{t-p+2})$ is potentially non-stationary process whose fourth moment can be shown as bounded in a similar manner as the proof for \eqref{y1}; its last $K$ element $(\Upsilon_2 \tilde y_{t-p+1})$ has bounded fourth moment shown in \eqref{y2} in Lemma \ref{lemmacov}. 

    In summary, \eqref{ux1}, \eqref{uy1} and \eqref{ux2}, \eqref{uy2} are bounded above by a constant almost surely because all of these terms have zero mean and finite variance.
    
\textbf{Proof of} \eqref{xe0} (Similarly for \eqref{xe1} and \eqref{xe2}):
Expand $e_{t,h}=\sum_{i=0}^{h-1}\Psi_{1\bullet,i}u_{t+h-1}$. Given $\sum_{i=0}^{h-1}\|\Psi_{i}\|^2 =O_p(1)$ by \eqref{sumirf02} of Lemma \ref{lemmabound} and $\|\bar T^{-1/2} \sum_{t=p}^{T-h} x_t u_{t+k}' \|=O_p(1)$ by \eqref{ux0} for the case of $\delta=0$, $\|\bar T^{-1/2} \sum_{t=p}^{T-h} x_t e_{t+k,h} \|  \leq \sum_{i=0}^{h-1}\|\Psi_{i}\|^2 \|\bar T^{-1/2} \sum_{t=p}^{T-h} x_t u_{t+h-i,h}' \| =O_p(1)$. Similarly, \eqref{xe1}/ \eqref{xe2} can be proved by using the results of \eqref{sumirf12} and \eqref{ux1} / \eqref{sumirf22} and \eqref{ux2}.
  
\end{proof}

\newpage
\subsection{Lemma A.6}
\label{prooflemmaa6}

\begin{proof}
\textbf{Proof} of \eqref{zzx0}:
 \begin{align}
    \begin{split}
        &\| \bar T^{-1/2}\sum_{t=p}^{T-h} 
(\hat z_{t} - 
 z_{t} ) x_{t}'G_{\delta,T}'  \| \\
 =& 
 \sum_{i=0}^{p-1}\| \bar T^{-1/2}\sum_{t=p}^{T-h}(\hat{u}_{t-i}- {u}_{t-i})x_{t}'G_{\delta,T}' \| \\
=&  \sum_{i=0}^{p-1} \|\bar T^{-1/2} \sum_{t=1}^{T}u_t x_t' G_{\delta,T}'(\sum_{t=1}^{T} G_{\delta,T}x_t x_t'G_{\delta,T}')^{-1} \sum_{t=p}^{T-h}G_{\delta,T}x_{t-i}x_{t}'G_{\delta,T}' \|\\
\leq &    (\frac{\bar T}{T})^{1/2} \sum_{i=0}^{p-1} 
\| T^{-1/2}\sum_{t=1}^{T}u_t x_t'G_{\delta,T}' \| 
\| ( T^{-1}\sum_{t=1}^{T}G_{\delta,T} x_t x_t'G_{\delta,T}')^{-1} \| 
\| \bar T^{-1}  \sum_{t=p}^{T-h}G_{\delta,T}x_{t-i}x_{t}'G_{\delta,T}' \|\\
=&O_p(1).
    \end{split}
\end{align}
By Lemma \ref{lemma2moments}, $\| T^{-1/2}\sum_{t=1}^{T}u_t x_t'G_{\delta,T}' \| $ and $\| \bar T^{-1}  \sum_{t=p}^{T-h}G_{\delta,T}x_{t-i}x_{t}'G_{\delta,T}' \|$ are $O_p(1)$ terms for $\delta=0,1,2$. Also, $\| ( T^{-1}\sum_{t=1}^{T}G_{\delta,T} x_t x_t'G_{\delta,T}')^{-1} \|$ is bounded above by a constant almost surely by Assumption \ref{ass4}, and $\bar T/T<1$. Thus, \eqref{zzx0} is proved.

\textbf{Proof} of \eqref{ze0}:

\begin{align}
    \begin{split}
        & \frac{1}{(w'\Omega_{\beta,h} w)^{1/2}}\| \bar T^{-1/2}\sum_{t=p}^{T-h}  (\hat z_{t} - z_{t}) e_{t,h} \| \\
        =& \frac{1}{(w'\Omega_{\beta,h} w)^{1/2}}\sum_{i=0}^{p-1}\| \bar T^{-1/2}\sum_{t=p}^{T-h}  (\hat u_{t-i} - u_{t-i}) e_{t,h} \|\\
        =& \frac{1}{( Tw'\Omega_{\beta,h} w)^{1/2}} \sum_{i=0}^{p-1}\|(T^{-1/2}\sum_{t=p}^{T} u_{t} x_{t}' ) (T^{-1}\sum_{t=p}^{T} x_{t} x_{t}' )^{-1} (\bar T^{-1/2}\sum_{t=p}^{T-h} x_{t-i} e_{t,h} )\|\\
        =& \frac{1}{( Tw'\Omega_{\beta,h} w)^{1/2}} \sum_{i=0}^{p-1}\|(T^{-1/2}\sum_{t=p}^{T} u_{t} x_{t}' G_{\delta,T}') (T^{-1}\sum_{t=p}^{T}G_{\delta,T} x_{t} x_{t}' G_{\delta,T}')^{-1} \\
        & (T^{-1/2}\sum_{t=p}^{T-h}G_{\delta,T} x_{t-i} e_{t,h} )\|\\
        \leq &
        \frac{1}{( Tw'\Omega_{\beta,h} w)^{1/2}}  \sum_{i=0}^{p-1}\|T^{-1/2}\sum_{t=p}^{T} u_{t} x_{t}' G_{\delta,T}'\|
         \| (T^{-1}\sum_{t=p}^{T}G_{\delta,T} x_{t} x_{t}' G_{\delta,T}')^{-1} \| \\
        & \| T^{-1/2}\sum_{t=p}^{T-h}G_{\delta,T} x_{t-i} e_{t,h} \| \\
        \xrightarrow{p}&0
    \end{split}
\end{align}
By Lemma \ref{lemma2moments}, $\|T^{-1/2}\sum_{t=p}^{T} u_{t} x_{t}' G_{\delta,T}'\|$ is a $O_p(1)$ term in the limit for $\delta=0,1,2$; and $\| T^{-1/2}\sum_{t=p}^{T-h}G_{\delta,T} x_{t-i} e_{t,h} \| $ is a $O_p(1)$, $O_p(h)$, and $O_p(h^2)$ term in the limit for $\delta=0,1$, and $2$, respectively. Also, $\| ( T^{-1}\sum_{t=1}^{T}G_{\delta,T} x_t x_t'G_{\delta,T}')^{-1} \|$ is bounded above by a constant almost surely by Assumption \ref{ass4}, and $w'\Omega_{\beta,h} w$ is bounded below by a constant shown in Lemma \eqref{lemmavarscore}. In summary, when $\delta=0$, the product of all three norms is $O_p(1)$ and the convergence holds as $T\rightarrow \infty$; when $\delta=1$, the product of all three norms is bounded by a $O_p(h)$ term, given the condition imposed on horizon $h$, $\frac{h}{T}\max(1,\frac{h}{w'\Omega_{\beta,h} w})\xrightarrow{p}0$, the convergence holds; lastly, when $\delta=2$, the product of all three norms is bounded by a $O_p(h^2)$ term, given the condition imposed on horizon $h$, $\frac{h^3}{T}\max(1,\frac{h}{w'\Omega_{\beta,h} w})\xrightarrow{p}0$, the convergence holds. In summary, \eqref{ze0} is proved.
\end{proof}

\newpage
\subsection{Lemma A.7}
\label{prooflemmaa7}

\begin{proof}
Denote $\tilde \Psi_h =\Pi \Psi_h$. Without loss of generality, the proof will focus on the first element of $\tilde  y_t$. We first write $\tilde y_{1,t}$ as the summation of the product of impulse response functions and past innovations, and the boundedness will be proved case by case for $\delta\in \{0,1,2\}$.
\begin{align}
        \begin{split}
            & \max_{1\leq t\leq T }\mathbb E[ \tilde y_{1,t}^4]\\
            = & \max_{1\leq t\leq T }\mathbb E[(\sum_{l=0}^t \tilde\Psi_{1\bullet,l}u_{t-l})^4]\\
            \leq & \max_{1\leq t\leq T } \sum_{l=0}^t \| \tilde\Psi_{1\bullet,l}\|^4 \mathbb E[\|u_{t-l}\|^4] + 6 \sum_{l_1=0}^t\sum_{l_2\neq l_1} \| \tilde\Psi_{i\bullet,l_1}\|^2  \tilde\Psi_{i\bullet,l_2}\|^2\mathbb E[\|u_{t-l_1}\|^2 \|u_{t-l_2}\|^2]\\
            &(\text{By mean-independence assumption})\\
            \leq &  \mathbb E[\|u_{t}\|^4] 
            \max_{1\leq t\leq T } 3 \left( \sum_{l=0}^t \| \tilde\Psi_{1\bullet,l}\|^2 \right)^2.
        \end{split}
    \end{align}
    Since $\mathbb E[\|u_{t}\|^4]$ is finite, the boundedness of the fourth moment depends on the term $\max_{1\leq t\leq T } \left( \sum_{l=0}^t \| \tilde\Psi_{1\bullet,l}\|^2 \right)^2$.

    \textbf{Case 1:} $\delta=0$.  
    
    In this case, \eqref{sumirf02} in Lemma \ref{lemmabound} has shown that $\sum_{h=0}^\infty \| \Psi_{h}\|^2$  is bounded by a constant. As $\tilde\Psi=\Pi \Psi$ and $\Pi$ is a constant matrix, $\max_{1\leq t\leq T }\mathbb E[ \tilde y_{1,t}^4]$ is bounded above by a constant.

    \textbf{Case 2:} $\delta=1$.   
    
    In Appendix \eqref{prooflemmairf}, we have shown that $\tilde\Psi_h=\sum_{i=0}^{h}P_1^i J_1\mathbf B_1^i J_1'$ when $\delta=1$. As $P_1$ is diagonal matrix, it yields,
    \begin{align}
    \label{b190}
        \begin{split}
             \| \tilde\Psi_{1 \bullet,h}\| \leq & \sum_{l=0}^h |\rho_{1,1}|^{h-l} \|\mathbf B_{1}^l\|\\
            \leq & \sum_{l=0}^h | \rho_{1,1} |^{h-l} c_1 (1-\epsilon)^l \\
            =  & \max(|\rho_{1,1}|, 1-\epsilon/2)^h
            c_1 \sum_{l=0}^h  \frac{| \rho_{1,1} |^{h-l}}{\max(|\rho_{1,1}|, 1-\epsilon/2)^h} (1-\epsilon)^l \\
            \leq &\max(|\rho_{1,1}|, 1-\epsilon/2)^h
            c_1 \sum_{l=0}^h \left(\frac{1-\epsilon}{\max(|\rho_{1,1}|, 1-\epsilon/2)}\right)^l  \\
            &(\text{as }\frac{| \rho_{1,1} |^{h-l}}{\max(|\rho_{1,1}|, 1-\epsilon/2)^h} \leq \max(|\rho_{1,1}|, 1-\epsilon/2)^{-l})\\
            \leq  &\max(|\rho_{1,1}|, 1-\epsilon/2)^h
            c_1 \frac{1}{1-\frac{1-\epsilon}{\max(|\rho_{1,1}|, 1-\epsilon/2)} }\\
            \leq &    c_1(\frac{2}{\epsilon} -1)\max(|\rho_{1,1}|, 1-\epsilon/2)^h\\
            &(\text{as }\frac{1-\epsilon}{\max(|\rho_{1,1}|, 1-\epsilon/2)} \leq \frac{1-\epsilon}{1-\epsilon/2}=1-\frac{\epsilon}{2-\epsilon})
        \end{split}
    \end{align}
    Then, we show the boundedness,
    \begin{align}
    \label{b189}
        \begin{split}
            \max_{1\leq t\leq T }  \sum_{l=0}^t \| \tilde\Psi_{1\bullet,l}\|^2 
            \leq & \sum_{l=0}^T \| \tilde\Psi_{1\bullet,l}\|^2  \\
            \leq & c_1^2 (\frac{2}{\epsilon} -1)^2 \sum_{l=0}^T  \max(|\rho_{1,1}|, 1-\epsilon/2)^{2l}
           \\
            &(\text{the upper bound of }\| \tilde\Psi_{1\bullet,l}\| \text{ when $\delta=1$ is given in } \eqref{b190}) \\
            \leq &
            c_1^2 (\frac{2}{\epsilon} -1)^2  \max\left( \min(\frac{1}{1-|\rho_{1,1}|^2},T),\frac{1}{1-(1-\epsilon/2)^2} \right)
            \\
            &\left(\sum_{l=0}^T  |\rho_{1,1}|^{2l}\leq \min(\frac{1}{1-|\rho_{1,1}|^2},T), \quad  \sum_{l=0}^T (1-\epsilon/2)^{2l}\leq \frac{1}{1-(1-\epsilon/2)^2} \right)\\
            \leq &\text{constant}* \min(\frac{1}{1-|\rho_{1,1}|},T)  \\
            & (\text{as } |\rho_{1,1}|\leq 1, \min (\frac{1}{1-|\rho_{1,1}|^2},T)\leq \min\left(\frac{1}{1-|\rho_{1,1}|}, T\right)\equiv g_{1,1} )\\
            \equiv & \text{constant}* g_{1,1}.
        \end{split}
    \end{align}
     Since $\Upsilon_1$ is a diagonal matrix whose the first element is $g_{1,1}^{-1/2}$, \eqref{b189} yields that $ \max_{1\leq t\leq T } \left( \sum_{l=0}^t \| \tilde\Psi_{1\bullet,l}\|^2 \right)^2$ divided by $g_{1,1}^2$ produces a constant term. Thus, it proves that $\max_{1\leq t\leq T }\mathbb E[\|\Upsilon_{1} \tilde y_{t}\|^4] =O_p(1)$.

    \textbf{Case 3:} $\delta=2$.  
    In Appendix \eqref{prooflemmairf}, we have shown that $\tilde\Psi_h=\sum_{i=0}^{h}P_2^i\left( \sum_{j=0}^{h-i}P_1^i J_1\mathbf B_1^{h-i-j} J_1'\right)$ when $\delta=2$. As $P_1,P_2$ are diagonal matrices, it yields
    \begin{align}
    \label{b191}
        \begin{split}
            & \| \tilde\Psi_{1 \bullet,h}\| \\
        \leq & 
        \sum_{l_1=0}^{h}
        |\rho_{1,2}|^{l_1} \sum_{l_2=0}^{h-l_1} |\rho_{1,1}|^{l_2}
        \| \mathbf B_2^{h-l_1-l_2}  \| \\
        \leq & 
        c_1(\frac{2}{\epsilon} -1) \sum_{l_1=0}^{h}
        |\rho_{1,2}|^{l_1} \max(|\rho_{1,1}|, 1-\epsilon/2)^{h-l_1} \\
        \leq & 
        c_1(\frac{2}{\epsilon} -1)
        \max(|\rho_{1,2}|, 1-\epsilon/4)^h 
        \sum_{l_1=0}^{h}
        \frac{|\rho_{1,2}|^{l_1}}{\max(|\rho_{1,2}|, 1-\epsilon/4)^h} \max(|\rho_{1,1}|, 1-\epsilon/2)^{h-l_1} \\
        \leq & 
        c_1(\frac{2}{\epsilon} -1)
        \max(|\rho_{1,2}|, 1-\epsilon/4)^h 
        \sum_{l_1=0}^{h}
        \left(
        \frac{\max(|\rho_{1,1}|, 1-\epsilon/2)}{\max(|\rho_{1,2}|, 1-\epsilon/4)}
        \right)^{l_1}\\
        &(\text{because } \frac{|\rho_{1,2}|^{l_1}}{\max(|\rho_{1,2}|, 1-\epsilon/4)^h} \leq \max(|\rho_{1,2}|, 1-\epsilon/4)^{l_1-h} )
        \end{split}
    \end{align}
   Then, the boundedness of $ \sum_{l=0}^T \| \tilde\Psi_{1\bullet,l}\|^2$ will be discussed under the following two circumstances. The purpose that the proof is illustrated in two cases is when $\rho_{1,1}$ is close to unit circle, the definition of the parameter space restrict $|\rho_{1,2}|=1$; on the other hand, when $\rho_{1,1}$ is bounded away from the unit circle, two roots $\rho_{1,1},\rho_{1,2}$ can be arbitrarily close to each other.
   
    Circumstance 1: $|\rho_{1,1}|\leq  1-\epsilon/2$. The last expression of \eqref{b191} is less than
    \begin{align}
    \label{b.205}
        c_1(\frac{2}{\epsilon} -1) (\frac{4}{\epsilon} -1)\max(|\rho_{1,2}|, 1-\epsilon/4)^h,
    \end{align}
    since $\sum_{l_1=0}^{h}
        \left(
        \frac{\max(|\rho_{1,1}|, 1-\epsilon/2)}{\max(|\rho_{1,2}|, 1-\epsilon/4)}
        \right)^{l_1} \leq \sum_{l_1=0}^{\infty} \left(\frac{1-\epsilon/2}{1-\epsilon/4}\right)^{l_1}=\frac{4}{\epsilon}-1$. As $c_1(\frac{2}{\epsilon} -1) (\frac{4}{\epsilon} -1)$ is constant and $\sum_{l_1=0}^\infty (1-\epsilon/4)^{2l_1}$ is bounded by a constant from above, it yields
        \begin{align}
        \begin{split}
            \sum_{l=0}^T \| \tilde\Psi_{1\bullet,l}\|^2 \leq & 
            \sum_{l=0}^T
            c_1(\frac{2}{\epsilon} -1) (\frac{4}{\epsilon} -1)\max(|\rho_{1,2}|, 1-\epsilon/4)^l \\
            &(\text{use the result of \eqref{b.205}})\\
           \leq &           
            \text{constant} * \min\left( \frac{1}{1-|\rho_{1,2}|}, T \right)^2 \\
            &\left(\text{as }\sum_{l=0}^T  |\rho_{1,2}|^{l}\leq \min(\frac{1}{1-|\rho_{1,2}|^2},T), \quad  \sum_{l=0}^T (1-\epsilon/4)^{l}\leq \frac{4}{\epsilon} \right)\\
            = & \text{constant} * g_{i,2}.
        \end{split}
        \end{align}
        The last inequality is because $g_{i,2} =\min\left( \frac{1}{1-|\rho_{1,1}|}, T \right)^2\min\left( \frac{1}{1-|\rho_{1,2}|}, T \right) $ and the term $\left( \frac{1}{1-|\rho_{1,1}|}, T \right)$ is bounded by a constant from above given $|\rho_{1,1}|\leq  1-\epsilon/2$ in this circumstance. Note that the ‘\text{constant}’ term in the last expression may not be identical the the ‘\text{constant}’ term in the second last expression, as it shall take account of the constant upper bound for $\left( \frac{1}{1-|\rho_{1,1}|}, T \right)$.  
        In summary,  $ \max_{1\leq t\leq T } \left( \sum_{l=0}^t \| \tilde\Psi_{1\bullet,l}\|^2 \right)^2$ divided by $g_{1,2}^2$ produces a constant term under $\delta=2$ and  $|\rho_{1,1}|\leq  1-\epsilon/2$.

       Circumstance 2: $|\rho_{1,1}|\geq  1-\epsilon/2$. Given the definition of the parameter space, $|\rho_{1,2}|=1 $ once $|\rho_{1,1}|\geq  1-\epsilon/2$.  Thus, the last expression of \eqref{b191} is less than
       \begin{align}
        c_1(\frac{2}{\epsilon} -1) \sum_{l_1=0}^{h} |\rho_{1,1}|^{l_1}.
    \end{align}
    It yields
    \begin{align}
    \label{b195}
        \begin{split}
            &  \sum_{h=0}^T \| \tilde\Psi_{1 \bullet,h}\|^2 \\
            \leq & 
             c_1^2(\frac{2}{\epsilon} -1)^2 \sum_{h=0}^T  (\sum_{l_1=0}^{h} |\rho_{1,1}|^{l_1})^2  \\
            \leq &
            c_1^2(\frac{2}{\epsilon} -1)^2
             \sum_{h=0}^T
            \min\left(\frac{1-|\rho_{1,1}|^{h+1}}{1-|\rho_{1,1}|},h
            \right)^2  \\
            \leq & 
            c_1^2(\frac{2}{\epsilon} -1)^2
             \sum_{h=0}^T
            \min\left(\frac{1}{(1-|\rho_{1,1}|)^2},h^2
            \right)  \\
            =& c_1^2(\frac{2}{\epsilon} -1)^2 
            \min\left(\frac{T}{(1-|\rho_{1,1}|)^2}, O_p(T^3)
            \right)
        \end{split}
    \end{align}
    In this case, $|\rho_{1,1}|\geq  1-\epsilon/2$ and $|\rho_{1,2}|=1 $. It yields $g_{i,2} \equiv \min\left( \frac{1}{1-|\rho_{1,1}|}, T \right)^2\min\left( \frac{1}{1-|\rho_{1,2}|}, T \right)=T \min\left( \frac{1}{1-|\rho_{1,1}|}, T \right)^2  = \min\left( \frac{T}{(1-|\rho_{1,1}|)^2}, T^3 \right)$. Given the last expression of \eqref{b195}, it induces that $\left( \sum_{h=0}^T \| \tilde\Psi_{1 \bullet,h}\|^2 \right)^2 / g_{i,2}^2 = O_p(1)$. Thus, we have shown that the fourth moment of $\tilde y_{1,t}$ scaled by $g_{1,2}^{-2}$ is bounded above by a constant when $\delta=2$.

    In summary, the proof of boundedness of the fourth moment of $\tilde y_t$ is complete for $\delta=0,1,2$.
\end{proof}

\newpage

\section{Appendix: Algorithm for simulation-based inference}
\label{section7simul}

In this section, we present the Monte Carlo and bootstrap methodologies for hypothesis testing and constructing confidence intervals concerning two-stage GIR estimates.

\subsection{Monte Carlo test (LMC and MMC) and confidence interval}

In this subsection, we conduct simulation-based tests, namely the Local Monte Carlo test (LMC) and Maximized Monte Carlo test (MMC). The LMC test is an asymptotic Monte Carlo test relying on a consistent point estimator, whereas the MMC test, also an asymptotic Monte Carlo test, is predicated on a consistent set of estimators. The theoretical underpinnings of both LMC and MMC tests can be found in \cite{dufour2006monte}.

To execute these Monte Carlo tests, it is imperative to specify the distribution of the innovation process. While the most prevalent assumption is an i.i.d. Gaussian distribution, theoretically any innovation process that fully defines its distribution can be employed, such as the $t$-distribution with unknown degrees of freedom or the alpha-distribution, which accounts for heavy tails. It is noteworthy that if the distribution encompasses nuisance parameter(s), the consistent estimator(s) of those parameter(s) must be encompassed within the consistent set (or points). For the sake of computational expediency, we opt for the i.i.d. Gaussian distribution in this subsection.

The procedure for LMC test is outlined below:
\begin{enumerate}[1.]
    \item Fit the data with an unrestricted VAR($p$) model, yielding the estimates $\hat{\Phi}_i$ for $i=1,2,\cdots,p$ and $\hat{\Sigma}_u$.
    \item Choose $\delta$ and conduct two-stage estimation, yielding the two-stage estimate $\hat{\beta}_h^{LA(\delta)-2S}$ and its covariance matrix estimate $\hat{\Sigma}_{\beta,h}$. For the null hypothesis $\mathcal{H}_0: g(\beta_h)=0$, compute its Wald test statistic $\mathscr{W}_0$.
    \item For each Monte Carlo Simulation $n=1,2,\cdots,N$:
    \begin{enumerate}[a.]
        \item Generate VAR residuals $\hat{u}_{t}^*=\hat{\Sigma}_u^{1/2} \epsilon_t $ for $t=1,2,\cdots,T$, where $\epsilon_t\overset{i.i.d.}{\sim} N(0, I)$.
        \item Generate linear projection residuals, $\hat{u}_t^{(h)}=\sum_{i=0}^{h-1}\tilde\Psi_i \hat u_{t+h-i}^*$ for $t=p,p+1,\cdots,T-h$, where $\tilde \Psi_i$ is computed through the explicit formula \eqref{2.11} and $\hat{\Phi}_i$ from Step 1.
        \item Generate Monte Carlo data from the linear projection model \eqref{hregression}, using the restricted coefficient estimates $\hat{\beta}_h^{LA(\delta)-2S}$ which impose the constraints of the null hypothesis, $g(\beta_h)=0$, and $\hat{e}_{t,h}$ where $\hat{e}_{t,h}$ is the first element in $\hat{u}_t^{(h)}$.
        \item Apply the two-stage estimation method and compute the test statistics, denoted as $\mathscr{W}_b$.
    \end{enumerate}
    \item The simulated $p$-value, $\hat{p}_N[\mathscr{W}_0]$ is computed by
    \begin{align}
\hat{p}_{N}[x]=\left\{1+\sum_{n=1}^{N} \mathbb{1}\left[\mathscr{W}_{i}-x\right]\right\} /(N+1),
    \end{align}
    $\mathbb{1}[z]=1$ if $z \geqslant 0$ and $\mathbb{1}[z]=0$ if $z<0$.
    \item The null hypothesis is rejected at level $\alpha$ if the simulated $p$-value is less than $\alpha$.
\end{enumerate}

The aforementioned procedure for the LMC test mirrors the approach outlined in Section 6 of \cite{dufour2006short}. Should there be interest in constructing confidence interval for specific individual coefficient, we offer a test inversion approach. 
Initially, a numerical grid is established between two predetermined values, which serve as the boundaries for a particular coefficient. At each grid point, the null hypothesis, which states the coefficient is equal to the grid number, is tested. The confidence set encompasses those grid numbers for which the null hypothesis is not rejected at level $\alpha$. The minimum and maximum constants within that set establish the test-inversed $(1-\alpha)\%$ confidence interval.

It is noteworthy that the LMC test represents a special case of the MMC test, with the distinction lying in the treatment of nuisance parameters. Specifically, the LMC test considers consistent points for nuisance parameters, whereas the MMC test incorporates a consistent set for them. From a perspective of size control, if a hypothesis cannot be rejected using the LMC test, it will yield the same conclusion as the MMC test. Nevertheless, as elucidated by \cite{dufour2006monte}, computing the $p$-value using the MMC test may still offer valuable insights into the strength of evidence against the null hypothesis.

The MMC test entails a higher computational cost but affords more precise control over the test level. This arises because the MMC test incorporates a consistent set of nuisance parameters, unlike the LMC procedure described earlier, which relies on point estimates. The inclusion of the consistent set of nuisance parameters in the MMC test necessitates conducting the Monte Carlo test for each point in the consistent set, which can be computationally intensive, particularly when dealing with a large number of nuisance parameters. Despite this, we still present the procedure for the LMC test below for two primary reasons. Firstly, the LMC test can better regulate the test level, as point estimates may notably deviate from the true parameters, particularly in small samples. In certain tests, avoiding false rejections holds paramount importance. Secondly, with datasets featuring few nuisance parameters and ongoing advancements in computer speed, the MMC test may emerge as an attractive option.

The procedure for the LMC test is as follows:
\begin{enumerate}[1.]
    \item Formulate a restricted subset of the parameter space as the consistent set. Practically, utilize the consistent points obtained from the LMC test as the starting values, and construct a box by extending a few units on each side of the estimates, meanwhile restricting the parameters of interest according to the null hypothesis.
    \item Calculate the simulated $p$-values for each point in the consistent set, potentially by employing a grid. The maximized $p$-value corresponds to the simulated $p$-value of the MMC tests.
    \item Reject the null hypothesis at level $\alpha$ if the simulated $p$-value of the MMC tests is less than $\alpha$.
\end{enumerate}
It is reasonable to anticipate a significant increase in computational cost with the dimensionality of the parameter space. Conversely, the MMC test demonstrates superior ability to control the test level compared to the LMC test. The approach for constructing confidence intervals based on the MMC test closely resembles that of the LMC test, with the exception of substituting the LMC $p$-value with the MMC $p$-value.

\subsection{Bootstrap implementation}
In this subsection, we introduce a parametric bootstrap approach for the two-stage GIR estimates. Our implementation is grounded in the wild bootstrap algorithm, as proposed by \cite{gonccalves2004bootstrapping}, which accommodates autoregressive models with conditional heteroskedasticity.

For the construction of confidence intervals, we adopt equal-tailed percentile-$t$ intervals. The bootstrap algorithm employed for the two-stage estimates, aimed at constructing a $(1-\alpha)\%$ confidence interval, is delineated below. This algorithm closely adheres to the procedure advocated by \cite{montiel2021local}. Our inference results rely on the asymptotic variance matrix estimation formula introduced in Proposition \ref{unitheo}. The scalar parameter of interest is designated as $w'\beta_h$, where $w\in \mathbb R^{pK\times 1}$.

\begin{enumerate}[1.]
    \item Choose $\delta$ and compute the GIR estimates $\hat \beta_{h}^{LA(\delta)-2S}$, along with the covariance matrix $\hat{\Sigma}_{\beta,h}w$.
    \item Estimate the VAR($p$) model to obtain the VAR coefficient estimates and the residuals $\hat{u}_t$. (While not mandatory, it is recommended to utilize the coefficient bias-correction proposed by \cite{pope1990biases})
    \item Compute the recursive VAR-based GIR estimates, $\hat{\beta}_h^{(RC)}$, using the VAR coefficient estimates from Step 2 and the formula provided in Equation \eqref{recursive}.
    \item For each bootstrap iteration $b=1,2,\cdots,B$:
    \begin{enumerate}[(i)]
        \item Generate bootstrap residuals $\hat{u}_t^*=\epsilon_t\hat{  u}_t $, $t=1,2,\cdots,T$, where $\epsilon_t \overset{i.i.d.}{\sim}N(0,1)$.
        \item Generate bootstrap data $  y_t^*$, $t=1,2,\cdots,T$, through the VAR($p$) model with the coefficients estimates from Step 2, the simulated residual $\hat{  u}_t^*$, and the initial observations $(  y_0^*, \cdots ,   y_{-p+1}^*)$ which are a block of $p$ consecutive observations drawn randomly from the sample.
        \item Apply the two-stage estimation model and compute the GIR estimates $\hat{\beta}_{h,b}^{LA(\delta)-2S}$ and covariance matrix estimate $\hat{\Sigma}_{\beta,h,b}$.
        \item Store $\hat{T}^*_b =w'(\hat{\beta}_{h,b}^{LA(\delta)-2S} - \hat{\beta}_h^{(RC)} ) / (w'\hat{\Sigma}_{\beta,h,b}w)^{1/2}$
    \end{enumerate}
    \item Compute the $\alpha /2$ and $1 - \alpha/2$ quantiles of the $B$ draws of $\hat{T}^*_b$. Denote them by $\hat{q}_{\alpha/2}$ and $\hat{q}_{1-\alpha/2}$, respectively.
    \item The bootstrap confidence interval for $w'\hat \beta_{h}^{LA(\delta)-2S}$ is 
    \begin{align*}
        \left[w'\hat \beta_{h}^{LA(\delta)-2S} -  \hat{q}_{1-\alpha/2} 
        (w'\hat{\Sigma}_{\beta,h}w)^{1/2},w'\hat \beta_{h}^{LA(\delta)-2S}- \hat{q}_{\alpha/2}(w'\hat{\Sigma}_{\beta,h}w)^{1/2} \right].
    \end{align*}
\end{enumerate}

Note that the confidence interval derived from the aforementioned algorithm pertains to a single coefficient. However, the coverage ratio of a joint confidence band, constructed based on the $(1-\alpha)\%$ pointwise bootstrap confidence intervals, is typically less than $(1-\alpha)\%$ due to sequential testing issues—unless the estimates are perfectly correlated, as discussed generally in \cite{lutkepohl2020constructing}. 

We opt for the "sup-t" band for constructing simultaneous confidence bands (\cite{montiel2019simultaneous}). This choice is motivated by the "sup-t" band's ability to maintain the narrowest width among simultaneous confidence bands while preserving the $(1-\alpha)\%$ level, compared to several potential alternatives such as Bonferroni, Sidak, or Projection-based methods (see discussion in Section 3 of \cite{montiel2019simultaneous}).

Since the Bootstrap implementation in the aforementioned Step 4 produces a sequence of $\hat{T}_b^*$, representing the probability density function for the statistical pivot for $w'\hat{\beta}_h^{LA(\delta)-2S}$, we can derive the bootstrap distribution using the coefficient and covariance estimates from Step 1. Thus, Algorithm 2 in \cite{montiel2019simultaneous} can be implemented for constructing simultaneous confidence bands for multiple coefficients.

For example, in the case of a multi-horizon causality test where the null hypothesis involves $p$ coefficients in $\beta_h$, the simultaneous confidence band for these $p$ coefficients can be constructed as the method described. The null hypothesis of non-causality, which is these $p$ coefficients are equal to zero, can then be tested by verifying if these bands include zero or not. The null hypothesis will be rejected at level of $\alpha$ if at least one band does not include zero.
\newpage
\section{Appendix: Data source}
Here is the list of data source
\begin{itemize}
    \item CFNAI: Chicago Fed National Activity Index (CFNAI). Download from Federal Reserve Bank of Chicago website (https://www.chicagofed.org/research/data/cfnai/current-data)
    \item JNL: Jurado, Kyle, Ludvigson, Sydney C. and Ng, Serena, JLN 3-Month Ahead Macroeconomic Uncertainty [JLNUM3M], retrieved from FRED, Federal Reserve Bank of St. Louis; https://fred.stlouisfed.org/series/JLNUM3M, February 23, 2024.
    \item Unemp: The source code is: LNS14000000, from U.S. Bureau of Labor Statistics, Unemployment Rate [UNRATE], retrieved from FRED, Federal Reserve Bank of St. Louis; https://fred.stlouisfed.org/series/UNRATE, February 23, 2024.
    \item Inflation: Computed from monthly data of Consumer Price Index, $\text{inflation}_t = \text{CPI}_t / \text{CPI}_{t-1} -1$. The Consumer Price Index is from U.S. Bureau of Labor Statistics, Consumer Price Index for All Urban Consumers: All Items in U.S. City Average [CPIAUCSL], retrieved from FRED, Federal Reserve Bank of St. Louis; https://fred.stlouisfed.org/series/CPIAUCSL, February 23, 2024.
    \item FFR: Board of Governors of the Federal Reserve System (US), Federal Funds Effective Rate [FEDFUNDS], retrieved from FRED, Federal Reserve Bank of St. Louis; https://fred.stlouisfed.org/series/FEDFUNDS, February 23, 2024.
\end{itemize}

 \end{appendix}

\end{document}